\documentclass[english,dvips]{article}
\usepackage{babel,amsmath,latexsym,xspace}
\usepackage{graphicx,lscape,float,pictexwd}

\newcommand{\ma}[1]{\mbox{$#1$}}
\newcommand{\dickelinie}{\setplotsymbol ({\circle*{1.7}} [l])
   \plotsymbolspacing0.3mm }
\newcommand{\normalelinie}{\setplotsymbol ({\fiverm .})
   \plotsymbolspacing0.4pt }

\newcommand{\kreis}{
  \setplotarea x from -3 to 3, y from -3 to 3
  \circulararc 360 degrees from 0 2 center at 0 0 }

\newcommand{\tetraeder}{\setplotarea x from -3 to 3, y from -2 to 3 
                      \plot 0 2 0 0 1.73 -1 0 2 -1.73 -1 0 0 1.73 -1 -1.73 -1 /}

\newcommand{\pfeilLOhoch}
  {\arrow <1.75mm> [0.25,0.85] from -1.41 1.41 to -1.2 1.62 }
\newcommand{\pfeilLUrunter}
  {\arrow <1.75mm> [0.25,0.85] from -1.41 -1.41 to -1.2 -1.62 }

\newcommand{\pfeillMrechts}
  {\arrow <1.75mm> [0.25,0.85] from -0.66 0 to -0.36 0 }
\newcommand{\pfeilLmrechts}
  {\arrow <1.75mm> [0.25,0.85] from -1.33 0 to -1.03 0 }

\newcommand{\pfeiluIVlinks} {
    \arrow <1.75mm> [0.25,0.85] from  0.1 -1 to -0.05 -1 }

\newcommand{\pfeillVrunter} {
    \arrow <1.75mm> [0.25,0.85] from -0.61  0.95 to -0.95  0.35 }

\newcommand{\pfeilrVrunter} {
    \arrow <1.75mm> [0.25,0.85] from  0.61  0.95 to 0.95  0.35 }

\newcommand{\pfeiltlIVhoch} {
    \arrow <1.75mm> [0.25,0.85] from -0.823 -0.475 to -0.563  -0.325 }

\newcommand{\pfeiltrIVrunter} {
    \arrow <1.75mm> [0.25,0.85] from  0.563  -0.325 to 0.823 -0.475 }

\newcommand{\pfeiltoIVrunter} {
    \arrow <1.75mm> [0.25,0.85] from  0 1 to  0 0.7 }

\newcommand{\dickpfeilRUrunter}{\dickelinie
    \arrow <1.7mm> [0.25,0.85] from 1.41 -1.41 to 1.2 -1.62 
    \normalelinie}

\newcommand{\dickpfeilrMlinks}{\dickelinie
    \arrow <1.7mm> [0.25,0.85] from 0.66 0 to 0.36 0 
     \normalelinie}
\newcommand{\dickpfeilRmlinks}{\dickelinie
    \arrow <1.7mm> [0.25,0.85] from 1.33 0 to 1.03 0 
     \normalelinie}

\newcommand{\punktO}{\put{\ma\bullet} at 0 2 }
\newcommand{\punktU}{\put{\ma\bullet} at 0 -2 }

\newcommand{\punktlM}{\put{\ma\bullet} at -0.66 0 }
\newcommand{\punktrM}{\put{\ma\bullet} at 0.66 0 }

\newcommand{\strichLR}{\plot -2 0 2 0 / }

\newcommand{\eg}{e.\,g.\xspace}
\newcommand{\ie}{i.\,e.\xspace}
\newcommand\seteins{\mbox{{\small 1} \kern-0.68em 1}}
\newcommand{\tr}{\mbox{tr}}
\newcommand{\Tr}{\mbox{Tr}}
\newcommand{\eps}{\epsilon}
\newcommand{\la}{\lambda}
\newcommand{\ro}{\rho}
\newcommand{\si}{\sigma}
\newcommand{\ta}{\tau}

\newcommand{\graph}[2]{%
  \parbox[h]{#2cm}{%
    \includegraphics[width=#2cm]{#1.eps}%
  }%
}
\newcommand{\graphrot}[2]{%
  \parbox[h]{#2cm}{%
    \includegraphics[angle=-90,width=#2cm]{#1.eps}%
  }%
}

\begin{document}
\title{\bf Three-Loop Yang-Mills $\beta$-Function {\em via} 
the Covariant Background Field Method}

\author{
{\Large Jan-Peter B{\"o}rnsen}\\
\\
II. Institut f{\"u}r Theoretische Physik der Universit{\"a}t Hamburg\\ 
         Luruper Strasse 149, 22761 Hamburg, Germany\\
         email: \texttt{jan-peter.boernsen@desy.de}\\
\\
\\
{\Large Anton E.\,M. van de Ven}\\
\\
  Institute of Theoretical Physics, Utrecht University\\
  Leuvenlaan 4, 3584 CC Utrecht, The Netherlands\\ 
  email: \texttt{avdven@phys.uu.nl}
}
\maketitle
\begin{abstract}
We demonstrate the effectivity of the covariant background field method 
by means of an explicit calculation of the 3-loop $\beta$-function for a 
pure Yang-Mills theory. To maintain manifest background invariance throughout 
our calculation, we stay in coordinate space and treat the background field 
non-perturbatively. In this way the presence of a background field does not 
increase the number of vertices and leads to a relatively small number of 
vacuum graphs in the effective action. Restricting to a covariantly constant 
background field in Fock-Schwinger gauge permits explicit expansion of all 
quantum field propagators in powers of the field strength only. Hence, 
Feynman graphs are at most logarithmically divergent. At 2-loop order only a 
single Feynman graph without subdivergences needs to be calculated. At 3-loop 
order 24 graphs remain. Insisting on manifest background gauge invariance at 
all stages of a calculation is thus shown to be a major labor saving device.
All calculations were performed with {\em Mathematica} in view of its superior 
pattern matching capabilities. Finally, we describe briefly the extension of 
such covariant methods to the case of supergravity theories. 
\end{abstract}
%
%
\section{Introduction}
The essential quantity in quantum field theory is the effective action,
which for a given classical action sums up all quantum corrections and
from which the S-matrix can be found. Unfortunately, in practice this
functional can only be determined in perturbation theory. In
particular, for gauge theories the conventional effective action, in
contradistinction to the S-matrix, has the disadvantage of not being
gauge invariant. A possible solution to this problem is based on the
introduction of a background field~\cite{DW65,DW67,Ho71,tH73a}. The
resulting modified effective action is then gauge invariant with
respect to gauge transformations of the background field. This 
so-called background field method was invented to simplify theoretical
considerations and also quantum computations in gauge and gravitational
theories. Originally, the formalism was only applicable up to
\mbox{one-loop} order. The extension to higher loops was found 
in~\cite{tH75,DW81,Ab80,Bo81}. A proof of the renormalizability of
\mbox{Yang-Mills} theories in this formalism was given 
in~\cite{KZ75,LW95}, whereas~\cite{AGS83,BC99,FPQ00} demonstrate that
one obtains the same \mbox{S-matrix} as in the usual formalism. The 
\mbox{well-known} renormalization group functions of an arbitrary
nonabelian gauge theory at the \mbox{one-loop}~\cite{Po73,GW73,CEL74}
and \mbox{two-loop} level~\cite{Ca74,Jo74,MV83,MV84,MV85} were
reobtained in a simpler manner in the background field 
formalism~\cite{Ab82a,CM82,AGS83,vD82,IO82,JO82,vD84}. For calculations
in quantum gravitational theories and nonlinear sigma models, the
background field method is indispensable. By means of this method
Einstein gravity was shown to be \mbox{one-loop}~\cite{HV74} but not 
\mbox{two-loop}~\cite{GS85,GS86,Ve92} finite. Recently, the background
field method has found application in the standard model as 
well~\cite{Gr99,GHS01,GHS01b,DDW96,MPS02}.

The main purpose of this paper is to demonstrate the effectiveness of
the covariant background field method in the explicit calculation of
the three-loop \mbox{$\beta$-function} for a pure Yang-Mills theory.
The successful completion of this calculation gives us hope that the 
open issue of the three-loop renormalizability of supergravity may be
answered with similar methods. The three-loop renormalization group 
functions for QCD were first obtained with conventional field 
theoretical methods in~\cite{TVZ80,LV93}. Very recently these results 
were recovered and extended to include scalar fields and Yukawa 
couplings in~\cite{PGJ01}. Although these authors use the background 
field method, they rely upon momentum space methods which violate 
background gauge invariance at the intermediate level. Indeed, although
the background field method leads to a gauge invariant effective 
action, it is common practice to give up background field gauge 
invariance at intermediate levels of a computation~\cite{Ab80,Ab82a}. 
The reason for this can be traced to the desire to treat the 
interaction with the background field as a perturbation. One commonly 
chooses the background field covariant Feynman gauge, but in order to 
be able to define the gluon propagator and the Feynman rules one is 
led to split up each covariant derivative into an ordinary derivative 
plus a background gauge field term. As a consequence the background 
field appears explicitly in the Feynman rules and manifest background 
field gauge invariance is lost in the process. In addition this 
splitting up of covariant derivatives leads to a considerable increase
in the number of vertices as compared with their number in the absence 
of a background field. Nevertheless, a major gain is found in the fact
that it now suffices to calculate the two-point function of the 
background gauge field in order to determine the charge renormalization
and thus obtain the \mbox{$\beta$-function} with relative ease. The 
contributing Feynman graphs are at most quadratically divergent on 
dimensional grounds (see fig 1a). At the end of the calculation one 
adds up the various contributions and should obtain a transverse 
answer, i.e. proportional to the square of the background gauge field 
strength. This is usually seen as a valuable test on the correctness of
the calculations. However, the above procedure contradicts the spirit 
of the background field method and leads to unnecessary labor. This 
becomes especially important at higher loops in quantum gravitational 
theories.

\begin{figure}[h]
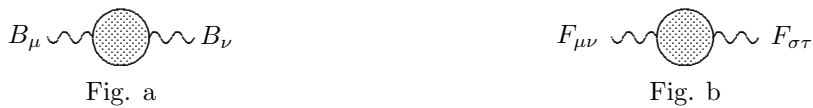

\setcoordinatesystem units <0.75 cm , 0.75 cm> point at 0 0
\centerline{
\beginpicture
\setplotarea x from -2.5 to 12.5, y from -1.5 to 1
\put {$B_\mu$} at -2.7 0
\put {$B_\nu$} at  0.7 0
\setquadratic
\plot -2.3 0 -2.2 -0.1 -2.1 0 -2.0  0.1 -1.9 0 -1.8 -0.1 -1.7 0 -1.6
       0.1 -1.5 0 /
\plot -0.5 0 -0.4 0.1  -0.3 0 -0.2 -0.1 -0.1 0  0.0  0.1  0.1 0  0.2
      -0.1  0.3 0 /
\setlinear
\circulararc 360 degrees from -0.5 0  center at -1 0
\setquadratic
\setshadegrid span <0.04cm> point at -1 0
\vshade -1.5 0 0 <,z,,> -1 -0.5 0.5 -0.5 0 0 /
\put {Fig. a} at -1.0 -1.0
\put {$F_{\mu\nu}$} at 7.1 0
\put {$F_{\sigma\tau}$} at  10.9 0
\setquadratic
\plot  7.7 0  7.8 -0.1  7.9 0  8.0  0.1  8.1 0  8.2 -0.1  8.3 0  8.4
       0.1  8.5 0 /
\plot  9.5 0  9.6 0.1   9.7 0  9.8 -0.1  9.9 0 10.0  0.1 10.1 0 10.2
      -0.1 10.3 0 /
\setlinear
\circulararc 360 degrees from 9.5 0  center at 9 0
\setquadratic
\setshadegrid span <0.04cm> point at 9 0
\vshade  8.5 0 0 <,z,,> 9 -0.5 0.5 9.5 0 0 /
\put {Fig. b} at 9.0 -1.0
\endpicture
} \caption[Comparison between the ordinary and the covariant background
           field method.]
          {Comparison between the ordinary (a) and the covariant (b)
           background field method.}
\label{Vergleich}
\end{figure}

In this paper we shall therefore insist on manifest background field
invariance and avoid the splitting up of background covariant
derivatives. This prohibits the otherwise standard transition to
momentum space and as a consequence all our calculations will be
performed in configuration space. Such a procedure has been advocated 
before in finding the finite part of the effective action for a pure 
Yang-Mills theory in a covariantly constant background field at 
\mbox{one-loop}~\cite{DR75,BMS77,Sa77} and at 
\mbox{two-loop}~\cite{BS78} order. It was also used in~\cite{VS86} to
determine the \mbox{two-loop} \mbox{$\beta$-function} for SQED in a 
very efficient way. The background field will appear only through
covariant derivatives in the gauge-fixed action and Feynman rules. The
number of vertices is thus the same as in the absence of the background
field. We shall work in the background covariant Feynman gauge and
obtain background field dependent gluon and ghost propagators via a 
heat kernel representation. Out of these exact propagators and the 
vertices we may construct a compact though formal expression for the
effective action at any particular loop order. In the chosen gauge 
there exists a Ward identity~\cite{ALN77,BV85} connecting the exact
gluon propagator $G_{\mu\nu}(x,x')$ and exact ghost propagator 
$G(x,x')$, namely
$$
D^\mu G_{\mu\nu}(x,x') + D_{\nu'} G(x,x') = 0
$$
We will use this identity to convert ghost graphs into gluon graphs in
the formal effective action and thus roughly half the number of graphs.
In order for the exact propagators to transform gauge covariantly at 
both endpoints, they contain in particular a Schwinger phasefactor
$$
\Phi(x,x')\ =\ P\,\exp\left(\int_x^{x'} dy^\mu B_\mu(y)\right) \quad ,
$$
where the symbol $P$ denotes path ordering. As this phasefactor is 
dimensionless it would seem to lead to quartically divergent graphs.
This can be avoided by choosing the so-called \mbox{Fock-Schwinger}
gauge $x^\mu B_\mu(x)=0$ (note that one is free to select different
gauges for the background and quantum fields). It is well-known that
in this gauge the gauge field can be expressed in terms of its own 
field strength. It was shown in~\cite{DS81,NSVZ84} that one may obtain in
the \mbox{Fock-Schwinger} gauge an explicit expansion of the propagator
in powers of the background field strength and its covariant 
derivatives. In fact, if one demands the background field strength to
be covariantly constant, then there exists a closed expression for the
associated heat kernel. The condition $D_\rho F_{\mu\nu}=0$ is 
analogous to Schwinger's choice of a constant electromagnetic field
strength for the electron~\cite{HE36,Sch51}. With these choices the
background field appears in the Feynman rules only through its field
strength and hence we obtain at most logarithmically divergent Feynman
graphs (see fig 1b). Here, each individual graph is background gauge 
invariant. This is a major simplification compared to the method 
described above where one finds quadratically divergent graphs and only
the sum of all graphs is gauge invariant.

To regularize the Feynman graphs we use dimensional regularization with 
(modified) minimal subtraction~\cite{HV72,BG72,As72,CM72}. Our 
renormalization procedure is based on the 
$R^\ast$-method~\cite{CT82,CS84}. This is a generalization of the 
\mbox{well-known} \mbox{$R$-method}~\cite{BP57} which not only 
eliminates all UV subdivergences, but also all \mbox{IR-divergences}
for any Feynman graph. Thus the \mbox{$R^\ast$-method} allows one to
renormalize each graph separately without introducing explicit
counterterms. It can be shown that the \mbox{$\beta$-function} is the
same in all $\overline{{\rm MS}}$ schemes based on dimensional 
regularization to {\em any} loop order~\cite{BL81}. We note that this
also holds at the level of individual renormalized graphs and this fact
constitutes a valuable check on our calculations.

All calculations in this paper were performed with 
{\em Mathematica}~\cite{MMA}. At first sight, this would not seem to be
the obvious choice as {\em Mathematica} is not optimized for speed or
very large expressions, its strength being in the domain of pattern
matching. However, with the necessary precautions, this programming
language can handle several thousand terms and finish a calculation in
a reasonable amount of time. Our manifestly covariant approach prevents
one from ever having to deal with more than a two-thousand terms and 
hence {\em Mathematica} works well.

The paper is organized as follows. In section~\ref{BFMinYM} we give a
brief introduction of the background field method applied to the 
\mbox{Yang-Mills} theory. Assuming that the field strength $F_{\mu\nu}$
is covariantly constant and that the quantum field is given in the 
Feynman gauge respectively the background field in the 
\mbox{Fock-Schwinger} gauge, the corresponding heat kernels and
propagators are deduced in section~\ref{PGs}. Pursuant to our covariant
approach the resulting Feynman rules are specified in section~\ref{FR}.
In section~\ref{YM1Loop} the \mbox{one-loop} calculation is presented.
It shows up that for determining the \mbox{UV-divergences} it suffices
to know the propagators. In section~\ref{YM2Loop}, we demonstrate that
the whole \mbox{two-loop} calculation can be done by evaluating only
one Feynman graph. Eventually we outline in section~\ref{YM3Loop} our
\mbox{three-loop} calculation done with the help of {\em Mathematica}.
Finally we give in section~\ref{CD} an outlook how \eg a 
\mbox{three-loop} calculation for supergravity may be accomplished by
using the proposed covariant background field method.

%
%
\section{The background field method in Yang-Mills
         theory}\label{BFMinYM}

We consider a nonabelian gauge theory with gauge fields
$A_\mu(x)=A^a_\mu(x)t_a$, $t_a$~being the generators of a
semisimple gauge group~$G$. The covariant derivatives 
$D_\mu =\partial_\mu + A_\mu$ transform homogeneously under a local
gauge transformation~$U(x)$, i.e.~$D_\mu \rightarrow U^{-1} D_\mu U$.
The same holds for the field strength $F_{\mu\nu}=[D_\mu,D_\nu]$ which
appears in the action 
\begin{equation}
S_{\rm cl} = -\,\frac{1}{4\,T(R)\,g^2}\int dx\ 
                \tr (F_{\mu\nu} F^{\mu\nu} )
           =  \,\frac{1}{4\,g^2}      \int dx\   
                     F^a_{\mu\nu} F_a^{\mu\nu}   \label{YMS}
\end{equation}
where $g$ is the gauge coupling constant and we normalized the
generators in the representation $R$ by 
$\tr(t_a t_b)= - T(R)\,\delta_{ab}$. We shall work in euclidean space
throughout. The equations of motion and Bianchi identities are given by
\begin{equation}
D^\mu F_{\mu\nu} = 0  \quad ,\quad  D_\rho F_{\mu\nu} + 
                   D_\mu F_{\nu\rho} + D_\nu F_{\rho\mu} = 0
\end{equation}
We split the field $A_\mu$ into a background field $B_\mu$ and a
quantum field $Q_\mu$ such that
\begin{equation}
A_\mu = B_\mu + Q_\mu
\end{equation}
This induces a similar splitting for the field strength
\begin{equation}
F_{\mu\nu}[A] = F_{\mu\nu} + D_\mu Q_\nu - D_\nu Q_\mu +[Q_\mu,Q_\nu]
\end{equation}
where on the right hand side $D_\mu$ and $F_{\mu\nu}$ represent the
background covariant derivative and field strength, respectively. From
here onward this notation will be left understood. An infinitesimal
gauge transformation of $A_\mu$ with parameter $\Lambda$ can of course
be distributed in many ways over $B_\mu$ and $Q_\mu$, but the two most
convenient choices are the ``quantum transformation"
\begin{equation}\label{QINV}
\delta B_\mu = 0{\phantom{D_\mu\Lambda}},\qquad 
\delta Q_\mu =            D_\mu\Lambda+[Q_\mu,\Lambda]\phantom{\quad .}
\end{equation}
and ``background transformation"
\begin{equation}\label{BINV}
\delta B_\mu ={\phantom 0}D_\mu\Lambda, \qquad
\delta Q_\mu ={\phantom{D_\mu\Lambda\ +}}[Q_\mu,\Lambda] \quad .
\end{equation}
The trick is to add now a gauge-fixing term which breaks the quantum
gauge invariance, but respects the background gauge invariance, \ie we
add to the classical action a term
\begin{equation}\label{SFIX}
S_{\rm fix} =\,  \frac{1}{2\alpha}\int dx\ G_a^{\ 2}
\end{equation}
where the gauge-fixing function $G$ transforms covariantly 
under~(\ref{BINV}). The associated Faddeev-Popov term is
\begin{equation}
S_{\rm FP} = 2\int dx\ \tr\, C^\ast\, 
                       \frac{\delta G}{\delta\Lambda}\, C  \quad ,
\end{equation}
where the variation of $G$ under a quantum gauge transformation is
meant and $C$ and $C^\ast$ are the ghost and antighost fields. The
total action $S_{\rm tot} =S_{\rm cl} +S_{\rm fix} +S_{\rm FP}$ then
appears in the generating functional for all Green functions
\begin{equation}
Z[B,J,\eta^\ast,\eta]\ =\ 
   N\int {\mathcal D}Q\, {\mathcal D}C^\ast\, {\mathcal D}C\,
   \exp(- S_{\rm tot} + J \cdot Q + \eta^\ast\cdot C + C^\ast\cdot\eta)
\end{equation}
with sources $J$, $\eta$ and $\eta^\ast$ and normalization factor~$N$ 
chosen such that~$Z[B,0,0,0]=1$. Note that the background field is not
coupled to the source~$J$. The generating functional for the connected
graphs is then given by
\begin{equation}\label{ZB2WB}
W[B,J,\eta^\ast,\eta] = \ln  Z[B,J,\eta^\ast,\eta]  \quad .
\end{equation}
Defining expectation values for all fields as follows
\begin{equation}
\bar Q      = \frac{\delta W}{\delta J}   \quad ,\quad
\bar C      = \frac{\delta W}{\delta \eta^\ast}   \quad ,\quad
\bar C^\ast = \frac{\delta W}{\delta \eta}  \quad ,
\end{equation}
one defines the effective action by
\begin{equation}\label{BGamma}
\Gamma[B,\bar Q,\bar C^\ast,\bar C ] = 
       J \cdot \bar Q + \eta\cdot\bar C^\ast +
       \bar C \cdot \eta^\ast -  W[B,J,\eta^\ast,\eta]
\end{equation}
The functionals $Z$ and $W$ will be invariant under background gauge
transformations, if all sources and ghost fields transform in the same
way as $Q$ does. The same holds for $\Gamma$ if one demands~$\bar Q$, 
$\bar C$ and $\bar C^\ast$ to transform as $Q$. A good gauge choice is
\eg the generalized Gervais-Neveu gauge
\begin{equation}
G = D^\mu Q_\mu + \beta\, Q^\mu Q_\mu   \quad ,
\end{equation}
but we will restrict to the background covariant Feynman gauge
$\alpha=1$,$\beta=0$ with associated Faddeev-Popov term
\begin{equation}\label{BSFP}
S_{\rm FP} \ =
  \ 2 \int dx\ \tr(C^\ast D^\mu (D_\mu C + [Q_\mu, C]) )\quad .
\end{equation}
If we now, as mentioned briefly in~\cite{JO82} decompose the current in
\begin{equation}
J_\mu \ =\ \bar J_\mu + j_\mu
\end{equation}
with
\begin{equation}
j_\mu \ =\ \left.\frac{\delta S_{tot}[B,Q]}
           {\delta Q^\mu}\right|_{Q\,=\,0}
\end{equation}
we can write the total action as
\begin{multline}
S_{tot}[B,Q] - J\cdot Q\\
= S_{cl}[B]
  +\frac{1}{2}\int dx\, \tr \left(Q^\mu \Delta_{\mu\nu}[B] Q^\nu\right)
  -\int dx \, \tr\left(C^\ast \Delta[B] C\right)\\
  \mbox{} - S_{int}[B,Q] - \bar J \cdot Q\label{EntS}
\end{multline}
with the definitions
\begin{eqnarray}
\Delta_{\mu\nu}[B] &=& -\left(\delta_{\mu\nu}D^2 +
                       2\,{\mathbf F}_{\mu\nu} \right)\\ 
\Delta[B] &=& - D^2 \label{DeltaToD2}\\
S_{int}[B,Q] &=&
  \int dx\,\tr\Bigl( \left(D_\mu Q_\nu\right) [Q^\mu,Q^\nu] +
  \frac{1}{4}[Q_\mu,Q_\nu] [Q^\mu,Q^\nu] \nonumber\\
    &&\phantom{\int dx\,\tr\Bigl(\,} + C^\ast D_\mu [Q^\mu, C]\Bigr)
\end{eqnarray}
where the field strength is given in the adjoint representation.
Following the definition of appendix~\ref{GT} this means that
$\mathbf F_{\mu\nu} Q^\nu$ corresponds to the commutator
$[F_{\mu\nu},Q^\nu]$.

Thus we get a new generating functional
\begin{eqnarray}
Z[B,\bar J,\eta^\ast,\eta] &=&
  \int{\mathcal D}Q\, {\mathcal D}C^\ast\, {\mathcal D} C\ 
  \exp\biggl[\nonumber\\
  &&\phantom{\int{\mathcal D}\,}
  - S_{cl}[B]- \frac{1}{2} \int dx\, Q^\mu\Delta_{\mu\nu}Q^\nu +
  \int dx \, C^\ast \Delta C \nonumber\\
  &&\phantom{\int{\mathcal D}\, }
   - S_{int}[B,Q] + \bar J\cdot Q +
  \eta^\ast \cdot C + C^\ast\cdot\eta \biggr]
\end{eqnarray}
depending on a new current~$\bar J$. The exponent is now free of any
term linear in the gauge field~$Q$, except the one coupling to the
current~$\bar J$.

Imposing the constraints $\bar Q = \bar C = \bar C^\ast = 0$,
the effective action of~(\ref{BGamma}) now reads
\begin{equation}\label{BGamma0}
 \Gamma[B,0,0,0]
\ =\ \left. - W[B,\bar J,\eta^\ast,\eta]\right|_{\delta
      W/\delta \bar J \,=\, \delta
      W/\delta \eta^\ast\,=\,\delta
      W/\delta \eta \,=\, 0}
\end{equation}
and is still invariant regarding the gauge transformations
of the background field given in~(\ref{BINV}). The
pertubative expansion of this effective action~(\ref{BGamma0})
contains only vacuum graphs, which are of course 1PI.
Thus, the number of emerging Feynman graphs is
drastically reduced as we will show later on.

The expansion of the effective action~(\ref{BGamma0}) in
terms of $\hbar$ is given by
\begin{equation} \label{G2S}
 \Gamma[B,0,0,0]
\ =\  S_{cl}[B]+ \left(\frac{1}{2}\ln\det\Delta_{\mu\nu}[B] -
                    \ln\det\Delta[B]\right) \hbar + O(\hbar^2)\quad.
\end{equation}
Hence the effective action~$\Gamma[B,0,0,0]$ equals the classical
action $S_{cl}[B]$, if we neglect all quantum effects by using the
approximation $\hbar \rightarrow 0$. To calculate the correction
linear in $\hbar$ which corresponds to the ordinary \mbox{one-loop}
calculation, we need to evaluate the logarithm of the determinant
of the operators $\Delta_{\mu\nu}$ and $\Delta$.
Higher corrections in $\hbar$ are represented by Feynman graphs,
which are built according to the Feynman rules of section~\ref{FR}.

As expected while expanding the new effective
action~$\tilde \Gamma[B,0,0,0]$ we meet the problem of
\mbox{UV-divergences}. Following the prescriptions of the
background field theory it suffices to redefine the
background field $B$ and the coupling constant $g$ by
\begin{equation}\label{g02g}
\begin{array}{rcccl}
B^\mu & \rightarrow & B^\mu_0 &=& Z_B^{1/2}\, B^\mu \\
g     & \rightarrow & g_0^2   &=& Z_g\, g^2\, \mu^\eps \quad .
\end{array}
\end{equation}

The quantum field $Q$ and the ghost fields~$c^\ast$
and $c$ however need not to be renormalized, since they
can only be found inside of the vacuum graphs and
as a result the renormalization constants of the vertices
would be cancelled with those of the propagators~\cite{Ab80}.

The renormalized field strength is given by
\begin{equation}
F_{\mu\nu} \ =\ Z_B^{1/2} \left(\partial_\mu B_\nu - 
                                \partial_\nu B_\mu +
                Z_g^{1/2} Z_B^{1/2} \left[B_\mu,B_\nu \right]\right),
\end{equation}
and has to be invariant in terms of the background field
transformation~(\ref{BINV}). Thus the two renormalization constants
have to satisfy the identity
\begin{equation} \label{Zg2ZB}
Z_g^{1/2} Z_B^{1/2} \ =\ 1\, .
\end{equation}

As already mentioned in~\cite{Ab82a} this is the reason why
we have only to evaluate the \mbox{two-point} functions of the 
background field $B_\mu$ in order to renormalize the Yang-Mills theory.
Preserving the covariance, as done in the presented calculations,
leads directly to \mbox{two-point} functions of the field strength
and has moreover the advantage that, in comparison to the common
procedure, the number of vertices does not increase.
The corresponding Feynman rules, given in section~\ref{FR},
contain two \mbox{3-vertices} and one \mbox{4-vertex}, whereas the
Feynman rules of~\cite{Ab82a} include four \mbox{3-vertices} and five
\mbox{4-vertices}. Hence the preservation of the covariance of the 
background field reduces the number of vertices and consequently the 
number of Feynman graphs.

In the case of the \mbox{two-loop} calculation the number of Feynman
graphs can be reduced once more from twelve in~\cite{Ab82a} to one
as shown in section~\ref{YM2Loop}.

A nice proof of the renormalizability of the Yang-Mills theory
in the background field formalism can be found in~\cite{LW95}.
Based on the fact that for a vanishing background field we
finally get the original Yang-Mills theory whose renormalizability
has already been proven, the authors L\"uscher and Weisz show,
by using the BRS-, the background field- and a \mbox{shift-symmetry},
that the \mbox{Yang-Mills} theory written in the background field
formalism is renormalizable as well.

%
%
\section{The exact propagators}\label{PGs}
For a given operator $\Delta$ the corresponding Green function
or propagator in~$d$~dimensions is defined by the wave equation
\begin{equation}
\Delta G(x,y) = \delta (x-y). \label{greens}
\end{equation}
The associated heat kernel $K$ satisfies the heat kernel
equation
\begin{equation}
\left(\frac{\partial }{ \partial \tau}+ \Delta\right)
  K(x,y;\tau)=0 \label{wlg}
\end{equation}
with the boundary condition
\begin{equation}
K(x,y;0) =\delta (x-y). \label{wlgab}
\end{equation}
The Green function $G$ is connected to its heat kernel $K$
via the equation
\begin{equation}
G(x,y) = \int_0^\infty d\tau\, K(x,y;\tau) \label{K2G}
\end{equation}
which therefore satisfies (\ref{greens}).

For the ordinary d'Alembertian $\Delta_0 = -\partial^2$ it is
straightforward to verify that
\begin{equation}\label{ymwk0}
K_0(x,y;\tau)
\ =\ \frac{1}{(4\pi\tau)^{d/2}}
  \exp\left(-\frac{1}{4\tau}(x-y)^2\right),
\end{equation}
which depends only on the coordinate
difference $x - y$, solves~(\ref{wlg}).
Thus, generalizing this result, one makes
for the heat kernel associated to the scalar operator~$\Delta$
of equation~(\ref{DeltaToD2}) the following ansatz
\begin{equation}\label{ymwk}
K(x,y;\tau)
\ =\ \frac{1}{(4\pi\tau)^{d/2}}\Phi(x,y){\mathbf L}(\tau)
  \exp\left(-\frac{1}{4\tau}(x-y)^\mu {\bf M}_{\mu\nu}(\tau) (x-y)^\nu
  \right),
\end{equation}
in which the phasefactor $\Phi(x,y)$ is defined by
\begin{equation}\label{phf}
\Phi(x,y) = P \exp\left(\int_x^y B^\mu(z)dz_\mu\right)\quad .
\end{equation}
Imposing the constraint
\begin{equation}\label{ymNb}
(D_\si F_{\mu\nu})\ \equiv\ [D_\si,F_{\mu\nu}]\ =\ 0
\end{equation}
implies
\begin{equation}\label{Fkom}
[F_{\rho\sigma},F_{\mu\nu}] = 0.
\end{equation}
Furthermore, Shore shows in~\cite{Sh81}, that the
constraint~(\ref{ymNb}) implies the following identities
\begin{eqnarray}
\Phi(x,y) F_{\mu\nu}(y) &=& F_{\mu\nu}(x) \Phi(x,y)\label{Phi1}\\
D_\mu(x) \Phi(x,y) &=& -\frac{1}{2} F_{\mu\nu}(x)(x-y)^\nu \Phi(x,y)
\label{Phi2a}\\
\Phi(x,y) \overleftarrow D_\mu(y) &=& 
 - \frac{1}{2} \Phi(x,y) F_{\mu\nu}(y)(x-y)^\nu\, .
\label{Phi2b}
\end{eqnarray}

Inserting the ansatz~(\ref{ymwk}) into the heat kernel 
equation~(\ref{wlg}) and taking into account the 
condition~(\ref{wlgab}) and the constraint~(\ref{ymNb}),
one obtains as solution
\begin{eqnarray}
K(x,y;\tau)
&=&  \frac{1}{(4\pi\tau)^{d/2}}\Phi(x,y)\,
     \left({\Tr}\,\frac{{\mathbf F}\tau}
     {\sinh{\mathbf F}\tau}\right)^{1/2}\nonumber\\
& &  \exp\left(-\frac{1}{4\tau}(x-y)^\mu {\mathbf F}_{\mu\sigma}\tau
    \left(\coth{\mathbf F}\tau\right)^{\sigma}_{\ \nu} (x-y)^\nu 
    \right) \label{Ksk}
\end{eqnarray}
in accordance with~\cite{BD75}, where $\Tr$ denotes the trace
over the Lorentz indices. The fraction following the $\Tr$ is
a formal expression, which is defined by its series expansion
\begin{equation}
\left(\frac{{\mathbf F}\tau}{\sinh{\mathbf F}\tau}\right)_{\mu\nu}
\ =\ \seteins\delta_{\mu\nu} -
     \sum_{k=1}^{\infty}\frac{2\left(2^{2k-1}-1\right)}{(2k)!}B_{2k}
     {\mathbf F}^{2k}_{\mu\nu} \ta^{2k},
\end{equation}
where ${\mathbf F}^{2k}_{\mu\nu}$ is a short notation for the
expression ${\mathbf F}_{\mu}^{\ \si_1}{\mathbf F}_{\si_1}^{\ \si_2}
\cdots{\mathbf F}_{\si_{2k-1}\nu}$. A detailed explanation of our 
notation is given in appendix~\ref{GT}.

Due to the identities~(\ref{Fkom}) and~(\ref{Phi1}) the
heat kernel of the vector operator $\Delta_{\mu\nu}$ can be
deduced from the heat kernel of the scalar operator $\Delta$.
To satisfy the heat equation
\begin{equation*}
\left(\delta_{\mu\lambda}\frac{\partial}{\partial \tau} -
      \delta_{\mu\lambda}D^2 -
      2{\bf F}_{\mu\lambda}\right) K^\lambda_{\ \nu}  = 0
\end{equation*}
and the corresponding boundary condition
\begin{equation*}
K_{\mu\nu}(x,y;0)\ =\ \seteins\delta_{\mu\nu} \delta(x-y)
\end{equation*}
it is sufficient to extend the heat kernel of $\Delta$
by an exponential factor. In detail we get
\begin{eqnarray}
K_{\mu\nu}(x,y;\tau)
&=& \left(\exp{2{\bf F}(x)\tau}\right)_{\mu\nu} K(x,y;\tau)\nonumber\\
&=& K(x,y;\tau) \left(\exp{2{\bf F}(y)\tau}\right)_{\mu\nu}, 
    \label{Kvk}
\end{eqnarray}
where, depending on whether we multiply the additional
exponential factor from the left or from the right, the
field strength in the exponential factor depends on~$x$ or~$y$.

The two heat kernels are connected by the following identity
\begin{equation}
D^\mu K_{\mu\nu} + K \overleftarrow D_\nu \ =\ 0, \label{WI}
\end{equation}
sometimes called Ward identity \cite{ALN77}. This identity holds even
without using the constraint (\ref{ymNb}) as shown in
\cite{BV85,NN88}.

Now, by using (\ref{K2G}), we can determine the
propagators corresponding to these heat kernels.
They satisfy the following heat equations
\begin{eqnarray}\label{Wgls}
\Delta G(x,y) &=& \seteins\delta(x-y)\\
\Delta_{\mu\lambda} G^\lambda_{\ \nu}(x,y) &=&
\seteins \delta_{\mu\nu} \delta(x-y)
\end{eqnarray}
and are connected by the identity
\begin{equation} \label{PgI}
D^\mu G_{\mu\nu} + G \overleftarrow D_\nu \ =\ 0
\end{equation}
which can be deduced from (\ref{WI}) by integrating
over $\tau$. This identity will be of great help,
since all ghost graphs contain scalar propagators
with a covariant derivative acting on it. By
transforming these into vector propagators the number
of graphs can be reduced about~50\%.

Applying further a covariant derivative from the right
to the identity (\ref{PgI}), we obtain
\begin{equation}\label{DGD2d}
D^\mu G_{\mu\nu}(x,y) \overleftarrow D^\nu
\ =\ -G(x,y)\overleftarrow D^2
\ =\ -D^2 G(x,y)
\ =\ \seteins\delta(x-y),
\end{equation}
which is another helpful identity, since it leads
to the cancellation of an exact ghost propagator.

Unfortunately, it is not possible to find
a closed representation for the propagators.
But as long as we are only interested in the
renormalization of the pure Yang-Mills theory
it is quite enough to take the expansion up to
quadratic order in the field strength. For
convenience we set $z=x-y$ and get the following
series expansion for the scalar propagator
\begin{eqnarray}\label{Skalarpropagator}
G(x,y)
&=& \int_0^\infty d\tau \, K(x,y;\tau)\nonumber\\
&=& \Phi(x,y)\,\int_0^\infty d\tau \, K_0(z;\tau)
    \left(1-\frac{\tau}{12}\, z^\mu {\mathbf F}_{\mu\nu}^2 z^\nu
    - \frac{\tau^2}{12}\,{\mathbf F}^2 \right)+O({\mathbf F}^3)
    \nonumber\\
&=& \Phi(x,y)\, G_0(z)
    - \frac{1}{12}\, z^\mu {\mathbf F}_{\mu\nu}^2 z^\nu\, G_1(z)
    - \frac{1}{12}\,{\mathbf F}^2 G_2(z) + O({\mathbf F}^3),
\end{eqnarray}
where $G_0$ represents the free propagator in coordinate space.
By convoluting $i$ functions $G_0$ one after another we get
the function $G_i$ multiplied by a factor $i!$ (see 
appendix~\ref{IGR}). The vector propagator is given by
\begin{eqnarray} \label{Vektorpropagator}
G_{\mu\nu}(x,y)
&=& \int_0^\infty d\tau \, K(x,y;\tau)
    \left(\exp{2{\bf F}\tau}\right)_{\mu\nu}\nonumber\\
&=& \int_0^\infty d\tau \, K(x,y;\tau)
    \left(\delta_{\mu\nu} + 2\,{\mathbf F}_{\mu\nu}\tau +
    2\,{\mathbf F}^2_{\mu\nu}\tau^2\right) + O({\mathbf F}^3)
    \nonumber\\
&=& \int_0^\infty d\tau \bigl[K(x,y;\tau)\delta_{\mu\nu} \nonumber\\
&&  \phantom{\int_0^\infty d\tau \bigl[}
    + K_0(z;\tau)\left(2\,\Phi(x,y) {\mathbf F}_{\mu\nu}\tau +
    2\,{\mathbf F}^2_{\mu\nu}\tau^2\right)\bigr] + O({\mathbf F}^3)
    \nonumber\\
&=& \delta_{\mu\nu} G(x,y) + 2\,\Phi(x,y){\mathbf F}_{\mu\nu} G_1(z)
    + 2\,{\mathbf F}_{\mu\nu}^2 G_2(z) + O({\mathbf F}^3)\ .
\end{eqnarray}
The phasefactor $\Phi$ depends on the background field $B$ and for this
reason depends on the field strength $\mathbf F$. Hence the phasefactor
can be set to one in all terms containing the field strength in 
quadratic order.

Using the identities in appendix~\ref{IGR} the explicit coordinate
difference can be replaced by derivatives of the Green functions. 
Doing so, the series expansion of the scalar propagator is given by
\begin{eqnarray}\label{ESG}
G(x,y)&=&\phantom{-} 
         \Phi(x,y)\,G_0(x-y) - \frac1{4}{\bf F}^2 G_2(x-y) \nonumber\\
      & &\mbox{} - \frac1{3}{\bf F}^2_{\sigma\tau} 
         \partial^\sigma\partial^\tau G_3(x-y) + O({\bf F}^3)\, ,
\end{eqnarray}
respectively in graphical notation by
\begin{eqnarray}\label{graphicalESG}
\hspace{-1.8cm}\raisebox{0.08cm}{
\graph{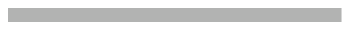}{3.5}}\hspace{-0.5cm}
& = & \phantom{-} 
      \Phi(x,y)\,\raisebox{0.02cm}{\rule[0.45mm]{1.6cm}{0.4pt}}
\hspace{-1.62cm} {\scriptstyle{\bullet}}
\hspace{1.4cm}   {\scriptstyle{\bullet}} - 
\frac{1}{4} {\bf F}^2  2 \,
\raisebox{0.02cm}{\rule[0.45mm]{1.6cm}{0.4pt}}
\hspace{-1.62cm} {\scriptstyle{\bullet}}
\hspace{ 0.36 cm} {\scriptstyle{\bullet}}
\hspace{ 0.36 cm} {\scriptstyle{\bullet}}
\hspace{ 0.36 cm} {\scriptstyle{\bullet}} \nonumber\\
&&\mbox{} - \frac1{3}{\bf F}^2_{\sigma\tau} 6 \ 
\raisebox{0.02cm}{\rule[0.45mm]{2.3cm}{0.4pt}}
\hspace{-2.40cm} {\scriptstyle{\bullet}}
\hspace{ 0.15cm}\stackrel{\sigma}{\scriptstyle{>}}
\hspace{ 0.15cm}\stackrel{\tau}{\scriptstyle{\rhd}}
\hspace{ 0.17cm}{\scriptstyle{\bullet}}
\hspace{ 0.17cm}{\scriptstyle{\bullet}}
\hspace{ 0.17cm}{\scriptstyle{\bullet}}
\hspace{ 0.17cm} {\scriptstyle{\bullet}}
 + O({\bf F}^3)\,.
\end{eqnarray}

The series expansion of the vector propagator
is given by
\begin{eqnarray} \label{EVG}
G_{\mu\nu}(x,y)&=&
\phantom{-} \Phi(x,y)\,\delta_{\mu\nu} \, G_0(x-y)
+ 2\, \Phi(x,y)\,{\bf F}_{\mu\nu} \, G_1(x-y) \nonumber\\
&&\mbox{} + \frac{1}{4}\left(8 {\bf F}^2_{\mu\nu} - {\bf F}^2 
  \delta_{\mu\nu}\right) G_2(x-y) \nonumber\\
&&\mbox{} - \frac1{3} {\bf F}^2_{\sigma\tau}\delta_{\mu\nu}
  \partial^\sigma\partial^\tau G_3(x-y) + O({\bf F}^3)\, ,
\end{eqnarray}  
whereas the graphical notation is of the shape 
\begin{eqnarray} \label{graphicalEVG}
\hspace{-2cm}\raisebox{0.08cm}{\graph{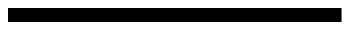}{4}}
\raisebox{-0.1cm}{$\hspace{-2.0cm}\scriptstyle{\mu}
\hspace{1.0cm}\scriptstyle{\nu}$}
\hspace{0.2cm} & = &
\phantom{-} \Phi(x,y)\,\delta_{\mu\nu}\, \raisebox{0.02cm}
{\rule[0.45mm]{1.4cm}{0.4pt}}
\hspace{-1.42cm}{\scriptstyle{\bullet}}
\hspace{ 1.25cm}{\scriptstyle{\bullet}} + 
2\, \Phi(x,y)\,{\bf F}_{\mu\nu}\,\raisebox{0.02cm}
{\rule[0.4mm]{1.5cm}{0.4pt}}
\hspace{-1.54cm}{\scriptstyle{\bullet}}
\hspace{ 0.62cm}{\scriptstyle{\bullet}}
\hspace{ 0.62cm}{\scriptstyle{\bullet}} \nonumber\\
&&\mbox{} + \frac{1}{4}\left(8 {\bf F}^2_{\mu\nu} - {\bf F}^2 
\delta_{\mu\nu}\right) 2 \,
\raisebox{0.02cm}{\rule[0.45mm]{1.6cm}{0.4pt}}
\hspace{-1.62cm}{\scriptstyle{\bullet}}
\hspace{ 0.36cm}{\scriptstyle{\bullet}}
\hspace{ 0.36cm}{\scriptstyle{\bullet}}
\hspace{ 0.36cm}{\scriptstyle{\bullet}} \nonumber\\
&&\mbox{} - \frac1{3} {\bf F}^2_{\sigma\tau}\delta_{\mu\nu} 6 \ 
\raisebox{0.02cm}{\rule[0.45mm]{2.3cm}{0.4pt}}
\hspace{-2.40cm} {\scriptstyle{\bullet}}
\hspace{ 0.15cm}\stackrel{\sigma}{\scriptstyle{>}}
\hspace{ 0.15cm}\stackrel{\tau}{\scriptstyle{\rhd}}
\hspace{ 0.17cm}{\scriptstyle{\bullet}}
\hspace{ 0.17cm}{\scriptstyle{\bullet}}
\hspace{ 0.17cm}{\scriptstyle{\bullet}}
\hspace{ 0.17cm} {\scriptstyle{\bullet}} + 
O({\bf F}^3)\, .
\end{eqnarray}

As shown above the graphical representation of the exact scalar 
propagator is illustrated by a thick gray line and the exact vector
propagator by a thick black line. Unfortunately the remaining
phasefactors prevent our expansions of the propagators to be 
translation invariant. If, however, we use a procedure given 
in~\cite{NSVZ84} we can restore the translation invariance for the
whole Feynman graph as shown in chapter~\ref{YM2Loop} 
and~\ref{YM3Loop}.

While analyzing the expansions (\ref{ESG}) and (\ref{EVG}) of the 
propagators, we realize that any possible divergences must reside
in the \mbox{Green functions}. Following the argumentation of 
appendix~\ref{IGR} the functions $G_i$ contain no 
\mbox{UV-divergences} but for $d=4-\eps$ the functions $G_i$ with
$i\geq 1$ have to be replaced by the functions $R_i$ in order to 
cancel the \mbox{IR-divergences}. As result in $4$ dimensions the 
function $R_1$ is the only one which harbors an IR pole with
constant residue. Therefore on the diagonal the scalar propagator
is given by
\begin{equation}
G(x,x)\ =\ 0
\end{equation}
whereas the vector propagator satisfies on the diagonal the equation
\begin{equation}\label{Gmn2Gnm}
G_{\mu\nu}(x,x)\ =\ -G_{\nu\mu}(x,x)\ =\ 
  \frac{1}{4\pi^2 \eps}{\mathbf F}_{\mu\nu}\, ,
\end{equation}
which will be needed to deduce the value of the tadpole defined in 
section~{\ref{YM2Loop}}.

After imposing the Feynman gauge ($\alpha=1$) and the
constraint~(\ref{ymNb}) we are still free to choose a gauge for the
background field $B_\mu(x)$. Instead of using the Feynman gauge
we select the \mbox{Fock-Schwinger} gauge \cite{Fo37,Sc73}
\begin{equation}
x^\mu B^a_\mu(x)=0\label{FSE}.
\end{equation}
Later on we will see that this gauge allows us
to transform every generated Feynman graph into logarithmic
divergent Feynman graphs. As a result the divergences of
these graphs may be represented by a Laurent series in coordinate
space containing no derivatives, respectively in momentum space
containing no dependence on any momentum.

As an aside, we note that by applying a Fourier transformation to
the Fock-Schwinger gauge we are lead to
\begin{equation*}
\partial^\mu \tilde B^a_\mu(p)=0\quad.
\end{equation*}
In this sense the \mbox{Fock-Schwinger} gauge in coordinate space and
the Lorentz gauge in momentum space are dual to each other.

As a result of using the \mbox{Fock-Schwinger} gauge, the background
field can be represented as a function of its own field
strength~\cite{DS81,NSVZ84}. To prove this, we take the derivative of
the \mbox{Fock-Schwinger} gauge (\ref{FSE})
\begin{equation*}
B_\nu(x)+x^\mu\partial_\nu B_\mu(x)= 0\quad
\end{equation*}
and replace afterwards $\partial_\nu B_\mu$ by
$-F_{\mu\nu}+\partial_\mu B_\nu+[B_\mu,B_\nu]$.
Since by definition of the \mbox{Fock-Schwinger} gauge
the expression $x^\mu[B_\nu(x),B_\mu(x)]$ vanishes,
we finally get
\begin{equation*}
(1+x^\mu\partial_\mu)B_\nu(x) = x^\mu F_{\mu\nu}(x).
\end{equation*}
Using the transformation $x=\lambda x$ and
the identity $\lambda \partial/\partial\lambda f(\lambda x)=
x^\mu\partial_\mu f(\lambda x)$ we are led to
\begin{equation*}
\left( 1 + \lambda \frac{\partial}{\partial\lambda} \right)
    B_\nu(\lambda x)
\ =\ \lambda x^\mu F_{\mu\nu}(\lambda x)
\end{equation*}
with solution
\begin{equation}\label{B2F}
    B_\nu(x)\ =\ \int_0^1\,d\lambda\,
    \lambda x^\mu F_{\mu\nu}(\lambda x)
\end{equation}
Since on the one hand the background field in the
\mbox{Fock-Schwinger} gauge is defined by its own field strength
and on the other hand the commutator of two field strength
tensors vanishes, the constraint (\ref{ymNb})
reduces to
\begin{equation*}
\partial_\rho F_{\mu\nu}=0\, ,
\end{equation*}
which means that the field strength is constant.
Finally this allows us to perform the integration in
equation~(\ref{B2F}) and we get a potential which depends
linearly on the field strength
\begin{equation}
B_\nu(x)=-\frac{1}{2}x^\mu F_{\mu\nu},
\end{equation}
from which we deduce the following phasefactor
\begin{equation}\label{PFSE}
\Phi(x,y)= \exp\left(\frac{1}{2} x^\mu F_{\mu\nu} y^\nu\right)\, ,
\end{equation}
which does of course satisfy the equations
(\ref{Phi1}-\ref{Phi2b}). Later on we will see
that this special form of the phasefactor is an
important requirement for the transformation of
all occuring Feynman graphs into graphs which are
at most logarithmically divergent.

%
%
\section{The vertices} \label{FR}
After the transformation $Q \longrightarrow g\,Q$
the Feynman rules for the Yang-Mills theory are given
by the following expressions.

The triple gluon vertex can be written as
\begin{eqnarray}\label{DV1}
    g\int dx\, &\bigl(&
      \mbox{}\phantom{-}\ f^{bcm} \delta_{\ro\ta} \delta(x-x_2) 
    \delta(x-x_3) D^{ma}_{\si} \delta(x-x_1) \nonumber\\
    &&\mbox{} - f^{bcm} \delta_{\ro\si} \delta(x-x_2) \delta(x-x_3) 
      D^{ma}_{\ta} \delta(x-x_1) \nonumber\\
    &&\mbox{} + f^{acm} \delta_{\si\ta} \delta(x-x_1) \delta(x-x_3) 
      D^{mb}_{\ro} \delta(x-x_2) \nonumber\\
    &&\mbox{} - f^{acm} \delta_{\ro\si} \delta(x-x_1) \delta(x-x_3) 
      D^{mb}_{\ta} \delta(x-x_2) \nonumber\\
    &&\mbox{} + f^{abm} \delta_{\si\ta} \delta(x-x_1) \delta(x-x_2) 
      D^{mc}_{\ro} \delta(x-x_3) \nonumber\\
    &&\mbox{} - f^{abm} \delta_{\ro\ta} \delta(x-x_1) \delta(x-x_2) 
      D^{mc}_{\si} \delta(x-x_3)
    \ \bigr)\nonumber\\
    &=&\mbox{}\phantom{-}\ \raisebox{0cm}{\hspace{-0.5cm}$$}
      \graph{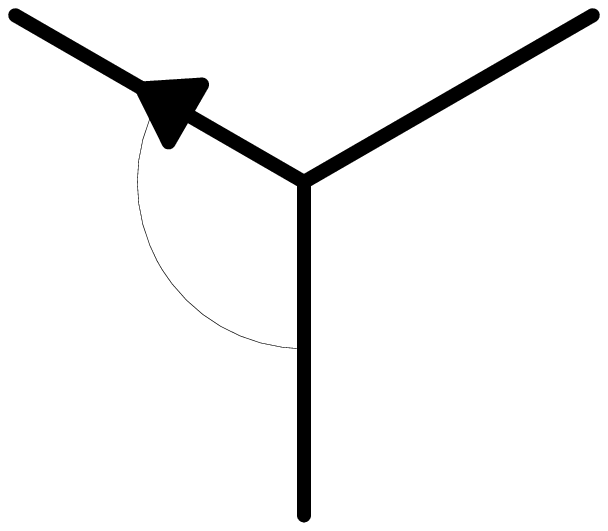}{3.0}
                           \raisebox{0.8cm}{\hspace{-2.6cm}$\rho,a$}
                           \raisebox{0.8cm}{\hspace{1.1cm}$\sigma,b$}
                           \raisebox{-1.2cm}{\hspace{-1.5cm}$\tau,c$}
                           \raisebox{0cm}{\hspace{0.8cm}$-$}
                           \raisebox{0cm}{\hspace{-0.5cm}$$}
      \graph{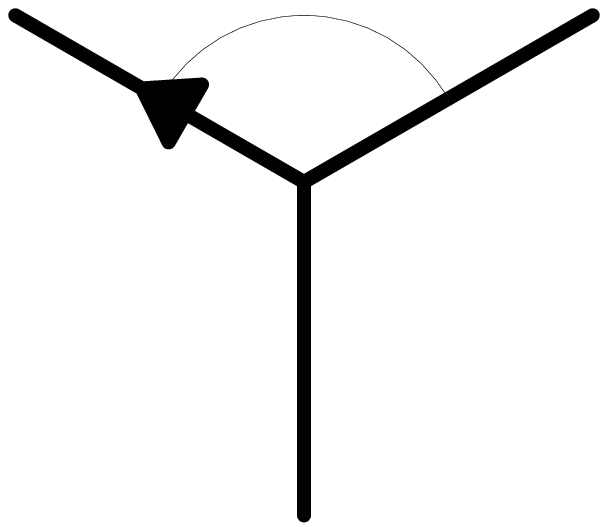}{3.0}
                           \raisebox{0cm}{\hspace{-0.5cm}$+$}
                           \raisebox{0cm}{\hspace{-0.5cm}$$}
      \graph{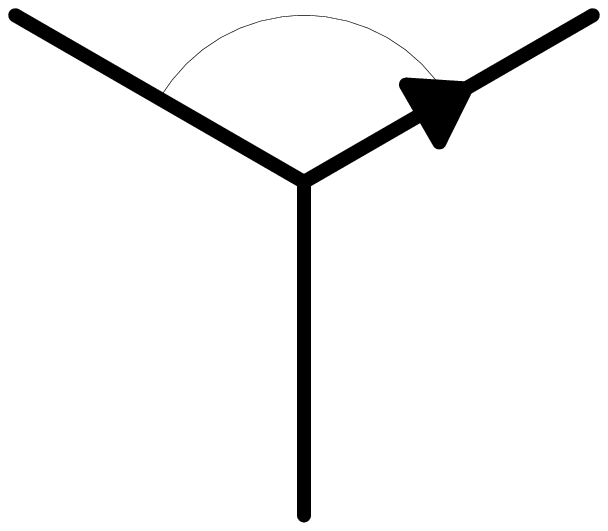}{3.0}
                           \nonumber\\
    &&\mbox{} - \raisebox{0cm}{\hspace{-0.5cm}$$}
      \graph{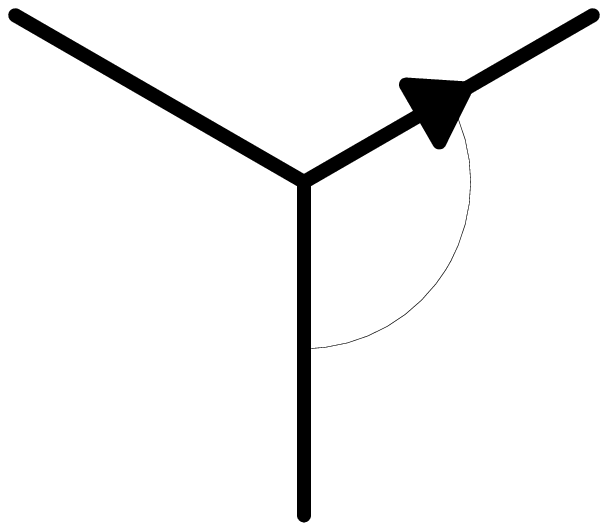}{3.0}
                           \raisebox{0cm}{\hspace{-0.5cm}$+$}
                           \raisebox{0cm}{\hspace{-0.5cm}$$}
      \graph{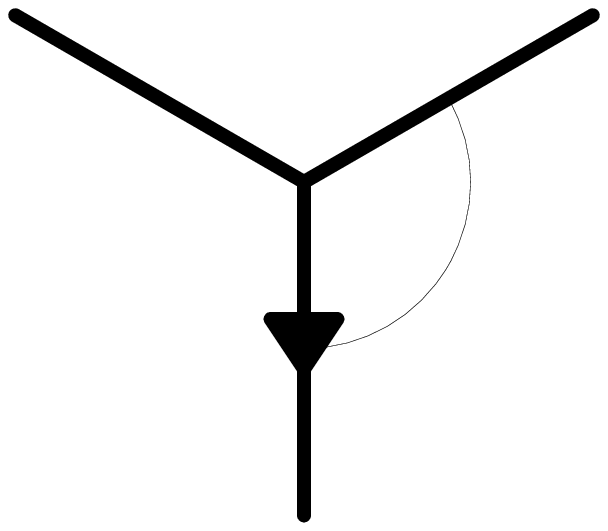}{3.0}
                           \raisebox{0cm}{\hspace{-0.5cm}$-$}
                           \raisebox{0cm}{\hspace{-0.5cm}$$}
      \graph{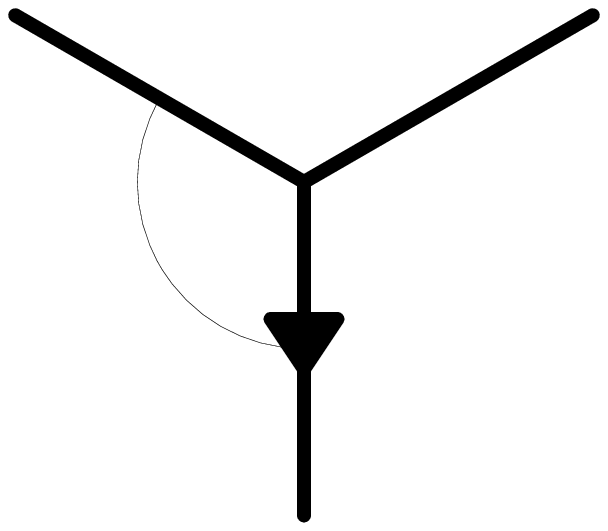}{3.0}
                           \raisebox{0cm}{\hspace{-0.5cm}$$},
\end{eqnarray}

The arrow indicates a covariant derivative operating on the line
on which it is drawn and the circular arc denotes the contraction
of its Lorentz index. As the lorentzian indices of the two
remaining propagators are always contracted the second circular
arc can be omitted.

Concerning the triple vertex containing two ghost fields, we have
to extend our graphical representation
\begin{eqnarray}\label{DV2}
 &&  g \int dx\, f^{bcm} \delta(x-x_2) \delta(x-x_3) D^{ma}_{\mu} 
                         \delta(x-x_1)\nonumber\\
&&   = \qquad            \graph{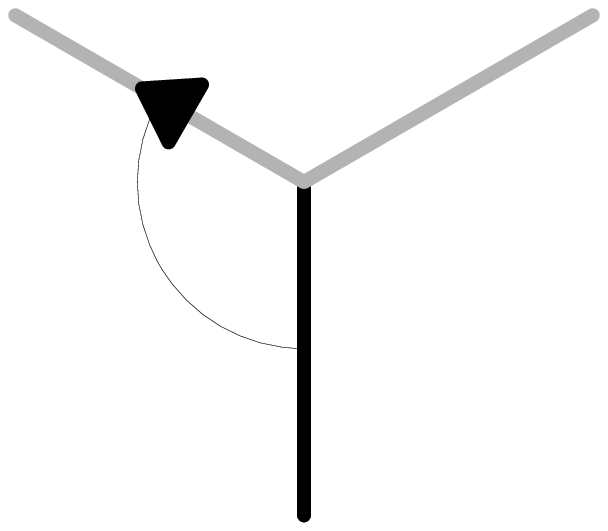}{3}
                         \raisebox{0.75cm}{\hspace{-2.4cm}$a$}
                         \raisebox{0.75cm}{\hspace{1.4cm}$b$}
                         \raisebox{-1.2cm}{\hspace{-1.1cm}$\mu,c$}
                         \raisebox{0cm}{\hspace{1cm}$$}\quad,
\end{eqnarray}
by an additional rule which says, that only those graphs are taken
into account where each gray line with a derivative is connected
to a gray line without.

The quadruple vertex is defined by
\begin{eqnarray}\label{VV}
  g^2&\int& dx\, \delta(x-x_1) \delta(x-x_2) \delta(x-x_3)
                 \delta(x-x_4)\times\nonumber\\
  &\bigl(&\mbox{}\phantom{+}\  
           f_{abm}f_{mcd}\left(\delta_{\mu\sigma}\delta_{\nu\tau}-
                               \delta_{\mu\tau}\delta_{\nu\sigma}
                               \right)+
           f_{acm}f_{mbd}\left(\delta_{\mu\nu}\delta_{\sigma\tau}-
                               \delta_{\mu\tau}\delta_{\nu\sigma}
                               \right) \nonumber\\
  &&  \mbox{}\! + 
           f_{adm}f_{mcb}\left(\delta_{\mu\sigma}\delta_{\nu\tau}-
                               \delta_{\mu\nu}\delta_{\sigma\tau}
                               \right)\ \ \bigr)\nonumber\\
                               \nonumber\\
  &=& \graph{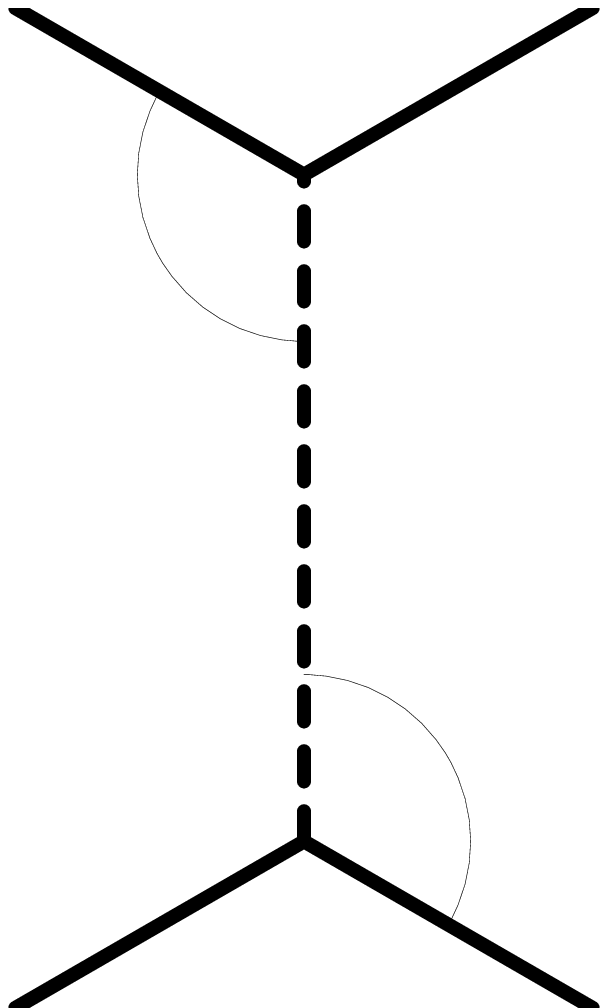}{2.5}\raisebox{1.5cm}{\hspace{-2.3cm}$\mu,a$}
                            \raisebox{1.5cm}{\hspace{1cm}$\nu,b$}
                            \raisebox{-1.5cm}{\hspace{-2.2cm}$\tau,d$}
                            \raisebox{-1.5cm}{\hspace{1cm}$\sigma,c$}-
      \graph{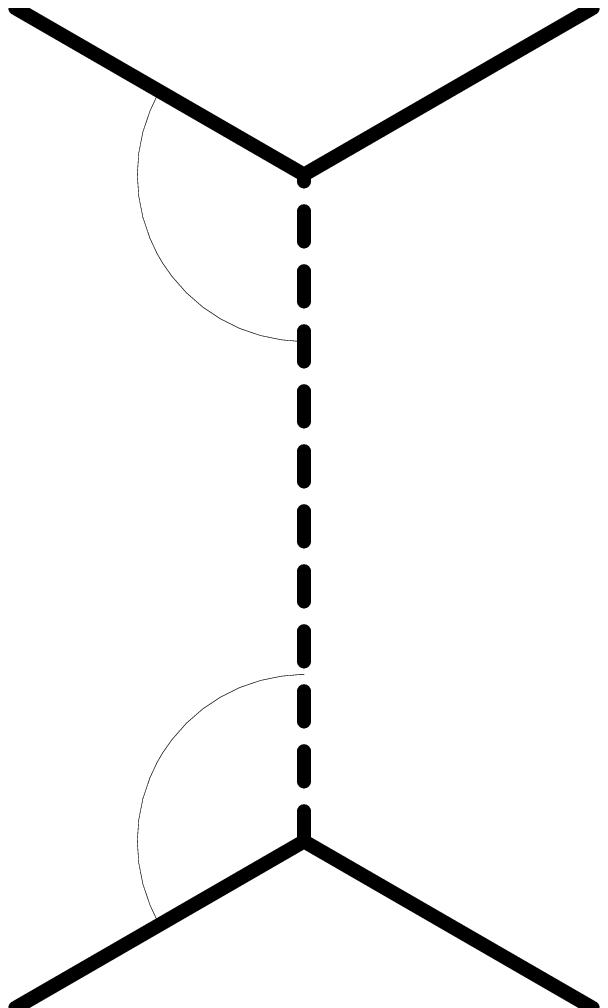}{2.5}
   +  \graph{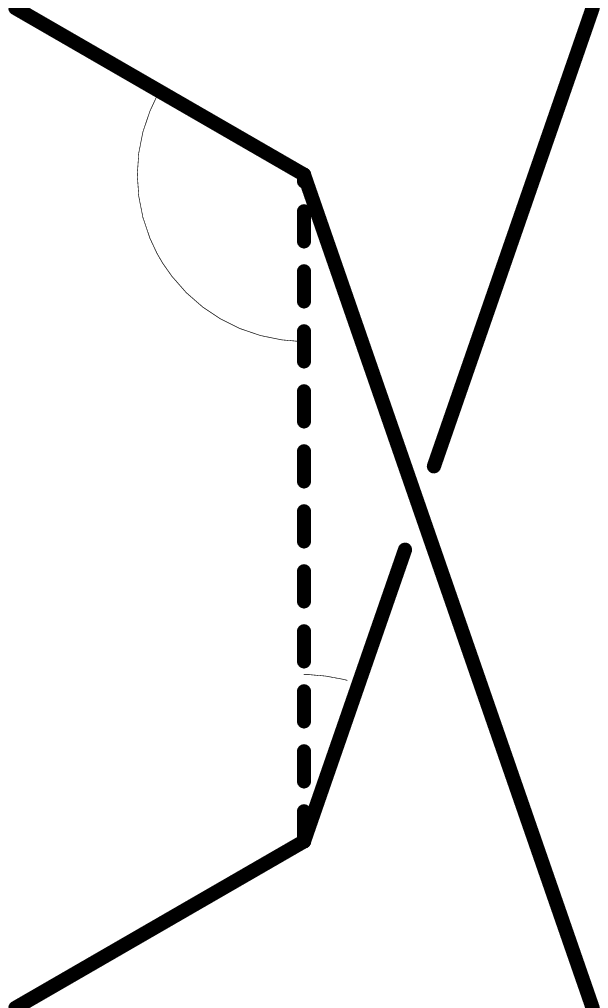}{2.5}\raisebox{1.5cm}{\hspace{-2.3cm}$\mu,a$}
                            \raisebox{1.5cm}{\hspace{1cm}$\nu,b$}
                            \raisebox{-1.5cm}{\hspace{-2.3cm}$\tau,d$}
                            \raisebox{-1.5cm}{\hspace{1cm}$\sigma,c$
                            \hspace{0.4cm}}-
      \graph{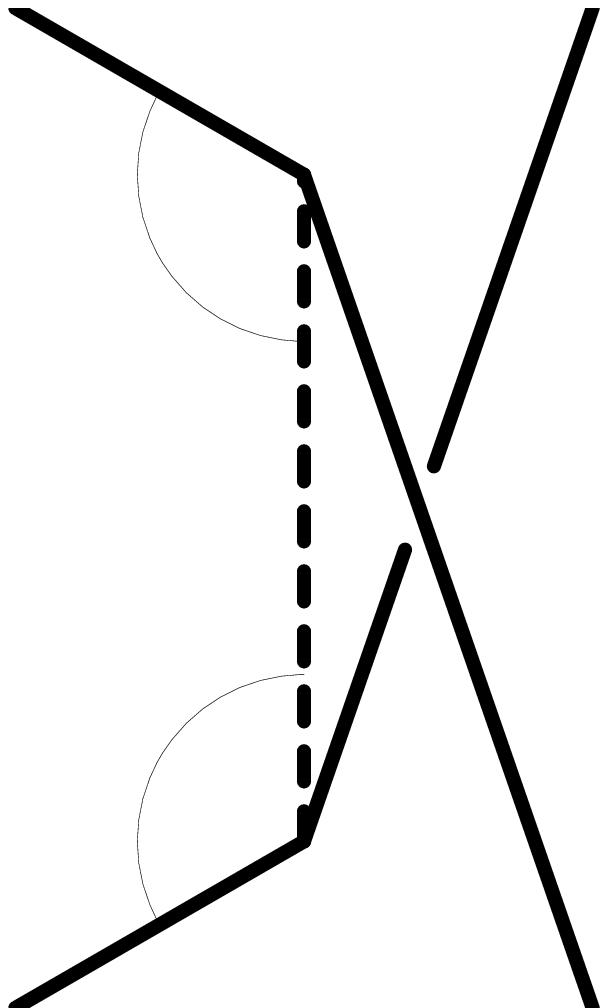}{2.5} \nonumber\\
  &&  \mbox{} + \hspace{0.5cm} 
      \graph{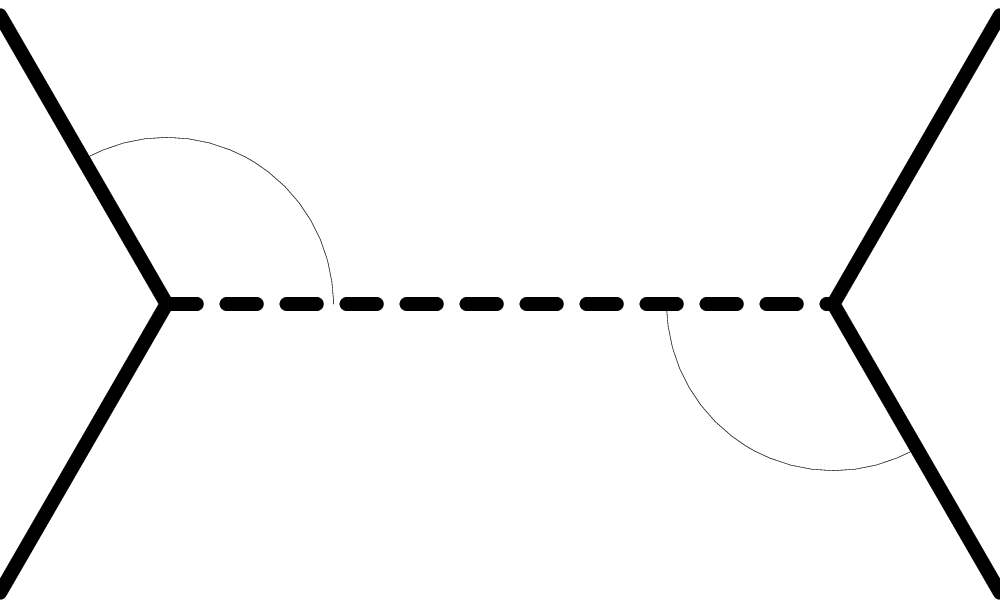}{2.5}\raisebox{1cm}{\hspace{-3cm}$\mu,a$}
                            \raisebox{1cm}{\hspace{2cm}$\nu,b$}
                            \raisebox{-1cm}{\hspace{-3.1cm}$\tau,d$}
                            \raisebox{-1cm}{\hspace{2cm}$\sigma,c$
                            \hspace{0.2cm}}-\hspace{0.4cm}
      \graph{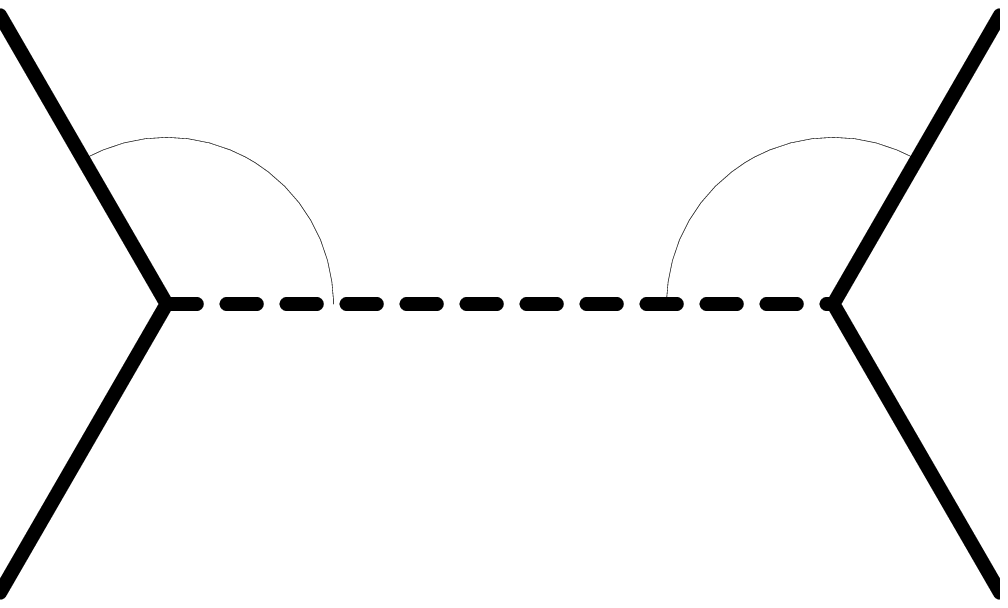}{2.5}\ .
\end{eqnarray}

Note, that regarding the graphical representation,
we have inserted a \mbox{$\delta$-function} represented by
a dashed line in the quadruple vertex. The two circular arcs
indicate, which indices of the two triple vertices are
contracted. Again the two remaining indices are contracted
as well. In those cases, where the two gluon lines of each
vertex are contracted, we omit the arcs entirely.

Thus, as an advantage of this representation, 
graphs containing only triple vertices are generated, which vastly
reduces the amount of possible topologies.

%
%
\section{One-loop}\label{YM1Loop}
It is well-known that the divergences in the order of \mbox{one-loop}
can be determined by the heat kernels (see \ie~\cite{Av91}). In the
case of pure Yang-Mills theory we are led to

\begin{eqnarray}
\Gamma_1
&=& \frac{1}{2}\,
    \ln \frac{\det\Delta_{\mu\nu}[B]}{\det\Delta_{\mu\nu}[0]} -
    \ln \frac{\det\Delta[B]}{\det\Delta[0]} \nonumber\\
&=& \lim_{s\rightarrow 0}\partial_s\Biggl(
    \frac{1}{2 \Gamma(s)}\int_0^\infty d\tau \,\tau^{s-1}
    \ \tr \int dx \nonumber\\
& & \quad\qquad\biggl[\phantom{2}\bigl(
    K^\mu_{\ \mu}(x,x;\tau|B) - K^\mu_{\ \mu}(x,x;\tau|0)\bigr)
    \nonumber\\
& & \quad\qquad \phantom{\biggl[} 
    - 2 \bigl(K(x,x;\tau|B) - K(x,x;\tau|0)\bigr)
    \ \biggr]\Biggr) \nonumber\\
&=& \frac{1}{2 (4 \pi)^{d/2}} \int_0^\infty d\tau \,
    \frac{1}{\tau^{1+d/2}}\ \tr \int dx\nonumber\\
& & \qquad 
    \Biggl[\left(\exp\left(-\frac{1}{2}
          \ln\left(\frac{\sinh{\bf F}\tau}{{\bf F}\tau}\right)
             \right)^\mu_{\ \mu}
          \left(\exp 2{\bf F}\tau\right)^\nu_{\ \nu} - 
                     d\,\seteins \right) \nonumber\\
& & \qquad\qquad\qquad\,\,\,\mbox{} -
       2 \left(\exp\left(-\frac{1}{2}\ln\left(\frac{
          \sinh{\bf F}\tau}{{\bf F}\tau}\right)\right)^\mu_{\ \mu} - 
          \seteins\right)
    \Biggr]\ .
\end{eqnarray}

As a matter of fact this integral cannot be solved. But since we are
only interested in the \mbox{UV-divergences} we expand it in terms of
$\tau$ and after the following integrations take the limit 
$\tau\rightarrow 0$.

Now with the help of
\begin{equation}
\exp\left(-\frac{1}{2}\ln\left(\frac{
          \sinh{\bf F}\tau}{{\bf F}\tau}\right)\right)^\mu_{\ \mu}
\ =\ \seteins + \frac{1}{12} {\bf F}^2 \tau^2 + \cdots
\end{equation}
and
\begin{equation}
\left(\exp 2{\bf F}\tau\right)^\nu_{\ \nu}
\ =\ d\, \seteins - 2 {\bf F}^2 \tau^2 + \cdots
\end{equation}
we eventually get
\begin{equation}
\Gamma_1^\text{div}
\ =\ \frac{22}{3 \epsilon} \frac{C_2}{16 \pi^2} S_{cl} + O(\eps^0) \ .
\end{equation}
which corresponds to the results of the literature~\cite{Po73,GW73}.

%
%
\section{Two-loop}\label{YM2Loop}
The contribution of the quadruple vertex (\ref{VV}) to the 
two-loop effective action is given by 

\begin{equation}\label{VV2}
\begin{array}{cccc}
\graph{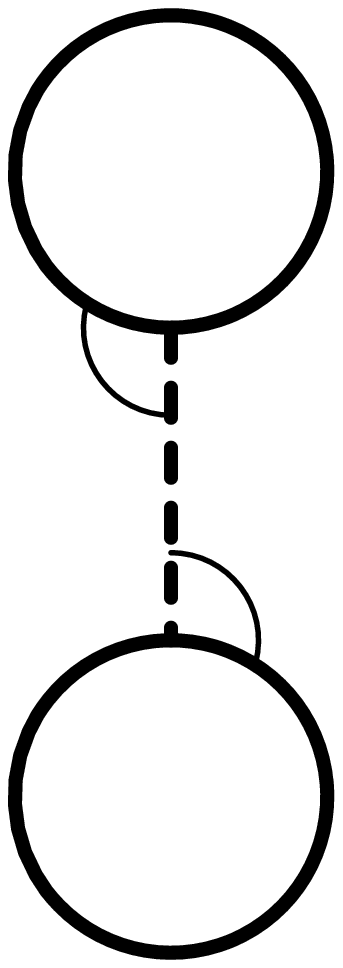}{2.5} & - & \graph{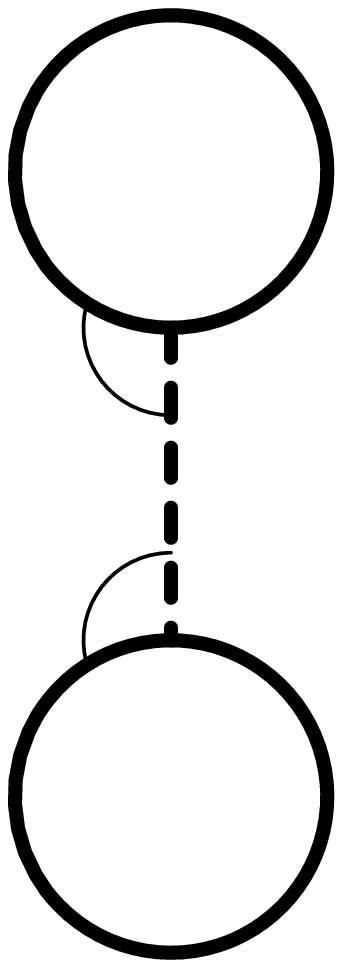}{2.5} & \\
\mbox{} \ +\!\!\!\!\graph{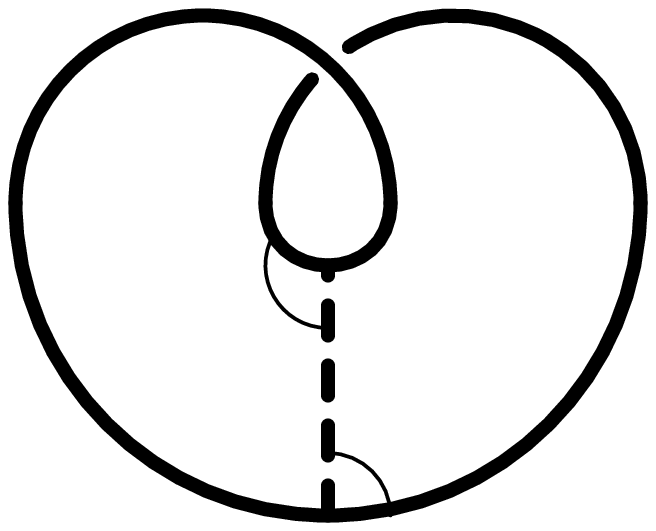}{3.0} & - & \graph{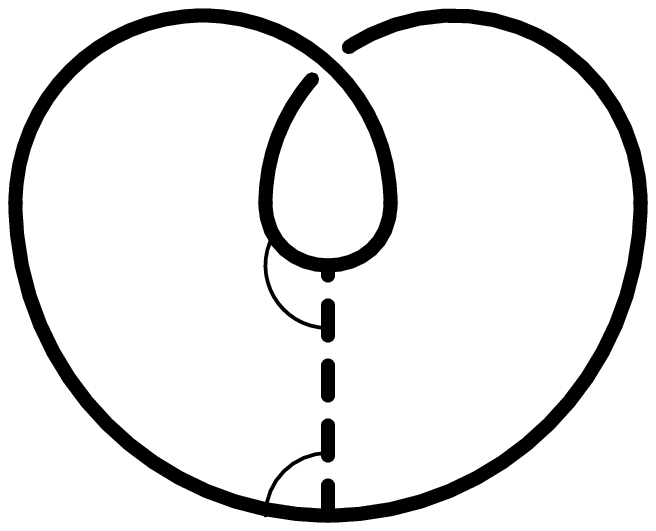}{3.0} & \\
\mbox{} +\graphrot{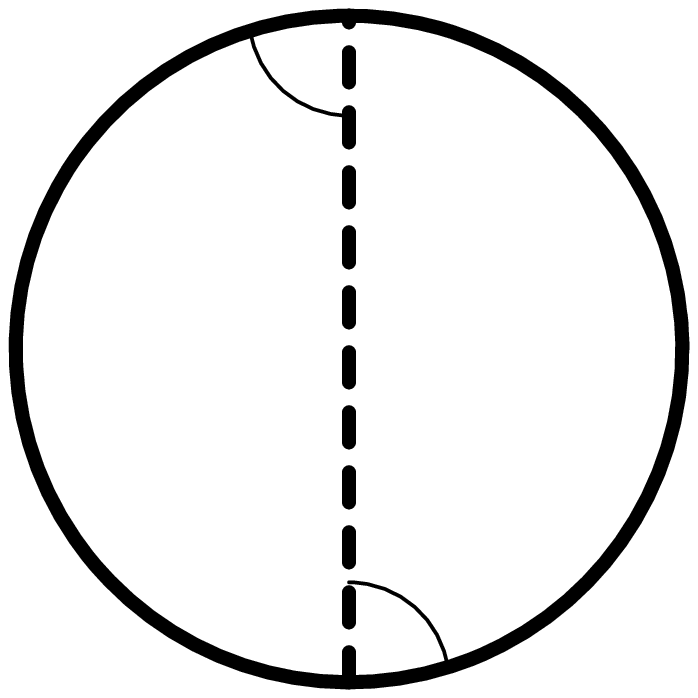}{2.5} & - & \graphrot{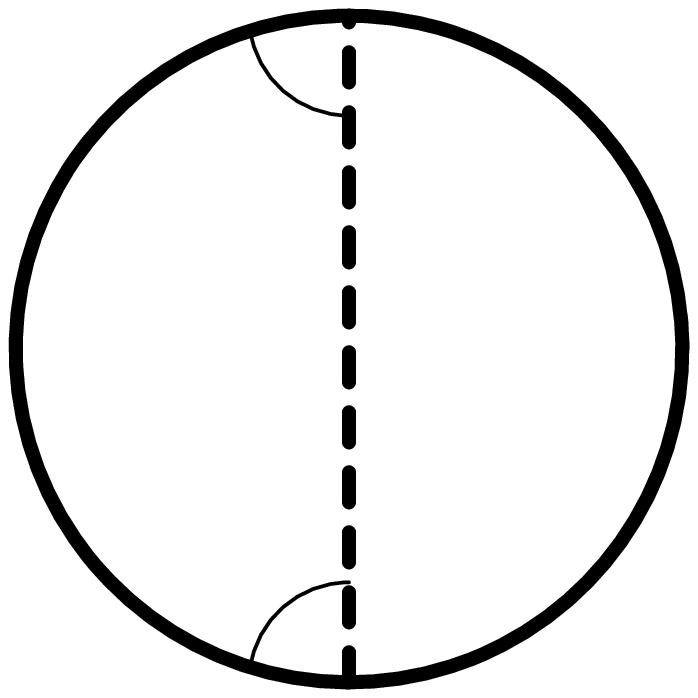}{2.5} & . \\
\end{array}
\end{equation}

By using the identity (\ref{Gmn2Gnm}), it can be shown, that
the second graph in the first row is similar to the
first graph in the first row. Due to the same identity
(\ref{Gmn2Gnm}) the expression $G^\mu_{\ \mu}(x,x)$
yields null and hence the first graph in the second
row and the second graph in the third row vanish.
If we remove the twist in the first graph of the
second row, we get an additional minus sign as a result
of the antisymmetry of the structure constant.
\begin{figure}[h]
\begin{equation*}
\graph{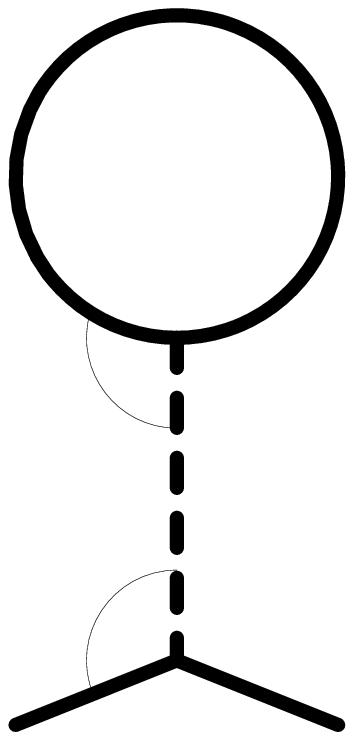}{3}
\ =\ 2\raisebox{-0.625cm}{
\graph{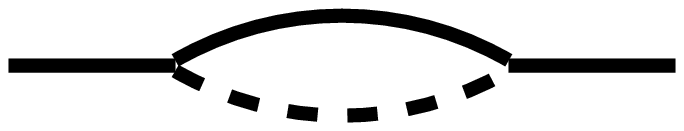}{3.5}}
\end{equation*}
\vspace{-2cm}
\caption{\label{Tadpole}Tadpole identity}
\end{figure}
Finally the two remaining graphs are shown to be equal
by means of the {\em tadpole-identity} illustrated in
figure~\ref{Tadpole}. Written as a formula we get
\begin{multline}
f_{abc}   G^{cb}_{\mu\nu}(x,x)\delta(x-x')
f_{ab'c'} G_{b'\bullet}^{\mu\,\bullet}(x',y')
          G_{c'\bullet}^{\nu\,\bullet}(x',z')\ =\\
       2\,G^{bb'}_{\mu\nu}(x,x')\delta(x-x')
f_{abc}   G_{c\,\,\bullet}^{\mu\,\bullet}(x,y')
f_{ac'b'} G_{c' \bullet}^{\nu\,\bullet}(x',z')
\end{multline}
which can be proved by using the Jacobi identity and
equation~(\ref{Gmn2Gnm}) only.

Thus, at least the contribution of the quadruple 
vertex shown in~(\ref{VV2}) can be reduced to 
a single graph
\begin{equation}\label{th1c2thc}
6 \graphrot{K2cb2}{2.5}\ =\ -\, 6 \graphrot{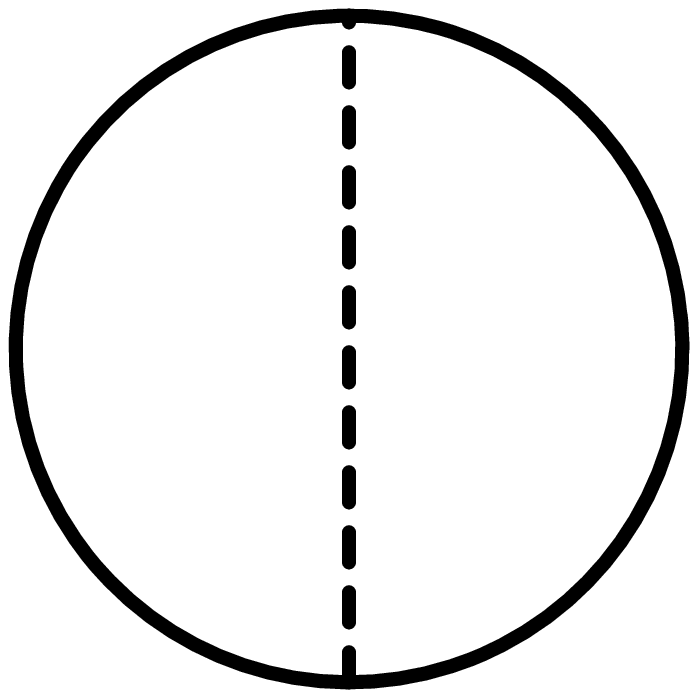}{2.5},
\end{equation}
where the minus sign on the right hand side is a result of 
equation~(\ref{Gmn2Gnm}). All graphs sharing the topology
of those in equation~(\ref{th1c2thc}) will, according to their
shape, from now on be called $\Theta$~graphs.

\begin{figure}[h]
\begin{eqnarray*}
\Gamma_2
&=&\phantom{+\ }\frac{3}{4}\graph{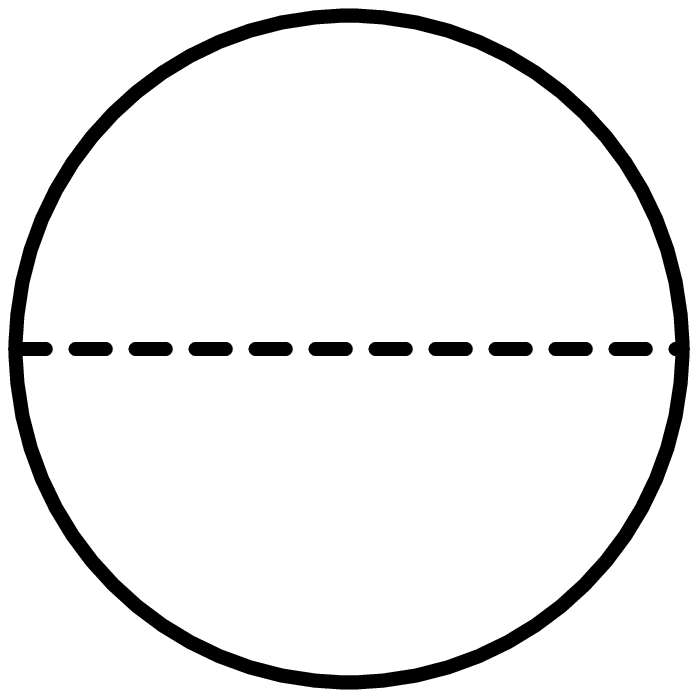}{2.5}\\
& &+\ \frac{1}{2}\quad\left(-\graph{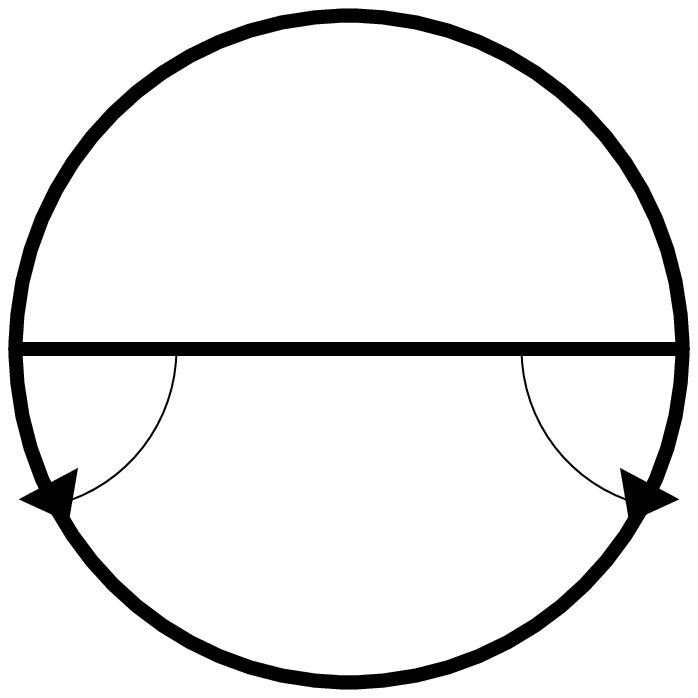}{2.5}\right.
   + \graph{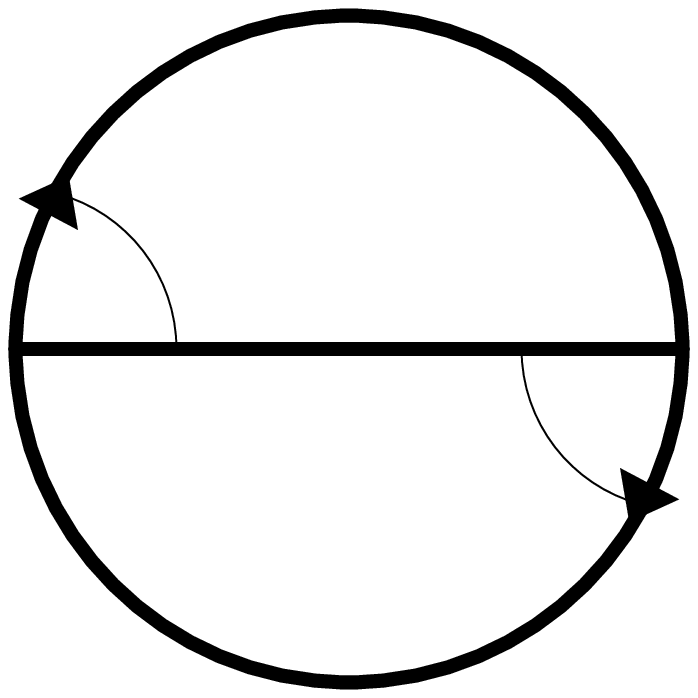}{2.5}\\
& &\qquad\qquad + \graph{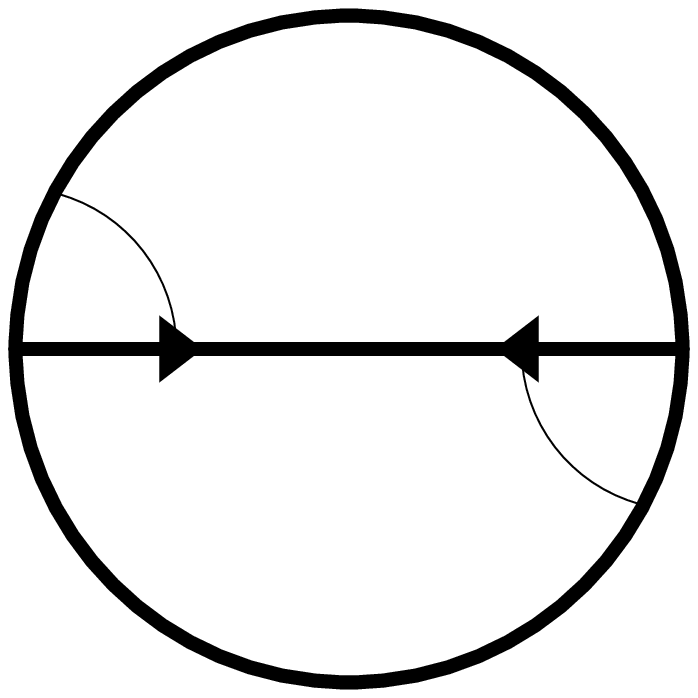}{2.5}
            - \, 2\graph{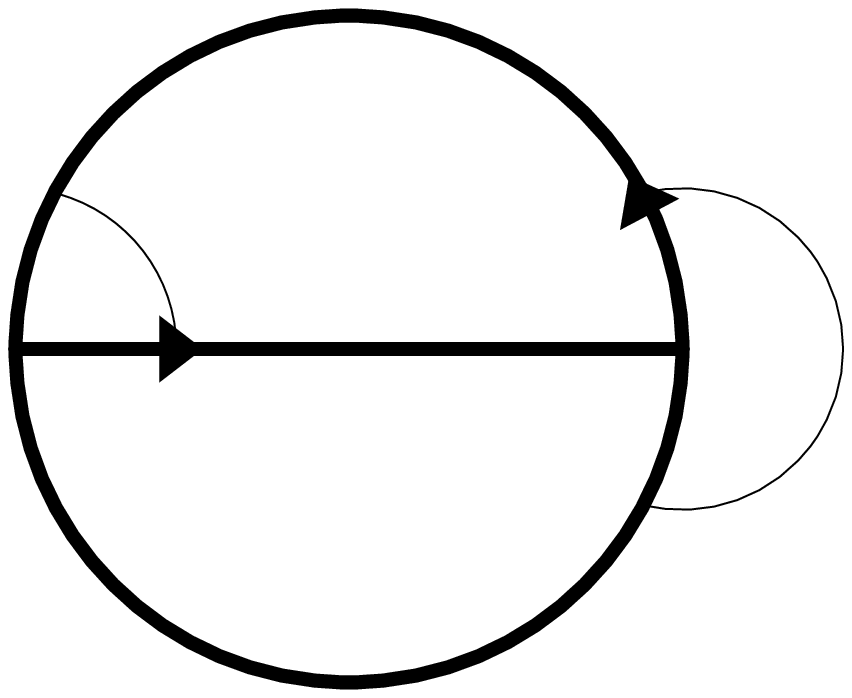}{2.5}
          + \left.\graph{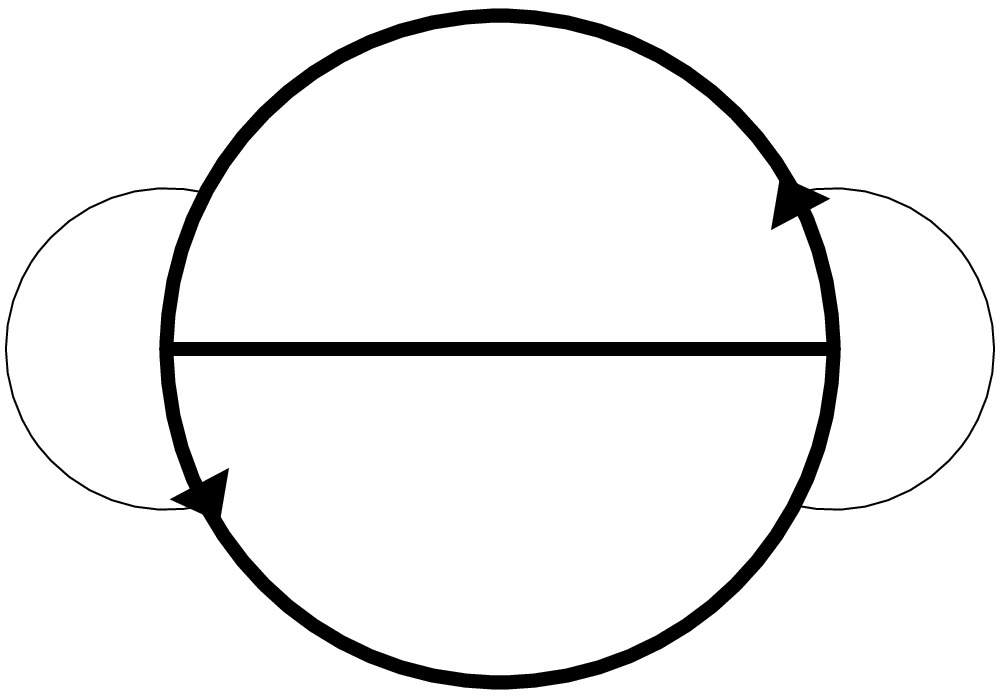}{2.5}\right)\\
& &-\ \frac{1}{2} \graph{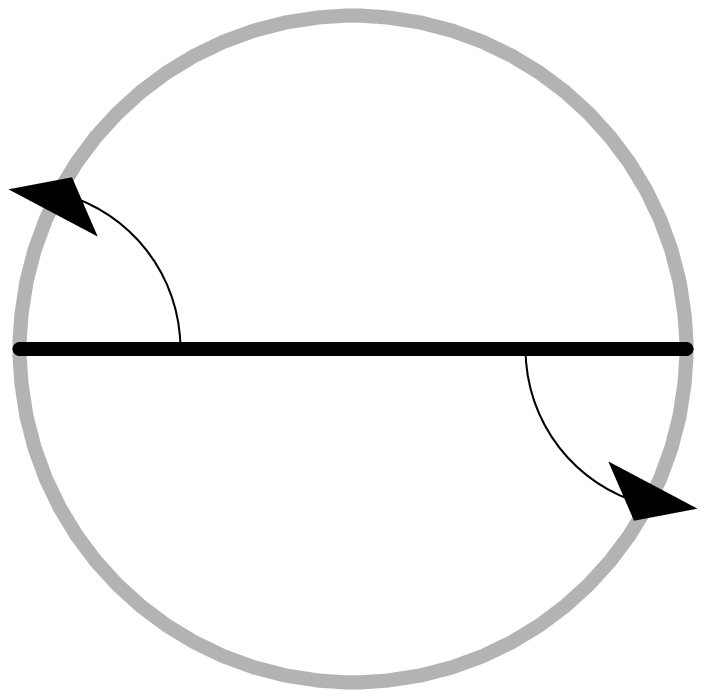}{2.5}
\end{eqnarray*}
\caption{\label{ur2loop}Two-loop effective action}
\end{figure}

Following~(\ref{DV1}) the contribution to the two-loop effective 
action of the two triple gluon vertices is given by 36 
$\Theta$~graphs. Fortunately the symmetry of the $\Theta$ graph 
allows us to replace the sum~(\ref{DV1}) of one vertex by one 
of its terms multiplied by 6 without loss of generality,
leaving us with only 6 graphs. Furthermore one of these graphs 
can - by exchange of the two coordinate space points - be transformed 
to a graph already existing. Adding finally the contribution of the
ghost vertices~(\ref{DV2}) we get the \mbox{two-loop} effective action
as shown in figure~\ref{ur2loop}.
Thus, using the tadpole identity and some symmetries we are left
solely with graphs of the $\Theta$ topologies. This simplification
gets even more important in higher loop calculations, since the
amount of possible reductions is increasing.

Before proceeding we will establish a notation, which
will be just as powerful as the graphical representation.
As very useful we consider the bracket notation
introduced in \cite{BMS77}.
This notation is free of group indices and the symmetries
are easily seen. It is defined by
\begin{equation}
(A,B,C)\ :=\ f_{{a_1}{b_1}{c_1}} f_{{a_2}{b_2}{c_2}}
                   A^{a_1a_2} B^{b_1b_2} C^{c_1c_2}
\ = \ -\beginpicture
    \setcoordinatesystem units <0.4 cm , 0.4 cm>
    \setplotarea x from -2 to 2, y from -1.5 to 1.5
    \circulararc 360 degrees from 0 1.5 center at 0 0
    \plot -1.5 0 1.5 0 /
    \plot -0.2  1.7 0.1  1.5 -0.2  1.3 /
    \plot -0.2  0.2 0.1    0 -0.2 -0.2 /
    \plot -0.2 -1.7 0.1 -1.5 -0.2 -1.3 /
    \put{$\scriptstyle{A}$}[b] at  0 1.9
    \put{$\scriptstyle{B}$}[b] at  0 0.4
    \put{$\scriptstyle{C}$}[t] at  0 -1.9
  \endpicture\quad ,
\end{equation}
where - due to clockwise representation of the two structure 
constants - the graphical representation holds a minus sign.

Thus, the symmetries are given by
\begin{equation}
(A,B,C)\ =\ (A,C,B) \ =\ (B,A,C)\quad ,
\end{equation}
which means that the three propagators of a $\Theta$ graph
are symmetric to each other.

The bracket notation corresponding to the graphical representation of
the effective action of figure \ref{ur2loop} is given by
\begin{eqnarray}\label{G2Kn}
\Gamma_2
&=& \int dx\, dy \Biggl[\phantom{+}\frac{3}{4}\, 
    (\seteins, G_{\mu\nu},G_{\mu\nu})\nonumber\\
& & \qquad\,+\,\frac{1}{2}\,
    \left(-\,(G_{\mu\nu},G_{\rho\sigma},D_\rho G_{\mu\nu} 
                                 \overleftarrow{D}_\sigma) +
    (D_\rho G_{\mu\nu},G_{\rho\sigma},G_{\mu\nu} 
                                 \overleftarrow{D}_\sigma)
		                         \right. \nonumber\\
& & \hspace{1.12cm}  \phantom{+\ \frac{1}{2}}
    \left. +\, (G_{\mu\nu},D_\mu G_{\rho\nu} 
                \overleftarrow{D}_\sigma,G_{\rho\sigma})-
        2\,(G_{\mu\nu} \overleftarrow{D}_\sigma,
            D_\mu G_{\rho\nu},G_{\rho\sigma})
				         \right. \nonumber\\
& & \hspace{1.12cm} \phantom{+\ \frac{1}{2}}
    \left. +\, (G_{\mu\nu} \overleftarrow{D}_\sigma,G_{\rho\nu},
                D_\mu G_{\rho\sigma}) 
                         \right) \nonumber\\
& & \qquad\,-\,\frac{1}{2}\, 
    (D_\mu G,G_{\mu\nu},G \overleftarrow{D}_\nu)\Biggr].
\end{eqnarray}
One should bear in mind, that up to now only the Feynman gauge
has been used.

But the effective action~(\ref{G2Kn}) can even be more simplified.
In a first step we transform with help of the socalled 
Ward~Identity~(\ref{PgI}) all ghost propagators into gluon propagators.

The graphical and the bracket notation are illustrated
by
\begin{eqnarray}\label{Geist2Gluon}
\graph{ThetaGeist}{2.5} &=& \graph{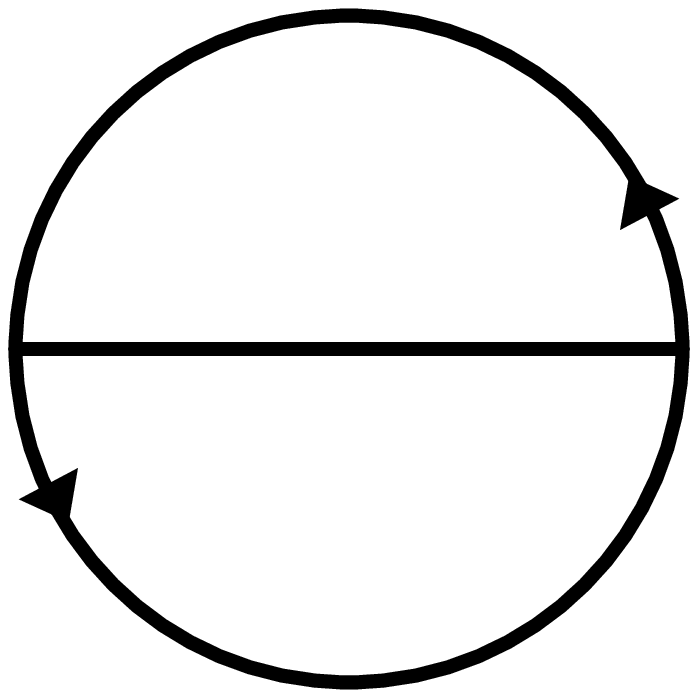}{2.5}\nonumber\\
  \Longleftrightarrow\qquad
  (D_\mu G,G_{\mu\nu},G \overleftarrow{D}_\nu) \ &=&\ 
  (G_{\mu\nu}\overleftarrow{D}_\nu, G_{\mu\sigma},
   D_\rho G_{\rho\sigma})\, .
\end{eqnarray}
The new emerged graph has a remarkable feature:
All covariant derivatives act on the propagator
with which their lorentzian index is contracted.
Such graphs will from now on be called {\em primary}.
Concerning $\Theta$ graphs with one derivative at
each vertex there exists still another primary
graph. It is generated by replacing the 
\mbox{$\delta$-function} according to equation~(\ref{DGD2d}) 
in the first graph of figure~\ref{ur2loop}. Thus we get
\begin{eqnarray}
\graph{UrThetaDelta}{2.5} &=& \graph{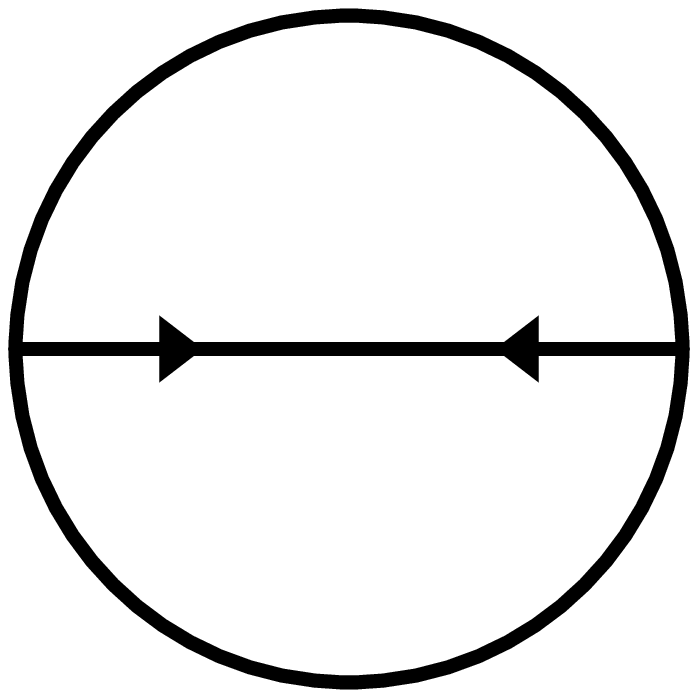}{2.5}\nonumber\\
\qquad\Longleftrightarrow\qquad
(G_{\mu\nu},\seteins,G_{\mu\nu})\ & =&\ 
(G_{\mu\nu},D_\rho G_{\rho\sigma} 
 \overleftarrow{D}_\sigma, G_{\mu\nu})\,.
\end{eqnarray}
Before introducing further simplifications we have to show
that the covariant derivatives follow the rules of
partial integration. To prove this statement we take the sum of three
vertices resulting from a partial integration and get
\begin{multline*}
(D_\mu G_1, G_2, G_3) + (G_1, D_\mu G_2, G_3) + 
(G_1, G_2, D_\mu G_3)\\
\begin{array}{l}
\ =\ \partial_\mu (G_1, G_2, G_3) + ([B_\mu, G_1], G_2, G_3) +
     (G_1, [B_\mu, G_2], G_3) + (G_1, G_2, [B_\mu, G_3])\\[1ex]
\ =\ \partial_\mu (G_1, G_2, G_3) + [B_\mu, (G_1, G_2, G_3)],
\end{array}
\end{multline*}
where the last equal sign stems from the identity
$D[C,E]+[C,D]E = [C,DE]$. Eventually this result
yields null, since the commutator vanishes due to the fact, that
the expression $(G_1, G_2, G_3)$ is a scalar and
the total derivative vanishes due to boundary conditions.

\begin{figure}[h]
\begin{eqnarray*}
\graph{UrTheta1}{2.5}
&=& 2\graph{ThetaDv12}{2.5}\!\! + \,
    2\graph{ThetaDv11}{2.5}\\
\graph{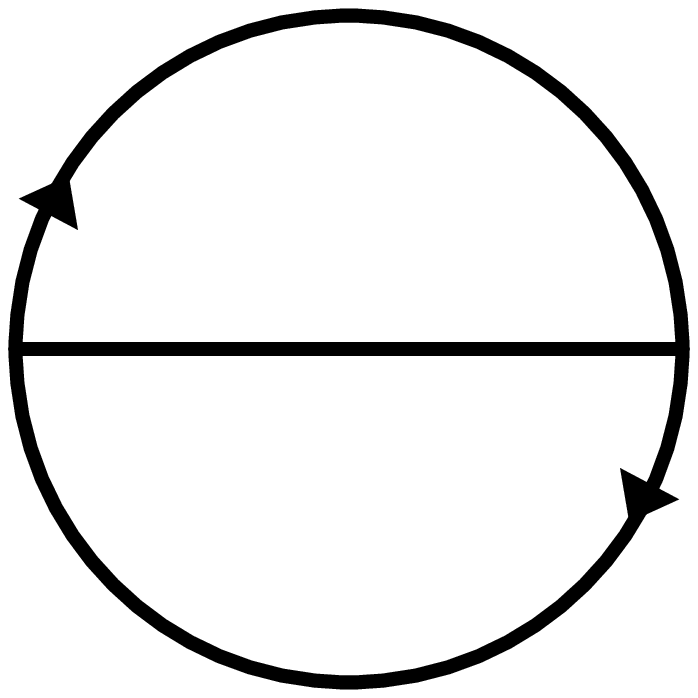}{2.5}
&=& \graph{ThetaDv21}{2.5}\!\! + \,
   2\graph{ThetaDv22}{2.5}+
    \graph{ThetaDv23}{2.5}
\end{eqnarray*}
\caption{Partial integration applied to the primary graphs}
\label{urthetagraphen}
\end{figure}

As shown in figure \ref{urthetagraphen} by partially
integrating each vertex of the two primary graphs
we get the remaining five graphs with correct coefficient,
but sometimes incorrect sign.

\begin{figure}[h]
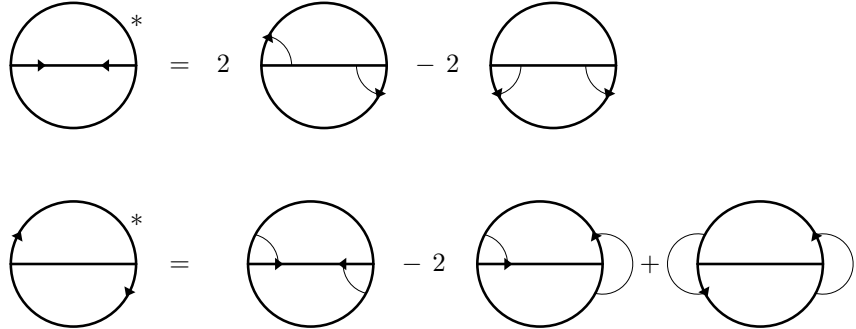

\begin{eqnarray*}
\graph{UrTheta1}{2.5}\raisebox{0.6cm}{\hspace{-0.5cm}$\ast$}
&=& 2\graph{ThetaDv12}{2.5}\!\! - \,
    2\graph{ThetaDv11}{2.5}\\
\graph{UrTheta2}{2.5}\raisebox{0.6cm}{\hspace{-0.5cm}$\ast$}
&=& \graph{ThetaDv21}{2.5}\!\! - \,
   2\graph{ThetaDv22}{2.5}+
    \graph{ThetaDv23}{2.5}
\end{eqnarray*}
\caption{Star operation applied to the primary graphs}
\label{sternurthetagraphen}
\end{figure}

The correct sign can be deduced from the number of arcs
which connect the derivative and the associated propagator
clockwise. For an odd number the sign is negative.
This procedure in what follows will be called star operation
and will be indicated by a $\ast$ symbol at the right upper side
of the graph. The graphical representation of the star operation
as applied to both primary $\Theta$ graphs is illustrated in
figure  \ref{sternurthetagraphen}.

Thus the $\Theta$ graphs of the Yang-Mills theory can be
divided in two classes.
By partially integrating the two vertices of the primary
graph with both derivatives operating on the same propagator,
we generate all graphs belonging to the first class.
The associated graphical representation can be found in the
first row of figure \ref{sternurthetagraphen}.
The second primary graph gained from the ghost graph by means
of (\ref{Geist2Gluon}), leads us by partial intergration
to the graphs of the second class (as shown in the second
row of figure \ref{sternurthetagraphen}).

Eventually the \mbox{two-loop} effective action
by means of the star operation, and the identities
(\ref{PgI}) and (\ref{DGD2d}) can be reduced
from originally seven graphs down to three as
illustrated in figure \ref{ur2loopneu}.

It should be emphasized that up to this moment we have
only used the Feynman gauge since, as mentioned before,
the two propagator identities
(\ref{PgI}) and (\ref{DGD2d}) are even valid without the
assumption $D_\rho F_{\mu\nu}=0$.

It should be mentioned that the gauge invariance of the
background field $B_\mu$ is preserved, and since all identities
used until now are non-pertubative as far as the background field
is considered, the new effective action can be used to calculate
the \mbox{two-loop} finite part as well.

\bigskip

\begin{figure}[h]
\begin{eqnarray*}
\Gamma_2&=&\quad\graph{UrThetaDelta}{2.5} \\
&&+\ \frac{1}{4}\quad\left(\graph{UrTheta1}{2.5}
                           \raisebox{0.6cm}{\hspace{-0.5cm}$\ast$}
                         - \graph{UrTheta1}{2.5}\right)\\
&&+\ \frac{1}{2}\quad\left(\graph{UrTheta2}{2.5}
                           \raisebox{0.6cm}{\hspace{-0.5cm}$\ast$}
                         - \graph{UrTheta2}{2.5}\right)\\
&=& \graph{UrThetaDelta}{2.5}-\graph{ThetaDv11}{2.5} - 
   2 \graph{ThetaDv22}{2.5}
\end{eqnarray*}
\caption{\label{ur2loopneu}Simplified two-loop effective action}
\end{figure}

\bigskip 

Assuming in addition  that the field strength is
covariantly constant~(\ref{ymNb}) we are now allowed
to use the expansion of the propagators~(\ref{ESG},~\ref{EVG}).
All terms of higher than quadratic order in the field strength
are dropped.

After considering the effect of the covariant derivative
on the phasefactor~(\ref{Phi2a},~\ref{Phi2b}) the latter can be
substituted by the identity in all terms containing the field
strength quadratically.

Imposing the \mbox{Fock-Schwinger} gauge on the background 
field $B_\mu$, we get the phasefactor defined in~(\ref{PFSE}) and a
field strength which is constant.

Due to the translation invariance we are allowed to fix a vertex
without restriction in general at the origin of coordinate 
space~\cite{NSVZ84,VS86}.  Thus all remaining phasefactors in the 
\mbox{two-loop} graphs can be neglected, meaning
\begin{equation}\label{Phi0y}
\Phi(0,y)\ =\ \seteins.
\end{equation}
Now all phasefactors have disappeared and we see
immediately that graphs
containing the field strength linearly do, for
symmetry reasons, not contribute to the result.
Graphs without field strength are neglected.

Finally only graphs provided with the following
properties are left:
They contain no phasefactor at all, but the field
strength in quadratic order, and they are logarithmically divergent.
Since the phasefactor has been the only term
depending on points in coordinate space as well as on
some group indices, the remaining graphs can be
decomposed in an ordinary Feynman graph and a graph
consisting of structure constants. A step by step
calculation of the second and third graph in the last
row of figure~\ref{ur2loopneu} is given in
appendix~\ref{2LoopGraph}. However a sketch
of this procedure applied to the last two graphs of
figure~\ref{ur2loopneu} is given by
\begin{eqnarray}
(G_{\mu\nu},G_{\rho\nu}\overleftarrow D_\sigma,D_\rho G_{\mu\sigma})
&=& \mbox{} \phantom{-}\ \,\! 
   (\Phi G_0,\Phi G_0 \overleftarrow D_\mu,D_\mu \Phi G_0)+
                       \cdots\label{EG2}\nonumber\\
&=& \mbox{} - (\Phi G_0,\Phi \partial_\mu G_0,\Phi\partial_\mu G_0)
    \nonumber\\
&&  \mbox{} + (\Phi G_0,\Phi {\bf F}_{\mu\rho}\partial_\rho G_1,
      \Phi {\bf F}_{\mu\sigma} \partial_\sigma G_1)+\cdots \nonumber\\
&=& \mbox{} \phantom{-}\ \,\!(G_0,{\bf F}_{\mu\rho}\partial_\rho G_1,
        {\bf F}_{\mu\sigma} \partial_\sigma G_1)+\cdots
\end{eqnarray}
The last equal sign stems from (\ref{Phi0y}). From the 16 graphs
which we get after the expansion of the propagators, we will
calculate as an example the graph which depends on the field
strength only via the covariant derivative and the phase
factor~(\ref{EG2}). The remaining graphs are transformed
to logarithmically divergent graphs containing the field
strength quadratically in a similar way.
\begin{eqnarray}
(G_0,{\bf F}_{\mu\rho}\partial_\rho G_1,{\bf F}_{\mu\sigma} 
                      \partial_\sigma G_1)
&=& G_0(x-y)\partial_\rho G_1(x-y)\partial_\sigma G_1(x-y)
   (\seteins,{\bf F}_{\mu\rho},{\bf F}_{\mu\sigma}) \nonumber\\
&=& -\graph{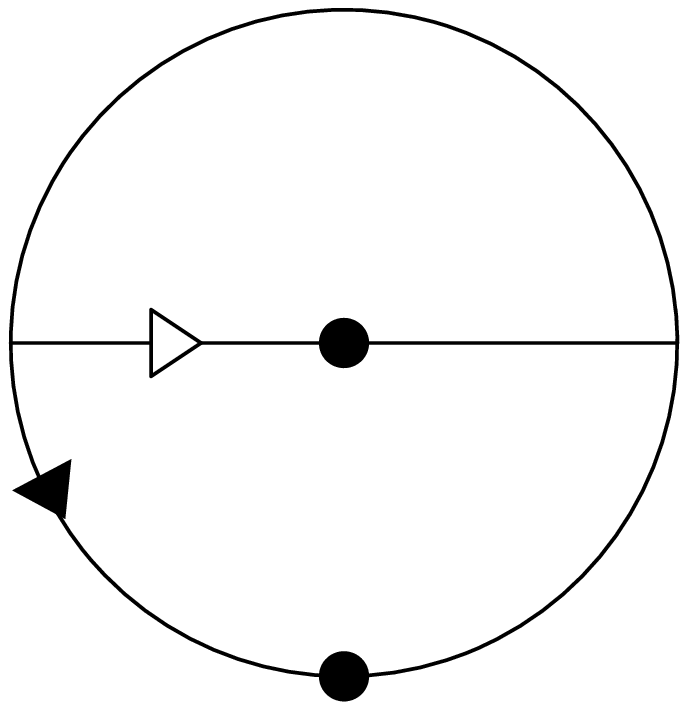}{2.5}
\raisebox{1mm}{
\beginpicture
    \setplotsymbol (.)
    \setcoordinatesystem units <0.8 cm , 0.8 cm>
    \setplotarea x from -1 to 1, y from -1.5 to 1.5
    \circulararc 360 degrees from 0 1 center at 0 0
    \plot -1 0  1 0 /
    \plot  0 -1  0 -0.8 /
    \plot  0  0  0  0.2 /
    \put{$\scriptstyle{F^b_{\mu\sigma}}$}[b] at 0 -0.7
    \put{$\scriptstyle{F^a_{\mu\rho}}$}[b] at 0 0.3
  \endpicture} \label{KlGr}\\
&=&\frac{1}{4\eps}\delta_{\rho\sigma}\delta(x-y) \quad
   \frac{1}{2} C_2^2 F^a_{\mu\rho} F^a_{\mu\sigma} \nonumber\\
&=&\frac{1}{8\eps} C_2^2 (F^a_{\mu\rho})^2 \delta(x-y) \nonumber
\end{eqnarray}

The first graph represents just an ordinary Feynman graph. The
corresponding divergences can be looked up in the table of
appendix~\ref{D2G}, whereas the value of the second graph consisting of
structure constants can be deduced by means of the rules as listed
in table~\ref{TabSt}. After determining the ordinary graph, the lacking
contractions have to be made. The following two identities
\begin{eqnarray*}
(\seteins,\seteins,{\bf F}^2) 
&=& \phantom{\frac12}C_2 \mbox{tr}({\bf F}^2)
\ =\ \phantom{\frac12} C_2^2 (F^a_{\mu\nu})^2 \\
(\seteins,{\bf F}_{\mu\nu},{\bf F}_{\mu\nu}) 
&=& \frac12 C_2\mbox{tr}({\bf F}^2)
\ =\ \frac12 C_2^2 (F^a_{\mu\nu})^2,
\end{eqnarray*}
suffice to obtain the result for all remaining \mbox{two-loop} graphs
made of structure constants.

Resisting the temptation to calculate the Feynman graphs immediately,
we show that by using several manipulations given in 
appendix~\ref{2LoopGraph} the number of graphs can be reduced to two
and finally even to one.

Thus the \mbox{two-loop} divergent part of the effective
action is given by
\begin{eqnarray}
\Gamma^\text{div}_2
&=&  (G_{\mu\nu},\seteins,G_{\mu\nu})-
     (G_{\mu\nu},G_{\rho\sigma},D_\mu G_{\rho\sigma} 
      \overleftarrow D_\nu)-
    2(G_{\mu\nu},G_{\rho\nu} \overleftarrow D_\sigma,
      D_\rho G_{\mu\sigma})\nonumber\\
&=&\frac{17}{12}(\seteins,G_1,G_1 {\bf F}^2) - 
   \frac{17}{6} (G_0,G_0,G_1 {\bf F}^2) +
    O(\eps^0)\label{2LFormel1}\\
&=&\frac{17}{12}\hspace{-0.5cm}\graph{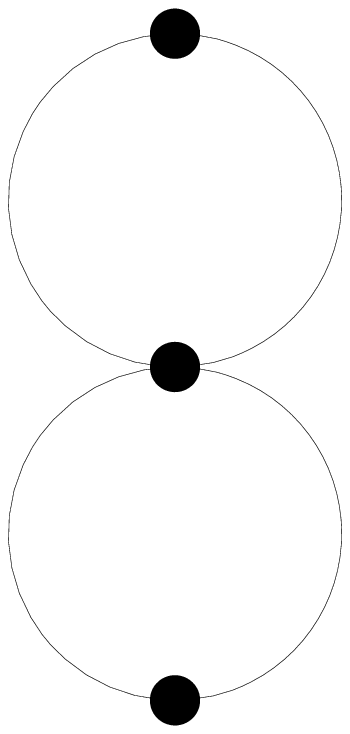}{2.5} \hspace{-0.5cm}
    (\seteins,\seteins,{\mathbf F}^2) -
    \hspace{0.2cm} \frac{17}{6}\graph{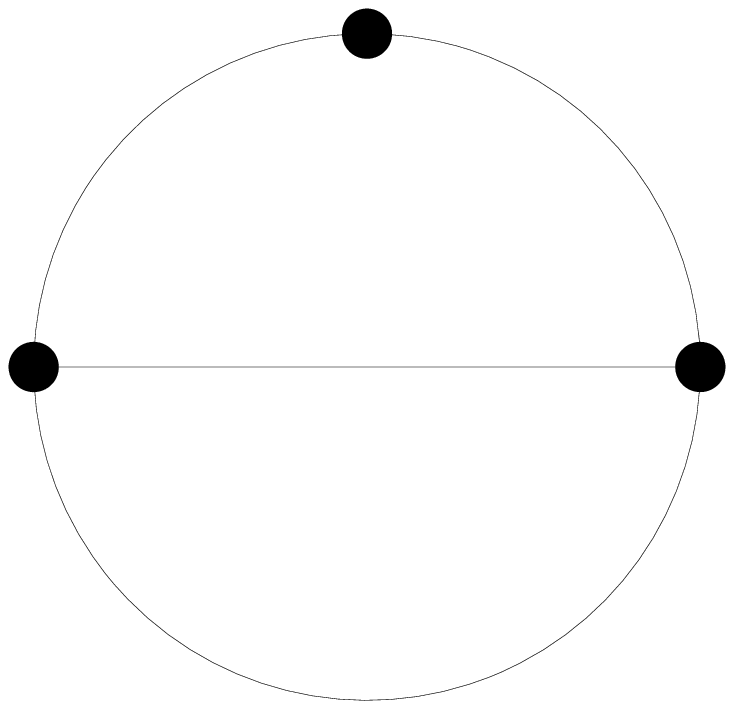}{2.5}
    (\seteins,\seteins,{\mathbf F}^2) +
    O(\eps^0)\label{2LBild1}\\
&=&-\frac{17}{6}(G_0,\partial_\mu G_1,\partial^\mu G_1 {\bf F}^2) +
    O(\eps^0)\label{2LFormel2}\\
&=&-\frac{17}{6} \graph{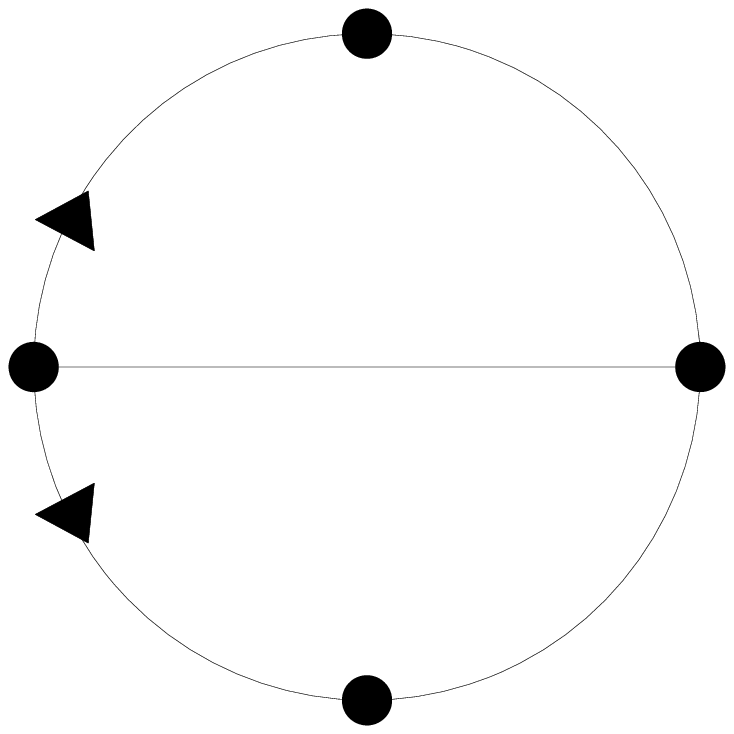}{2.5}
    (\seteins,\seteins,{\mathbf F}^2) +
    O(\eps^0)\label{2LBild2}\\
&=& \frac{34}{3 \eps}\left(\frac{C_2}{16 \pi^2}\right)^2 S_{cl} + 
    O(\eps^0)\, ,
\end{eqnarray}
whereas the transformation from the graphical
representation~(\ref{2LBild1}) to (\ref{2LBild2})
respectively from equation~(\ref{2LFormel1}) to (\ref{2LFormel2})
is accomplished by using the identity~(\ref{d12}). The resulting
\mbox{graph~(\ref{2LFormel2}, \ref{2LBild2})} has a remarkable property:
it contains no subdivergences and therefore has only a
simple pole. Thus just by looking at the final graph without
calculating any integral, we realize, that the quadratic
pole of the Yang-Mills theory vanishes at the order of two-loop.

%
%
\section{Three-loop}\label{YM3Loop}

In the following we will generalize the methods developed
in the last section such that they are suited
to determine divergences of Feynman graphs beyond \mbox{two-loop}.
Afterwards we will by means of these methods calculate the
\mbox{$\beta$-function} of the pure Yang-Mills theory up to 
\mbox{three-loop}.

After using the tadpole identity~(see figure \ref{Tadpole}) in 
three-loop the remaining graphs belong to one of the following 
two topologies. The graphs of the first topology will be called 
ball graphs, because of their shape. The bracket notation, 
respectively the graphical representation is given by:

\begin{eqnarray}
(A,B,C|K,L,M)_1 &:=&
                   f_{a_1 b_1 c_1} f_{a_2 b_2 c_2}
                   f_{a_3 b_3 c_3} f_{a_4 b_4 c_4}\nonumber\\
                && A^{a_1a_2} B^{b_2b_3} C^{c_2c_3}
                   K^{a_3a_4} L^{b_3b_4} M^{c_3c_4}
                   \rule[-3mm]{0mm}{8mm}\nonumber\\
 &=& \ \ \beginpicture
    \setcoordinatesystem units <0.4 cm , 0.4 cm>
    \setplotarea x from -2 to 2, y from -1.5 to 1.5
    \circulararc 360 degrees from 0 1.5 center at 0 0
    \plot -1.35 0.5 1.35 0.5 /
    \plot -1.35 -0.5 1.35 -0.5 /
    \arrow <1.75mm> [0.25,0.85] from  -1.5 0 to -1.5 0.2
    \arrow <1.75mm> [0.25,0.85] from  1.5 0 to 1.5 -0.2
    \arrow <1.75mm> [0.25,0.85] from  -0.1 -0.5 to  -0.3 -0.5
    \arrow <1.75mm> [0.25,0.85] from  -0.1 -1.5 to  -0.3 -1.5
    \arrow <1.75mm> [0.25,0.85] from  0.1 0.5 to  0.3 0.5
    \arrow <1.75mm> [0.25,0.85] from  0.1 1.5 to  0.3 1.5
    \put{$\scriptstyle{A}$}[b] at  -2 -0.2
    \put{$\scriptstyle{B}$}[b] at  0 1.9
    \put{$\scriptstyle{C}$}[b] at  0 0.8
    \put{$\scriptstyle{K}$}[b] at  2 -0.2
    \put{$\scriptstyle{L}$}[t] at  0 -0.8
    \put{$\scriptstyle{M}$}[t] at  0 -1.9
\endpicture
\raisebox{-0.5cm}\quad.
\end{eqnarray}

The ball graphs obey the following symmetries
\begin{eqnarray*}
(A,B,C|K,L,M)_1 &\stackrel{\mbox{}}{=} & (A,C,B|K,L,M)_1\\
&\stackrel{\mbox{flip vertical}}{=} &
 (\tilde{K},\tilde{B},\tilde{C}|\tilde{A},\tilde{L},\tilde{M})_1\\
&\stackrel{\mbox{flip horizontal}}{=} &
 (\tilde{A},\tilde{L},\tilde{M}|\tilde{K},\tilde{B},\tilde{C})_1,
\end{eqnarray*}
whereas propagators with a tilde have to be read backwards
($\tilde G_{\mu\nu}(x,y)=G_{\nu\mu}(y,x)$). Thus we are left with
16 possible representations for each ball graph.

Due to their symmetries the graphs of the second topology
are called tetrahedron graphs. We introduce the following
bracket notation respectively graphical representation:
\begin{eqnarray}
(A,B,C|K,L,M)_2 &:=& f_{a_1 b_1 c_1} f_{a_2 b_2 c_2}
                   f_{a_3 b_3 c_3} f_{a_4 b_4 c_4}\nonumber\\
                && A^{a_1a_2} B^{b_1b_4} C^{c_1c_3}
                   K^{a_4a_3} L^{b_3b_2} M^{c_2c_4}
                   \rule[0mm]{0mm}{5mm}\nonumber\\
 &=& \ \beginpicture
    \setcoordinatesystem units <0.5 cm , 0.5 cm>
    \setplotarea x from -1 to 1, y from -1.5 to 1.5
    \tetraeder
    \pfeiluIVlinks
    \pfeilrVrunter
    \pfeillVrunter
    \pfeiltlIVhoch
    \pfeiltrIVrunter
    \pfeiltoIVrunter
    \put{$\scriptstyle{L}$} at  0     -1.45
    \put{$\scriptstyle{A}$} at -1.1    1.05
    \put{$\scriptstyle{C}$} at  1.15   0.85
    \put{$\scriptstyle{M}$} at -0.3   -0.6
    \put{$\scriptstyle{K}$} at  0.6    0.1
    \put{$\scriptstyle{B}$} at -0.35   0.55
\endpicture.
\end{eqnarray}

Since the tetrahedron graphs show up the following
symmetries
\begin{eqnarray*}
(A,B,C|K,L,M)_2\stackrel{\mbox{cyclic}}{=}
(B,C,A|L,M,K)_2\stackrel{\mbox{inverted}}{=}
(C,B,A|\tilde{M},\tilde{L},\tilde{K})_2\\
\stackrel{\mbox{rotated}}{=}
(\tilde{L},M,\tilde{A}|\tilde{B},C,\tilde{K})_2,\\
\end{eqnarray*}
each graph has 24 representations.

\begin{landscape}
\begin{figure}
\begin{eqnarray*}
\Gamma^{(3)}_\text{Ball}
&=& \quad\frac{9}{4}\graph{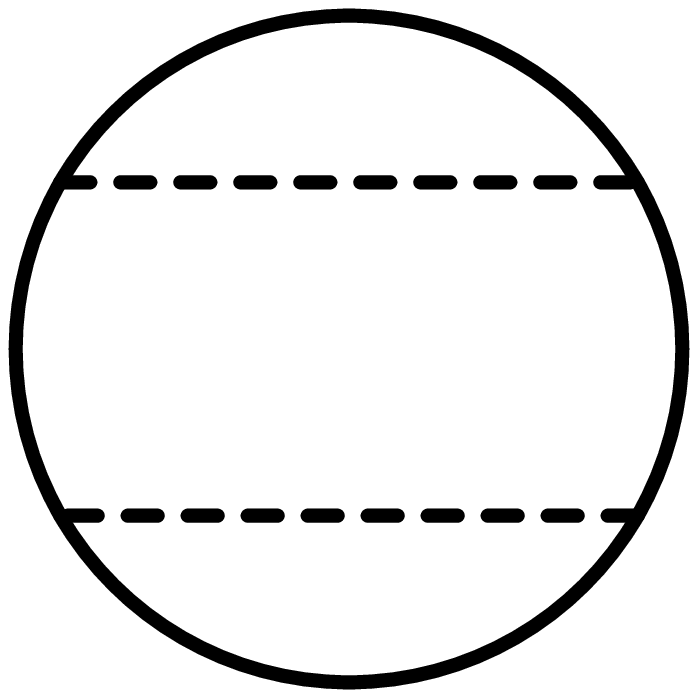}{2.5}
   + \frac{1}{8}\left(\graph{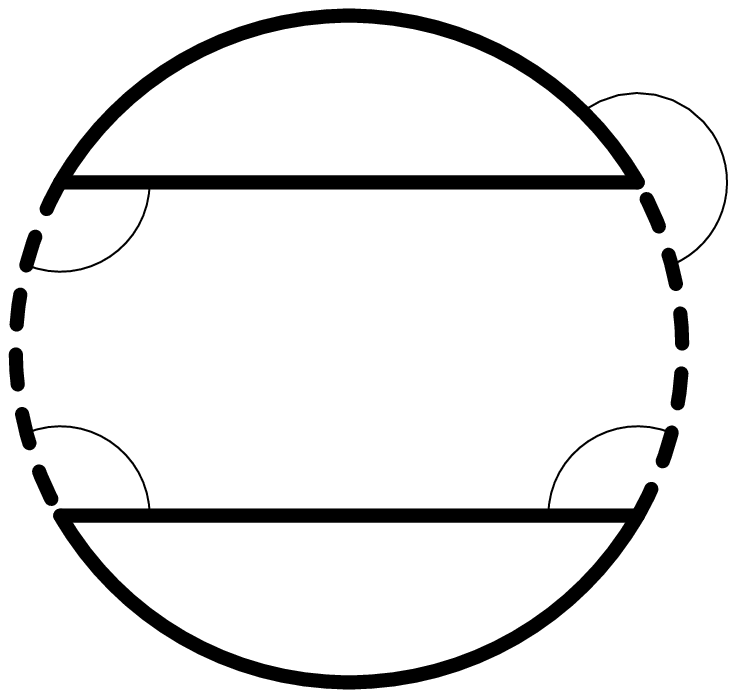}{2.5}
   - \graph{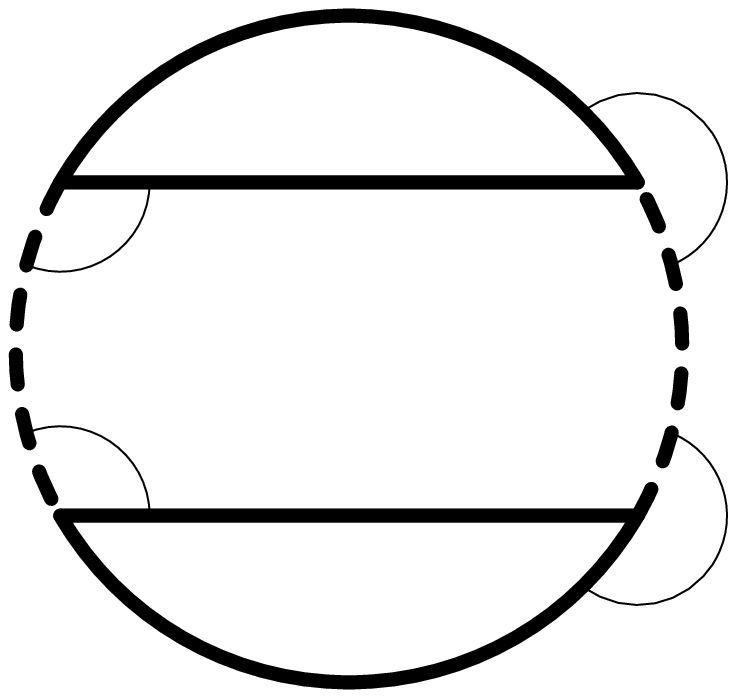}{2.5}\right)
   + \frac{3}{4}\left(
     2\graph{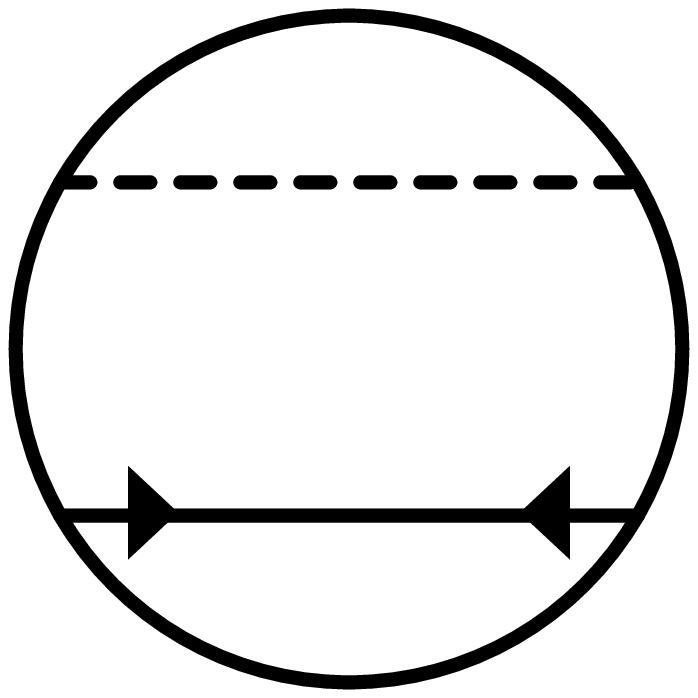}{2.5}
      \raisebox{0.6cm}{\hspace{-0.5cm}$\ast$}
   + 2\left(\graph{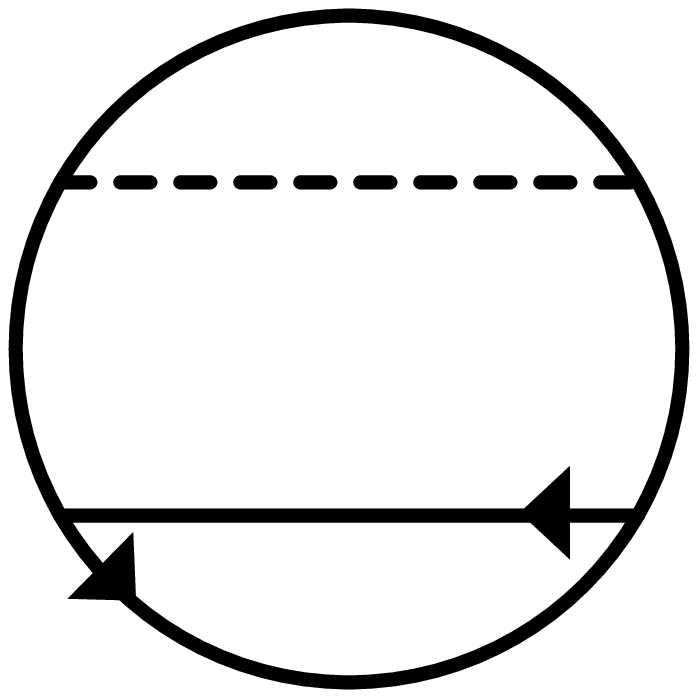}{2.5}
      \raisebox{0.6cm}{\hspace{-0.5cm}$\ast$\hspace{0.2cm}}
   -        \graph{BallSchneemann2}{2.5}\right)
     \right.\\
&&   \left. \mbox{} +
     4 \graph{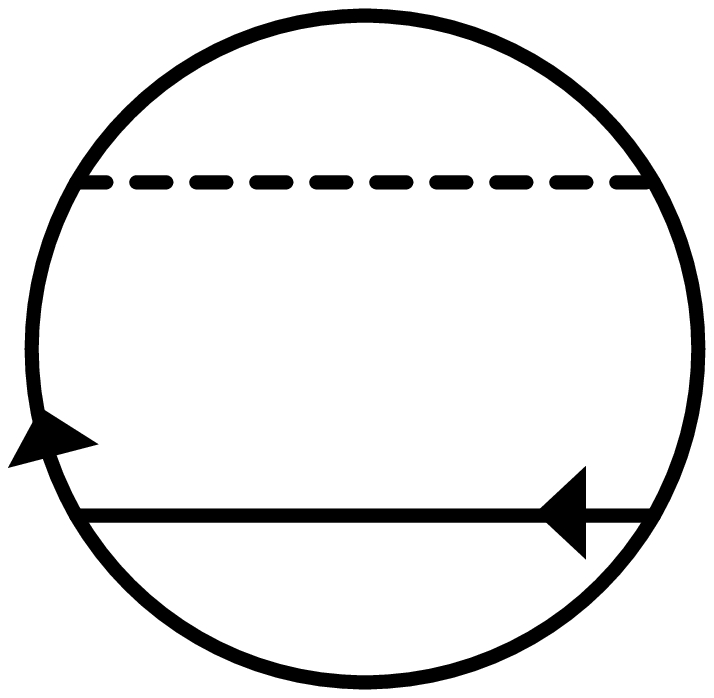}{2.5}
       \raisebox{0.6cm}{\hspace{-0.5cm}$\ast$}
   + \graph{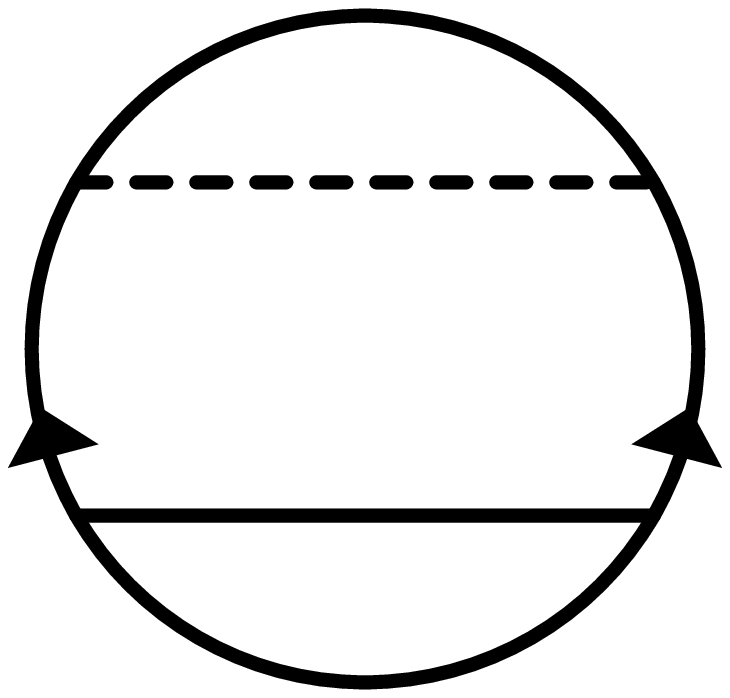}{2.5} 
     \raisebox{0.6cm}{\hspace{-0.5cm}$\ast$}\right)
   + \frac{1}{8}\left(2\graph{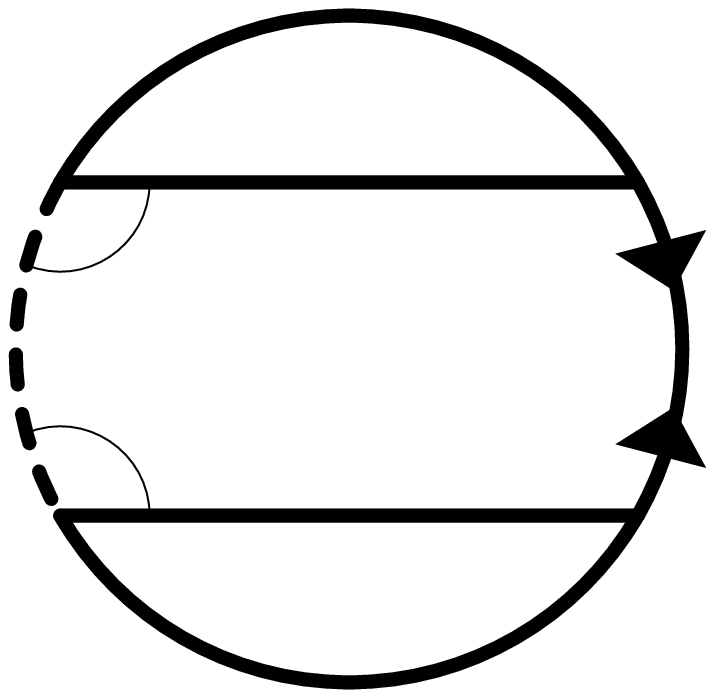}{2.5}
                        \raisebox{0.6cm}{\hspace{-0.5cm}$\ast$}
   +                  8\graph{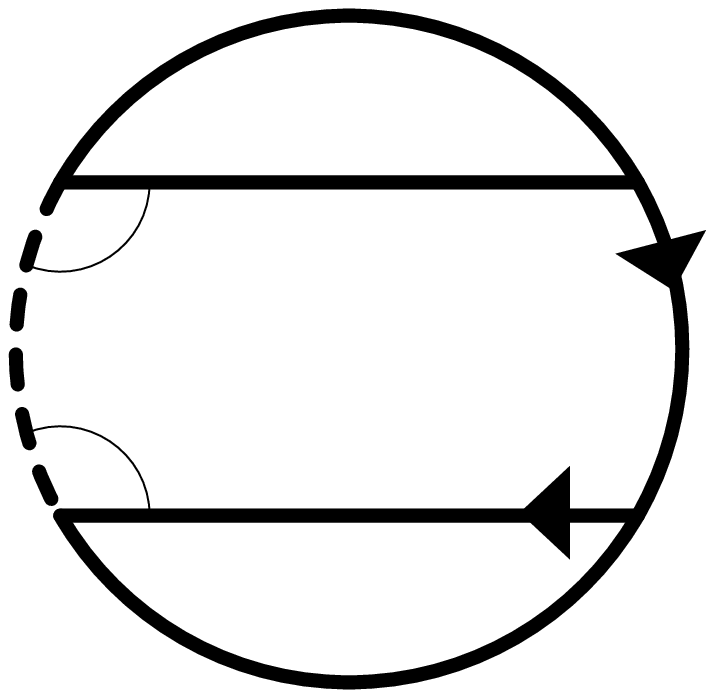}{2.5}
                       \raisebox{0.6cm}{\hspace{-0.5cm}$\ast$}
   +                  4\graph{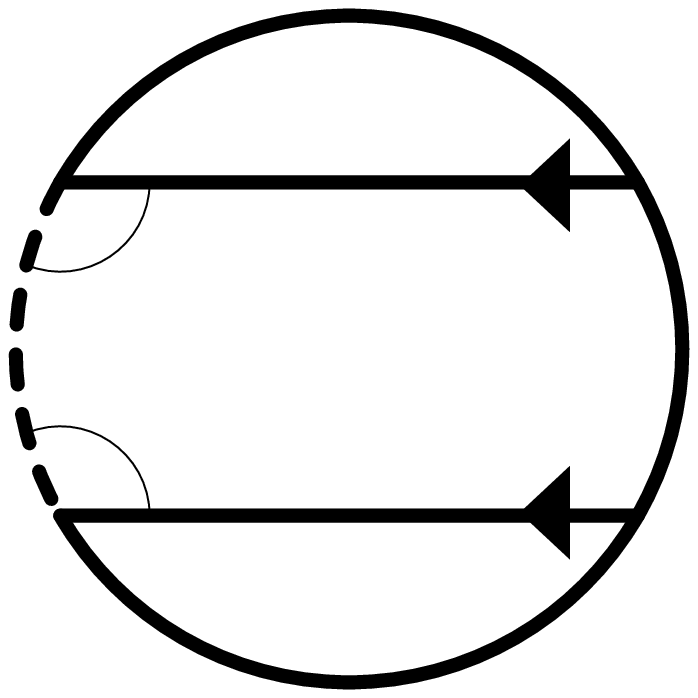}{2.5}
                       \raisebox{0.6cm}{\hspace{-0.5cm}$\ast$}
   +                  4\graph{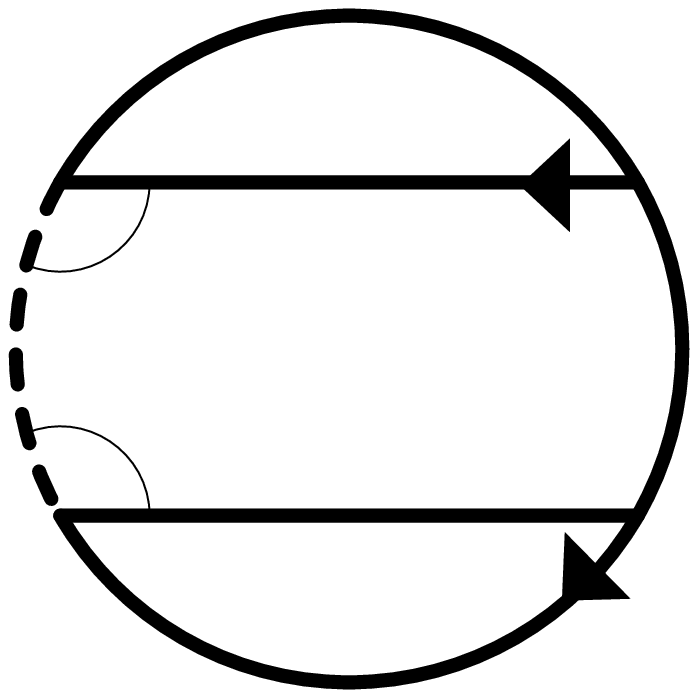}{2.5}
                       \raisebox{0.6cm}{\hspace{-0.5cm}$\ast$}
                \right)\\
&& \mbox{} +
   \frac{1}{16}\left(
      4\graph{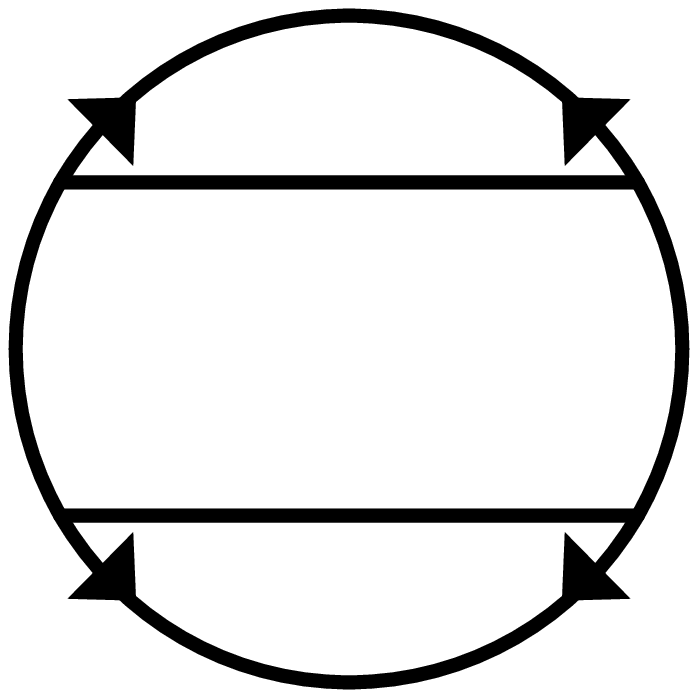}{2.5}
       \raisebox{0.6cm}{\hspace{-0.5cm}$\ast$}
   +  8\left( \graph{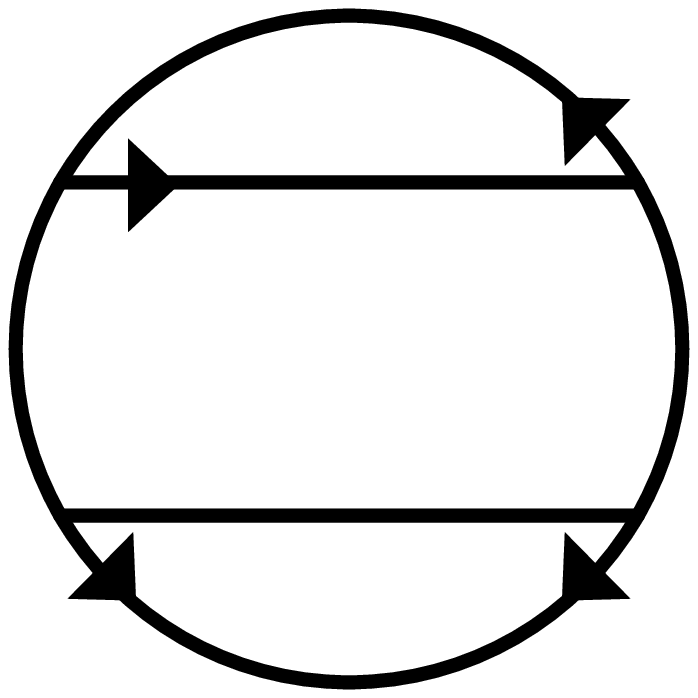}{2.5}
              \raisebox{0.6cm}{\hspace{-0.5cm}$\ast$\hspace{0.2cm}}
   -          \graph{Ball2}{2.5}\right)
   +  4\left( \graph{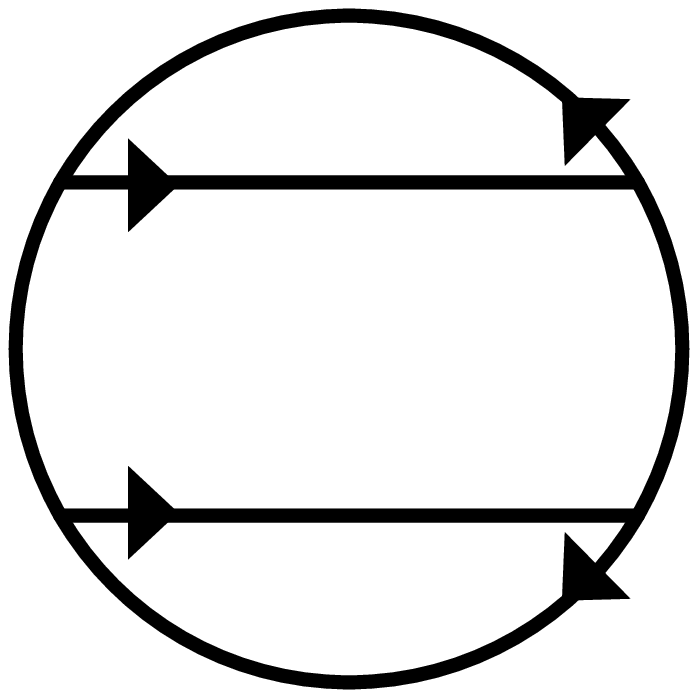}{2.5}
              \raisebox{0.6cm}{\hspace{-0.5cm}$\ast$}
   -         2\graph{Ball3}{2.5}
              \raisebox{0.6cm}{\hspace{-0.5cm}$\dagger$\hspace{0.2cm}}
   +          \graph{Ball3}{2.5}\right)
      \right.\\
&&   \left. \mbox{} +
      16\graph{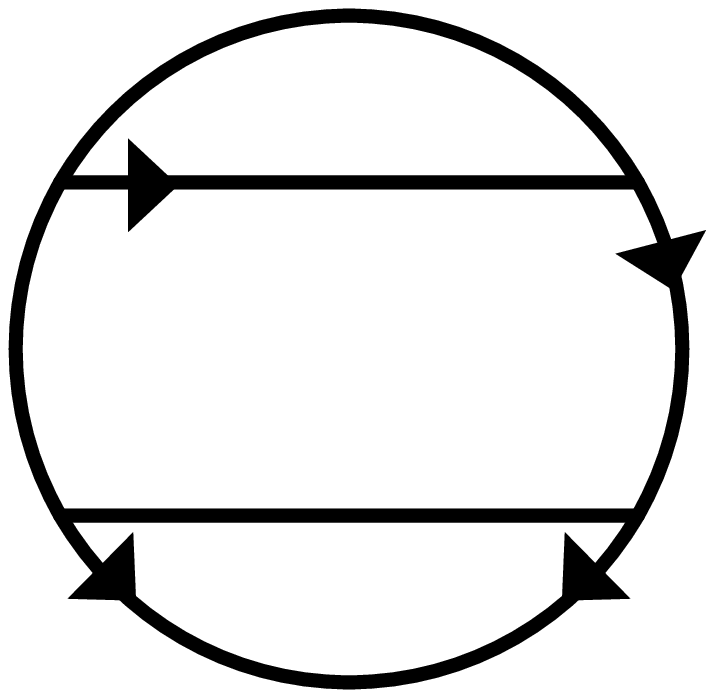}{2.5}
        \raisebox{0.6cm}{\hspace{-0.5cm}$\ast$}
   +  16\left(\graph{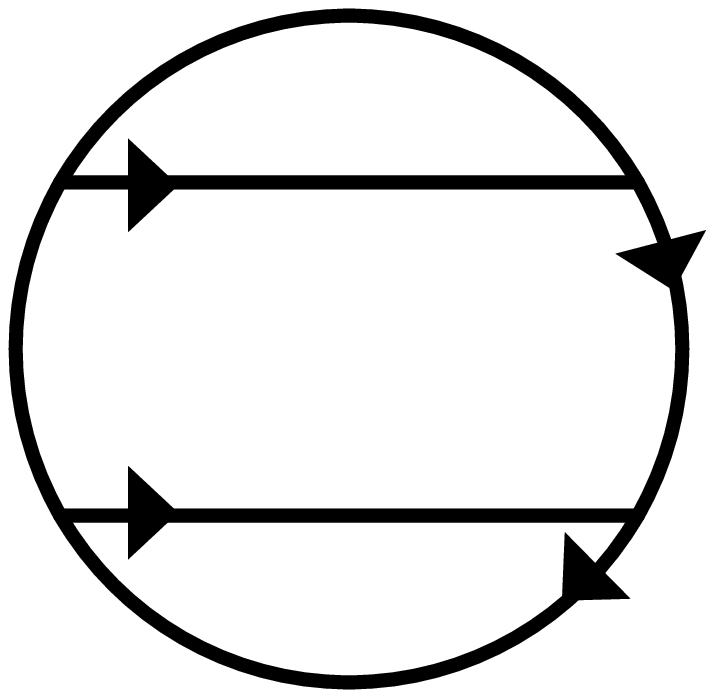}{2.5}
              \raisebox{0.6cm}{\hspace{-0.5cm}$\ast$\hspace{0.2cm}}
   -          \graph{Ball5}{2.5}\right)
   +   4\graph{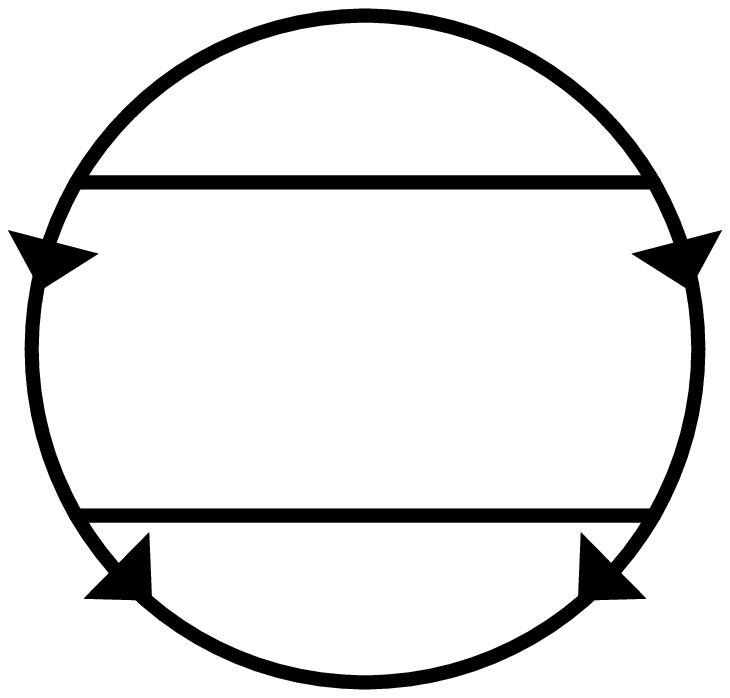}{2.5}
        \raisebox{0.6cm}{\hspace{-0.5cm}$\ast$}
   +   4\left( \graph{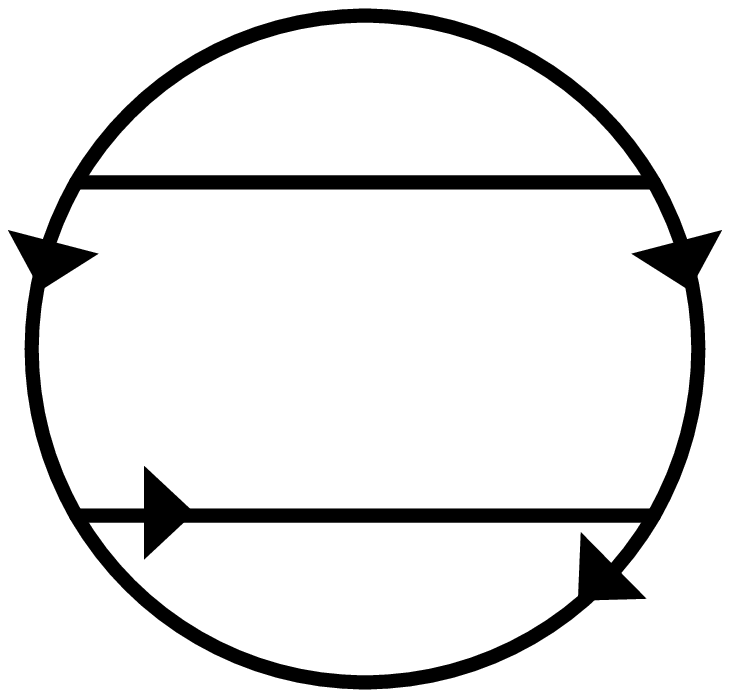}{2.5}
               \raisebox{0.6cm}{\hspace{-0.5cm}$\ast$\hspace{0.2cm}}
   -           \graph{Ball7}{2.5}\right)
     \right.\\
&&   \left. \mbox{} +
       8\left(\graph{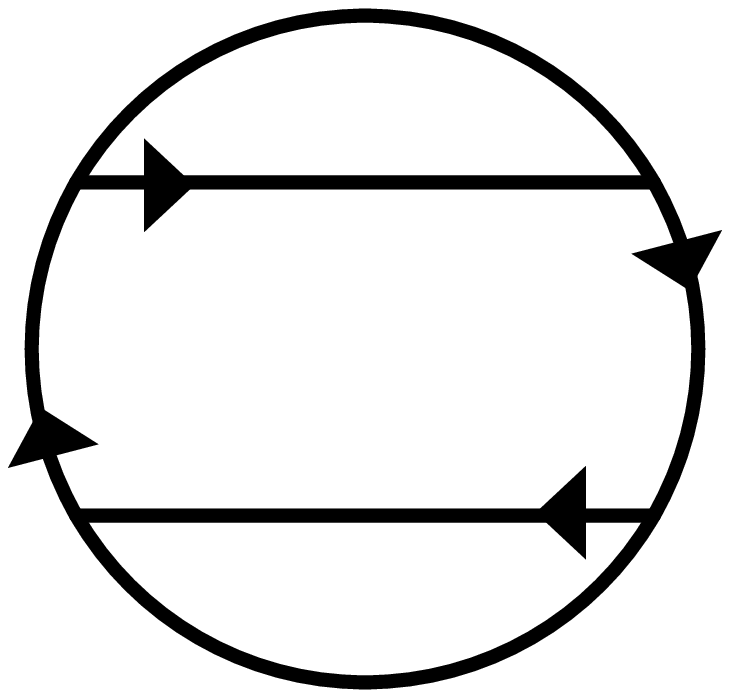}{2.5}
              \raisebox{0.6cm}{\hspace{-0.5cm}$\ast$\hspace{0.2cm}}
   -          \graph{Ball8}{2.5}\right)
   +   8\graph{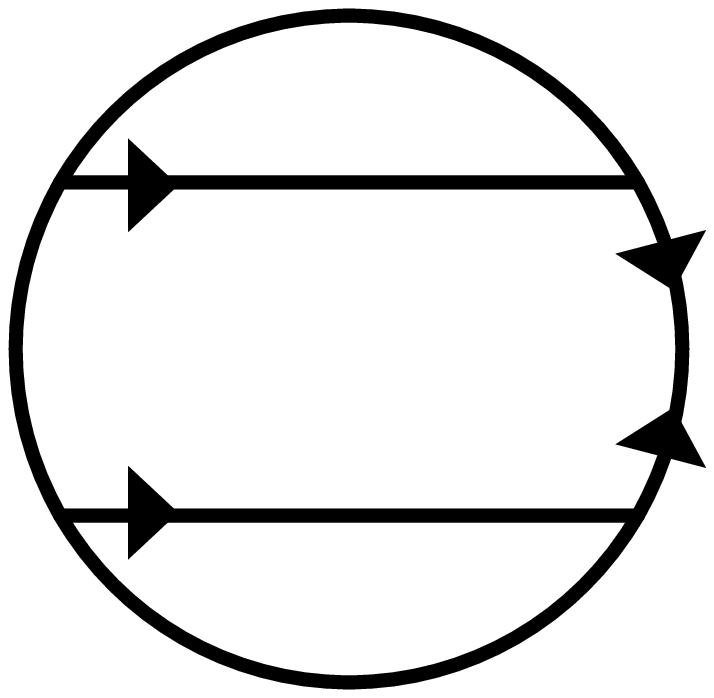}{2.5}
        \raisebox{0.6cm}{\hspace{-0.5cm}$\ast$}
   +   8\graph{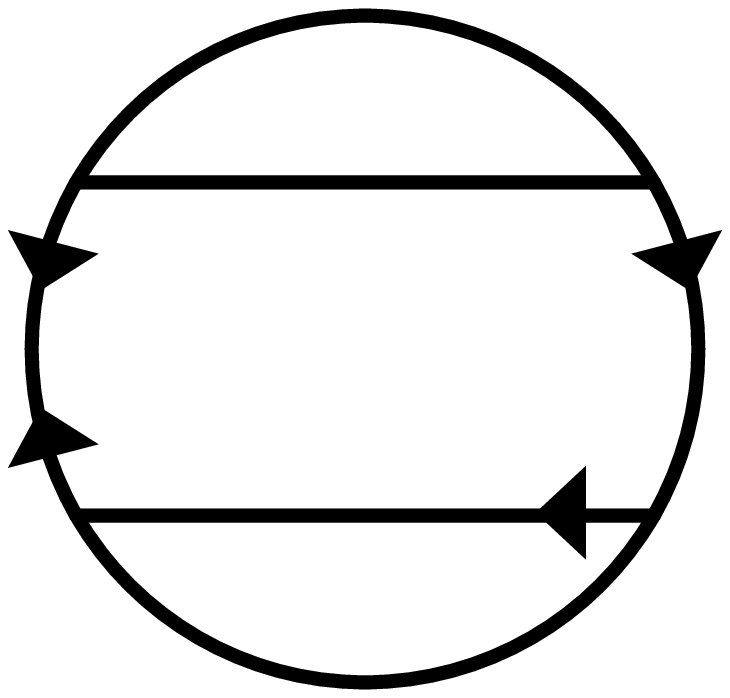}{2.5}
        \raisebox{0.6cm}{\hspace{-0.5cm}$\ast$}
   +    \graph{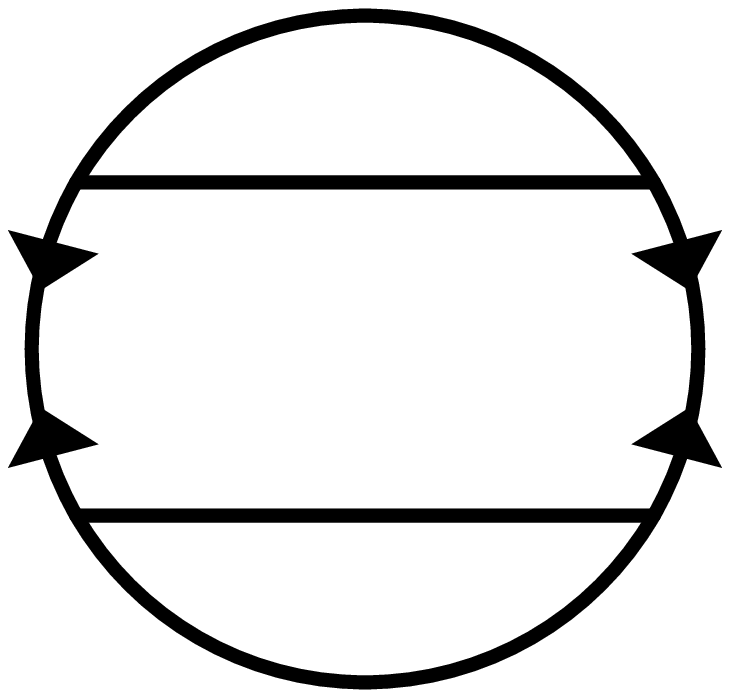}{2.5}
        \raisebox{0.6cm}{\hspace{-0.5cm}$\ast$}\right)\\
\end{eqnarray*}
\caption{Contribution of the ball graphs to the three-loop
         effective action}
\label{Ball3loop}
\end{figure}
\end{landscape}

\begin{landscape}
\begin{figure}
\begin{eqnarray*}
\Gamma^{(3)}_\text{Tetrahedron}
&=&\frac{1}{48}\left(
      6\hspace{-0.3cm} \graph{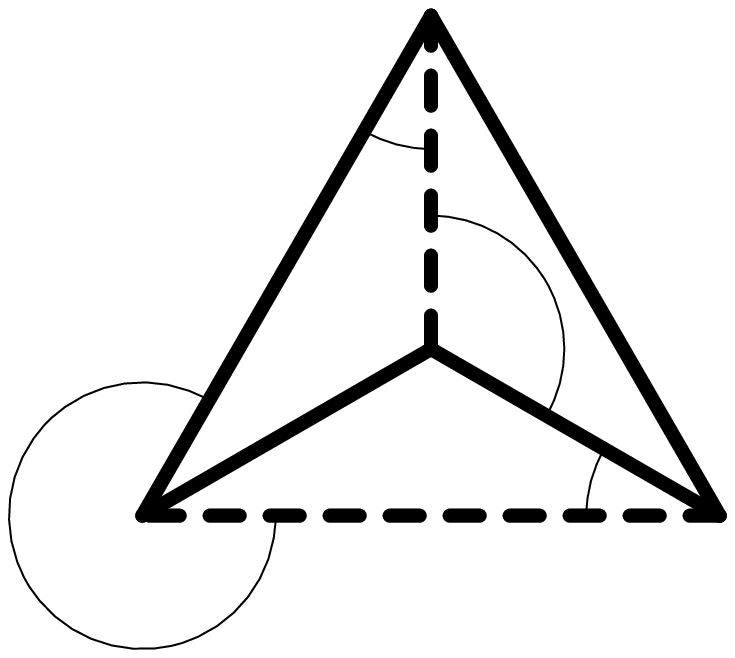}{2.8}\hspace{-0.3cm} +
      6\hspace{-0.3cm} \graph{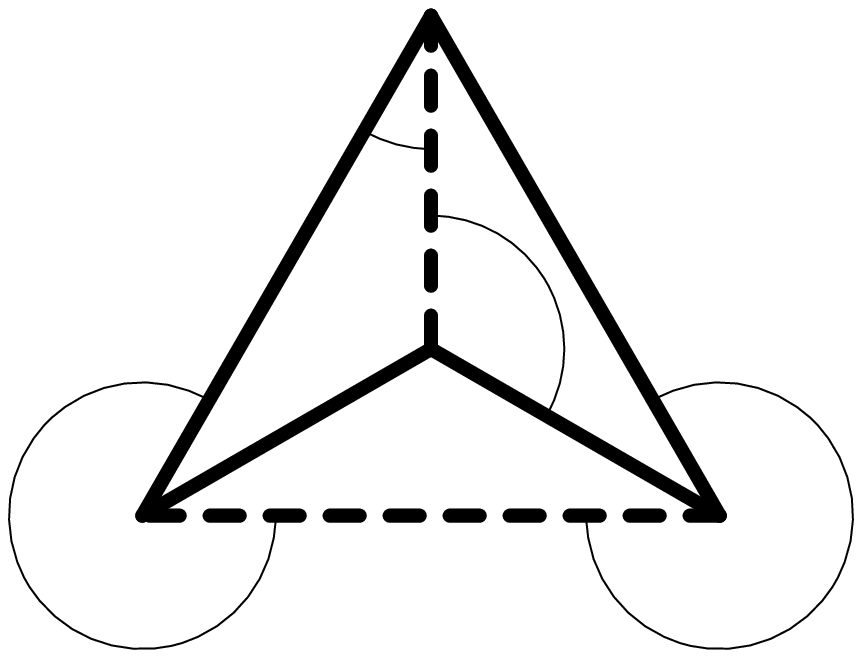}{2.8}\hspace{-0.3cm} +
      6\hspace{-0.3cm} \graph{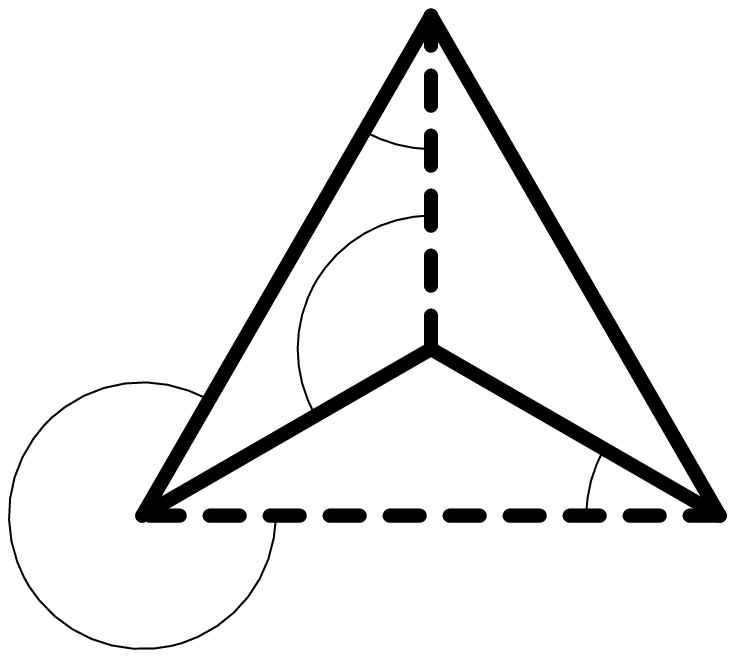}{2.8}\hspace{-0.3cm} + 
      6\hspace{-0.3cm} \graph{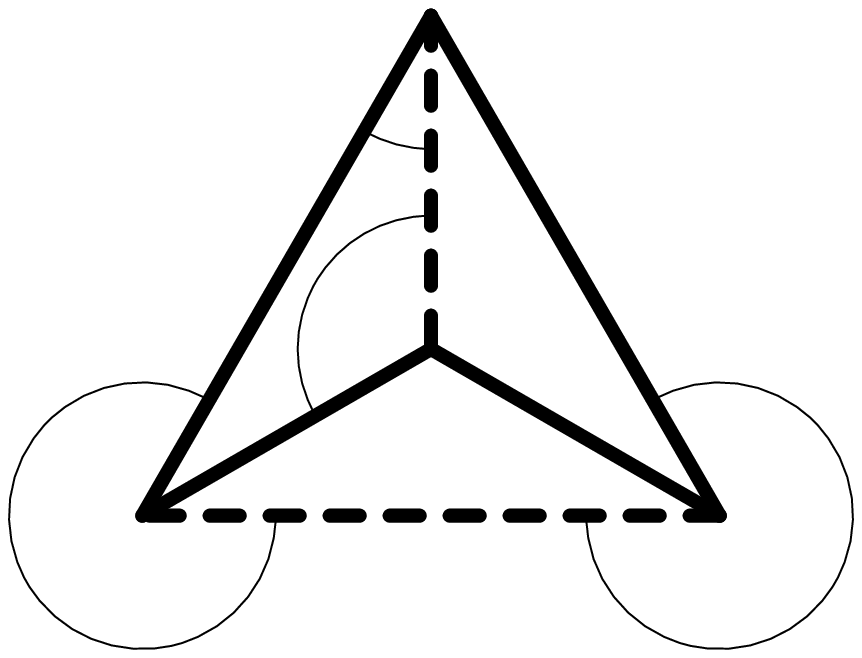}{2.8}\hspace{-0.2cm} 
      \right)\\
& &\mbox{} +
   \frac{1}{8}\left(
      2\hspace{-0.3cm} \graph{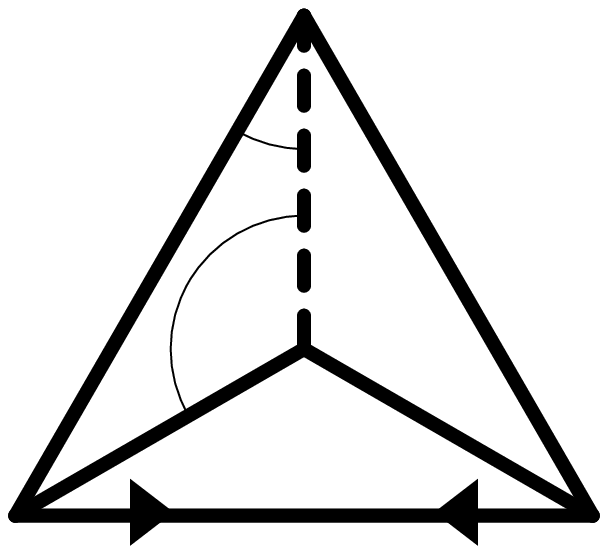}{2.8}
       \raisebox{0.9cm}{\hspace{-1cm}$\ast$\hspace{0.3cm}} +
      2\hspace{-0.3cm} \graph{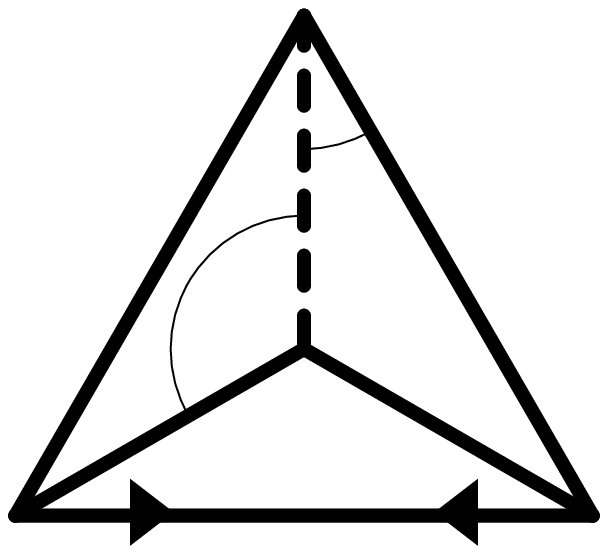}{2.8}
       \raisebox{0.9cm}{\hspace{-1cm}$\ast$\hspace{0.3cm}} +
      8\hspace{-0.3cm} \graph{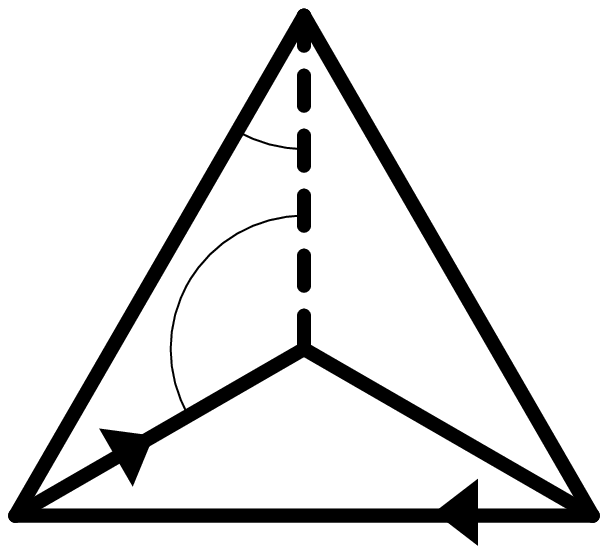}{2.8}
       \raisebox{0.9cm}{\hspace{-1cm}$\ast$\hspace{0.3cm}} +
      8\hspace{-0.3cm} \graph{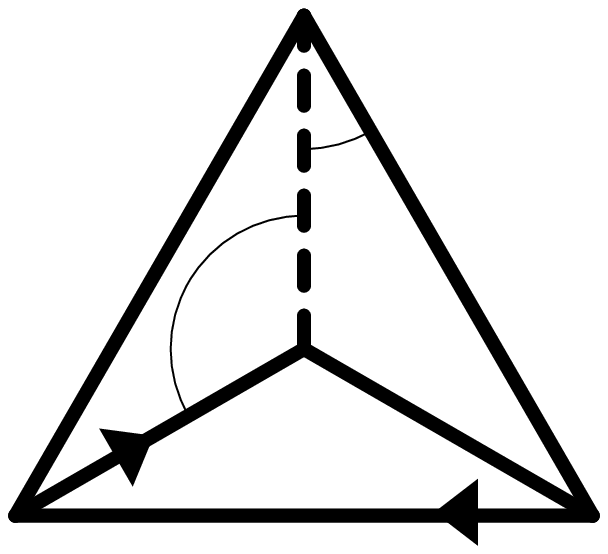}{2.8}
       \raisebox{0.9cm}{\hspace{-1cm}$\ast$\hspace{0.3cm}} +
      4\hspace{-0.3cm} \graph{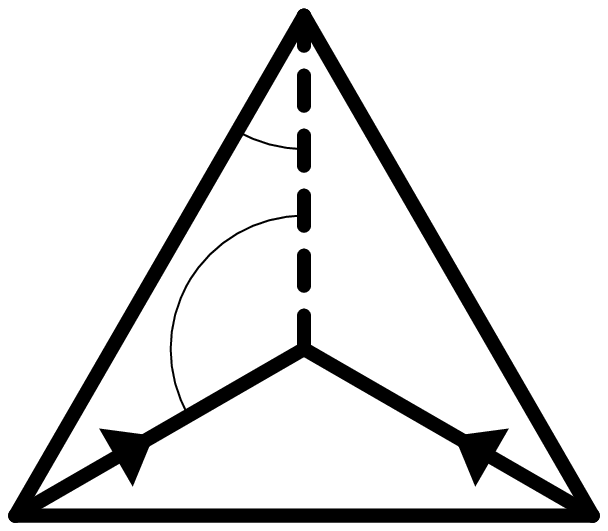}{2.8}
       \raisebox{0.9cm}{\hspace{-1cm}$\ast$\hspace{0.3cm}} +
      4\hspace{-0.3cm} \graph{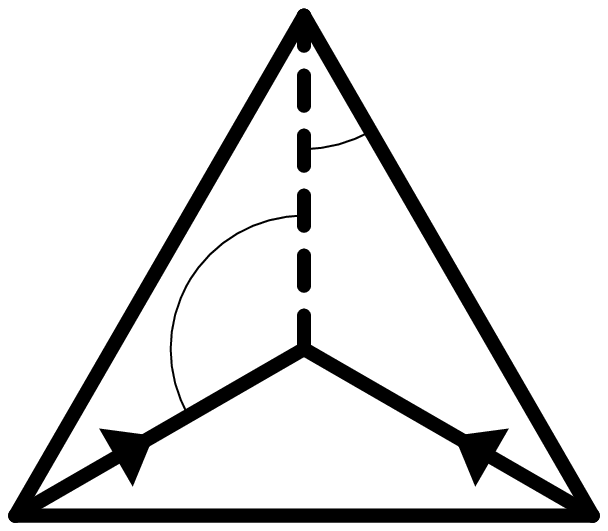}{2.8}
       \raisebox{0.9cm}{\hspace{-1cm}$\ast$\hspace{0.3cm}} 
      \right.\\
& &\qquad\left.\mbox{} +
      4\hspace{-0.3cm} \graph{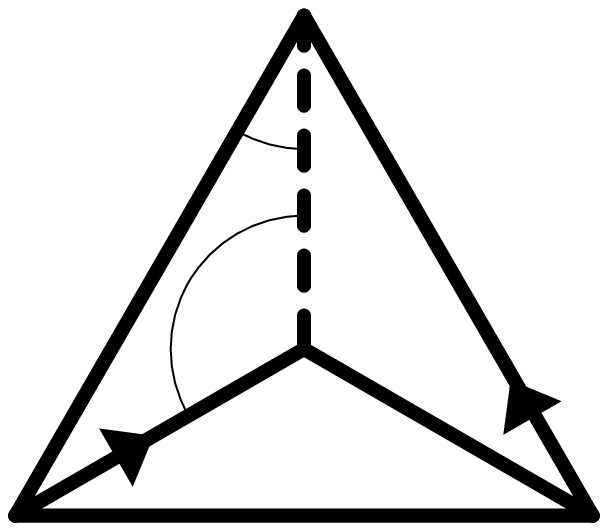}{2.8}
       \raisebox{0.9cm}{\hspace{-1cm}$\ast$\hspace{0.3cm}} +
      4\hspace{-0.3cm} \graph{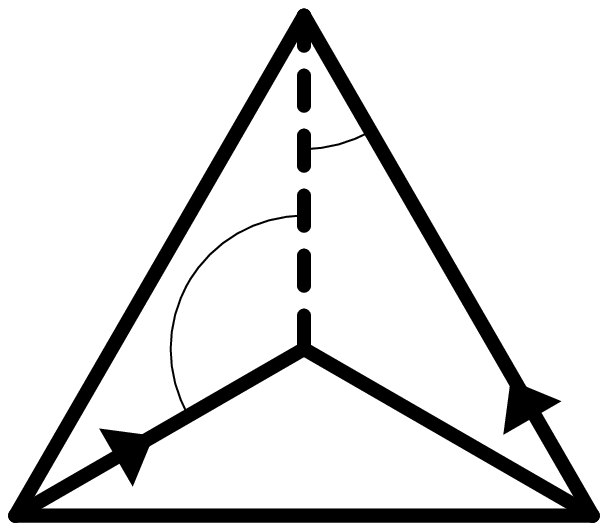}{2.8}
       \raisebox{0.9cm}{\hspace{-1cm}$\ast$} \quad\right)\\
& &\mbox{} +
   \frac{1}{24}\left(
      3\hspace{-0.3cm} \graph{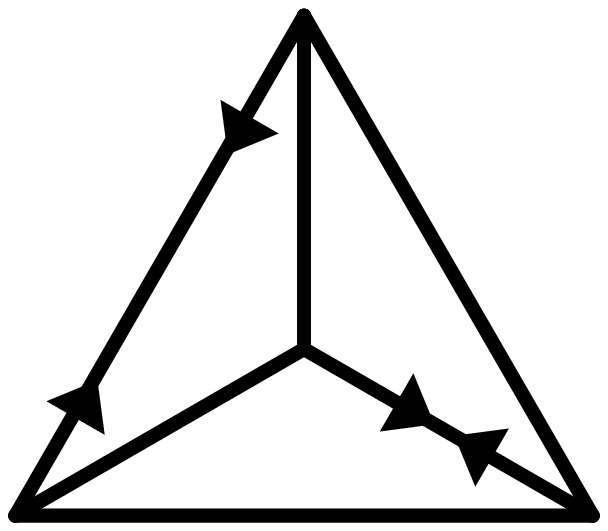}{2.8}
       \raisebox{0.9cm}{\hspace{-1cm}$\ast$\hspace{0.3cm}}+
      6\left(\hspace{-0.3cm} \graph{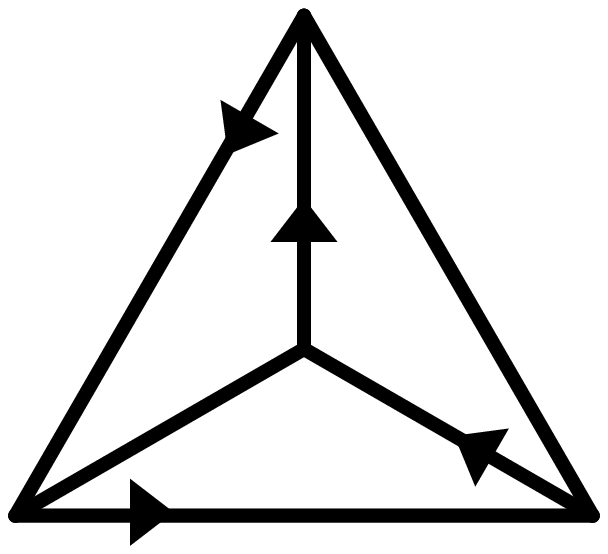}{2.8}
             \raisebox{0.9cm}{\hspace{-1cm}$\ast$\hspace{0.3cm}}-
             \hspace{-0.3cm} \graph{Tetraeder2}{2.8}\hspace{-0.3cm}
       \right)+
     12\hspace{-0.3cm} \graph{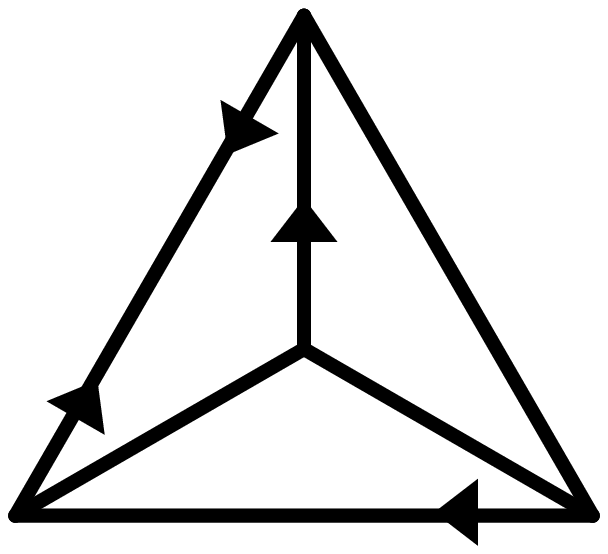}{2.8}
       \raisebox{0.9cm}{\hspace{-1cm}$\ast$\hspace{0.3cm}}+
     24\hspace{-0.3cm} \graph{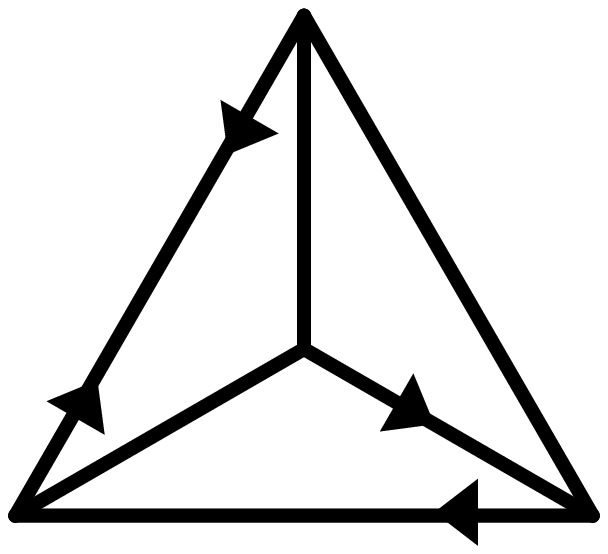}{2.8}
       \raisebox{0.9cm}{\hspace{-1cm}$\ast$\hspace{0.3cm}}+
     12\hspace{-0.3cm} \graph{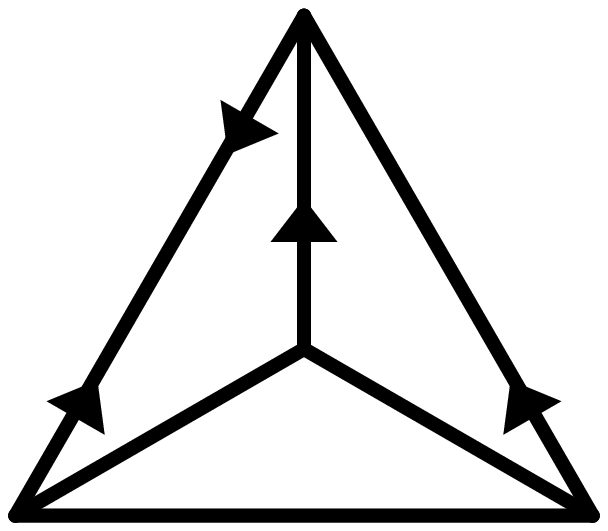}{2.8}
       \raisebox{0.9cm}{\hspace{-1cm}$\ast$\hspace{0.3cm}}
   \right.\\
& &\qquad\left.\mbox{} +
     24\left(\hspace{-0.3cm} \graph{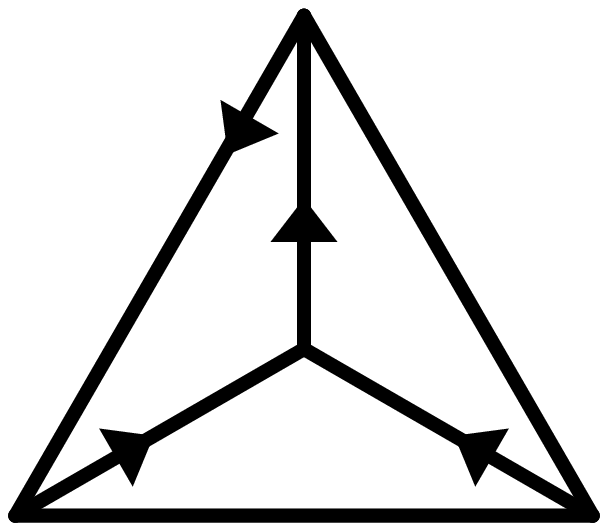}{2.8}
             \raisebox{0.9cm}{\hspace{-1cm}$\ast$\hspace{0.3cm}}-
	     \hspace{-0.3cm} \graph{Tetraeder6}{2.8}
	     \raisebox{0.9cm}{\hspace{-1cm}$\dagger$\hspace{0.5cm}}
       \right)\right)
\end{eqnarray*}
\caption{Contribution of the tetrahedron graphs to the three-loop
         effective action}
\label{Tetraeder3loop}
\end{figure}
\end{landscape}

After using the Ward identity~(\ref{PgI}) to eliminate
all ghost propagators the three-loop effective action
consists of the ball and tetrahedron graphs found in
figure~\ref{Ball3loop} and \ref{Tetraeder3loop}.

\begin{figure}[h]
\begin{eqnarray*}
     \hspace{-0.8cm}\graph{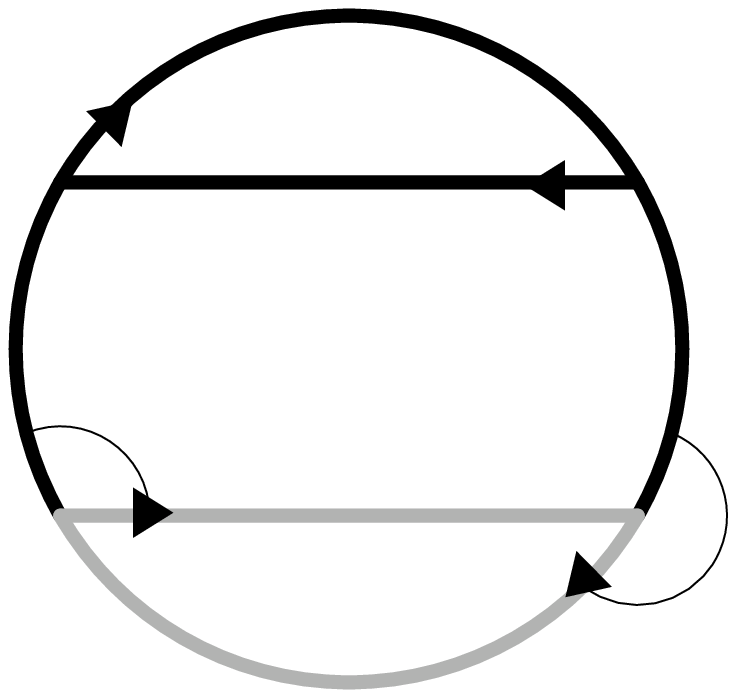}{2.8}\raisebox{0.6cm}
     {\hspace{-0.5cm}$\ast$\hspace{0.2cm}}
&=&  \begin{array}{c}
     \graph{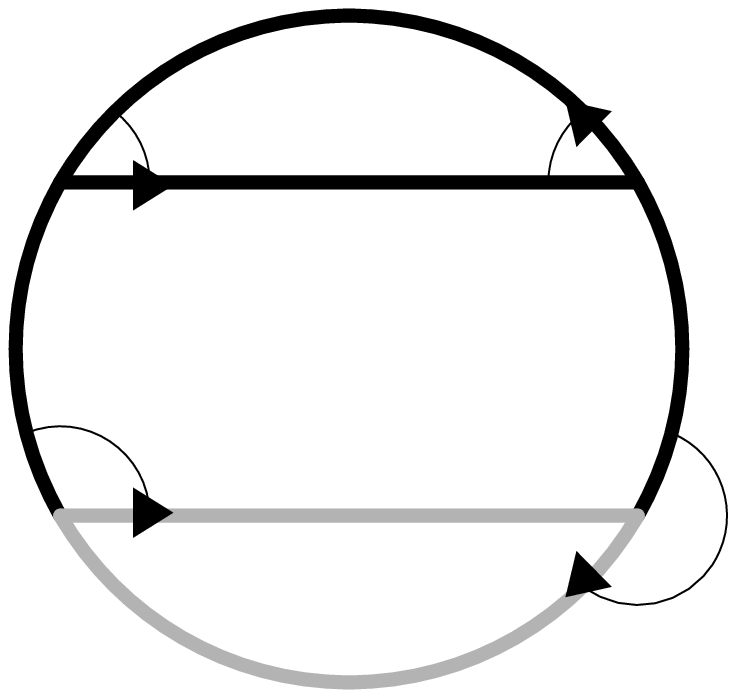}{2.8} \\[-0.5cm]
     \scriptstyle{\bigl(G_{\mu\nu},D_{\ro} G_{\nu\si},
     G_{\ro\ta} \overleftarrow{D}_{\si}|}\\
     \scriptstyle{\phantom{\bigl(}G_{\ta\la},
     G \overleftarrow{D}_{\mu},D_{\la} G\bigr)_1}
     \end{array}
    - 2
     \begin{array}{c}
     \graph{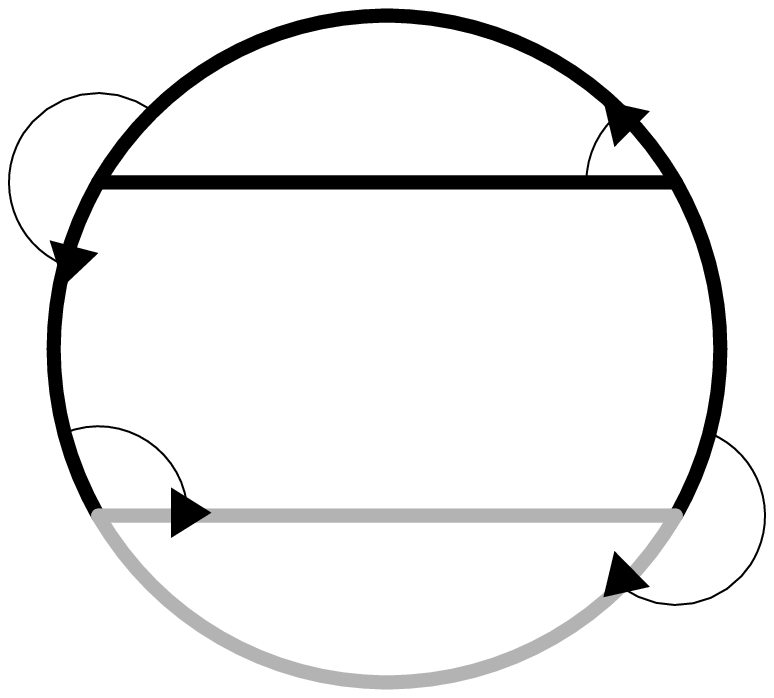}{2.8}\\[-0.5cm]
     \scriptstyle{\bigl(G_{\mu\nu} \overleftarrow{D}_{\ro}, 
     G_{\nu\si},G_{\ro\ta} \overleftarrow{D}_{\si}|}\\
     \scriptstyle{\phantom{\bigl(}G_{\ta\la},
     G D_{\mu},D_{\la} G\bigr)_1}
     \end{array}
     +
     \begin{array}{c}
     \graph{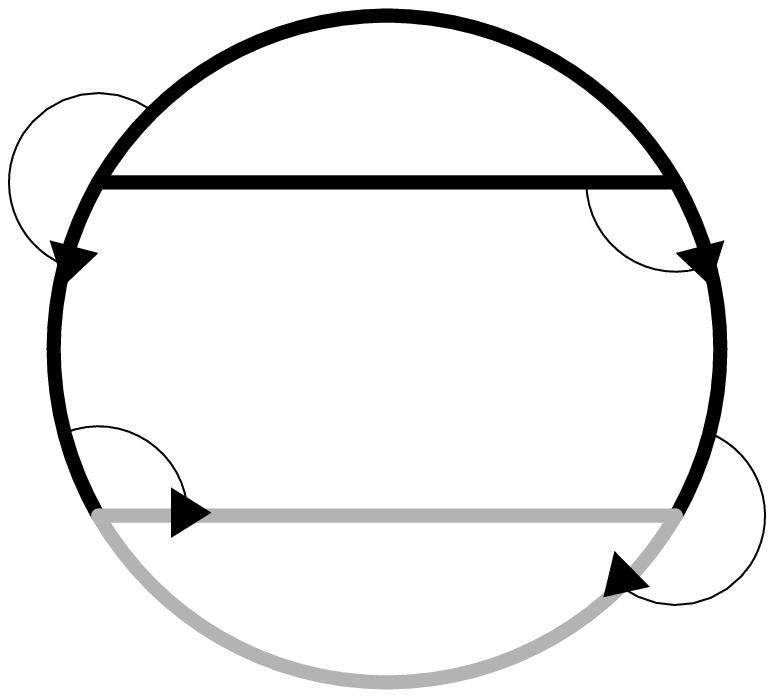}{2.8}\\[-0.5cm]
     \scriptstyle{\bigl(G_{\mu\nu} \overleftarrow{D}_{\ro}, 
     G_{\nu\si},G_{\ro\ta}|}\\
     \scriptstyle{\phantom{\bigl(}D_{\ta} G_{\si\la},
     G D_{\mu},D_{\la} G\bigr)_1}
     \end{array}\nonumber\\
&=&  \!\!\!\begin{array}{c}
     \graph{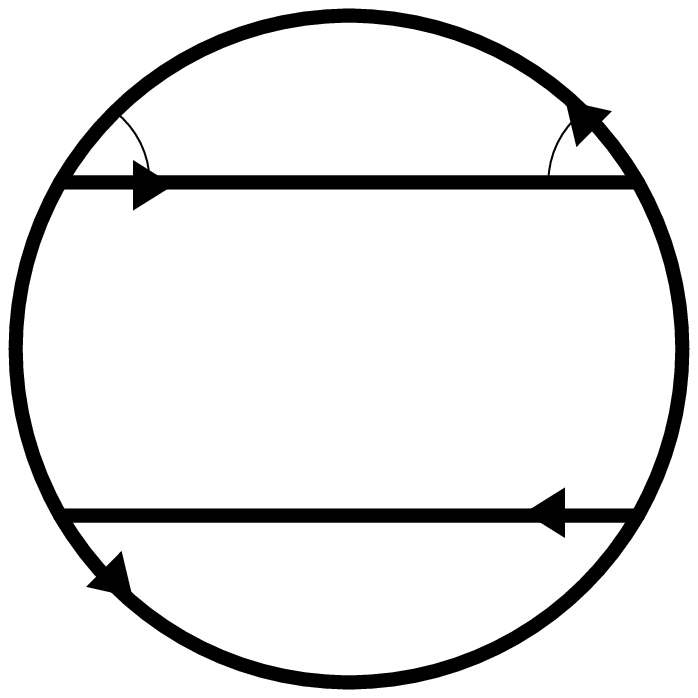}{2.8}\\[-0.5cm]
     \scriptstyle{\bigl(G_{\mu\nu},D_{\ro} G_{\nu\si},
     G_{\ro\ta} \overleftarrow{D}_{\si}|}\\
     \scriptstyle{\phantom{\bigl(}G_{\ta\la},D_{\eta} G_{\eta\mu},
     G_{\la\kappa}\overleftarrow{D}_{\kappa}\bigr)_1} 
     \end{array}\!\!\!\!\!\!\!
     - 2
     \begin{array}{c} 
     \graph{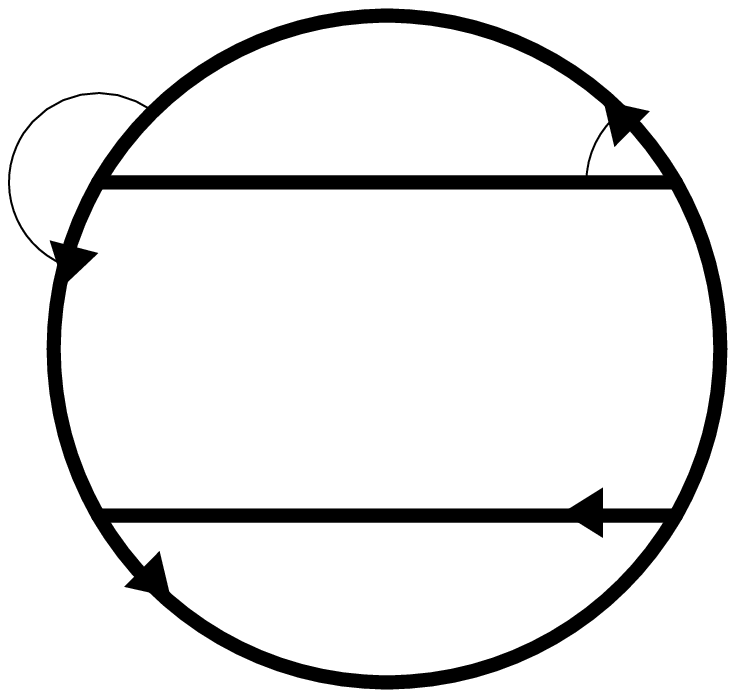}{2.8}\\[-0.5cm]
     \scriptstyle{\bigl(G_{\mu\nu} \overleftarrow{D}_{\ro}, G_{\nu\si},
     G_{\ro\ta} \overleftarrow{D}_{\si}|}\\
     \scriptstyle{\phantom{\bigl(}G_{\ta\la},D_{\eta} G_{\eta\mu},
     G_{\la\kappa}\overleftarrow{D}_{\kappa}\bigr)_1}
     \end{array} 
     +\!\!\!\!\!\!\!
     \begin{array}{c}
     \graph{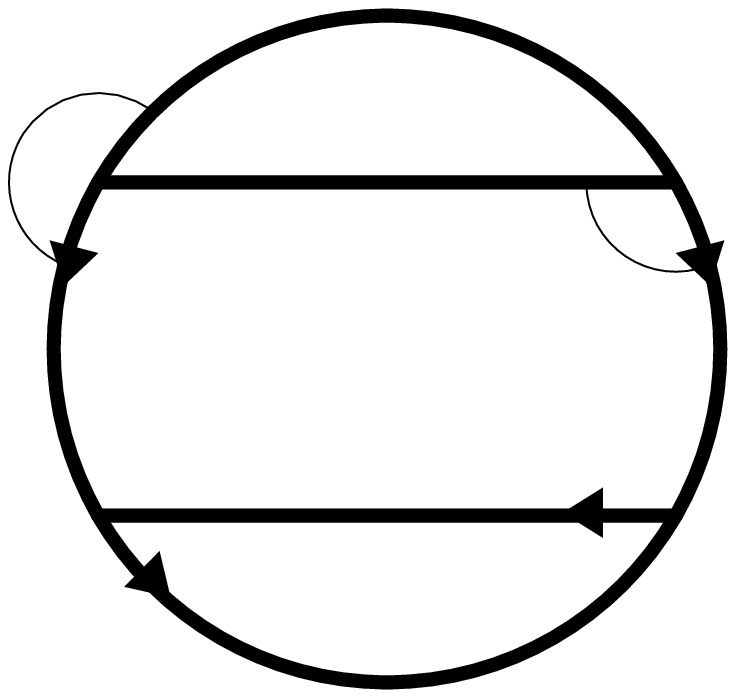}{2.8}\\[-0.5cm]
     \scriptstyle{\bigl(G_{\mu\nu} \overleftarrow{D}_{\ro}, G_{\nu\si},
     G_{\ro\ta}|}\\
     \scriptstyle{\phantom{\bigl(}D_{\ta} G_{\si\la},
     D_{\eta} G_{\eta\mu},
     G_{\la\kappa}\overleftarrow{D}_{\kappa}\bigr)_1}
     \end{array}\nonumber\\
&=&  \graph{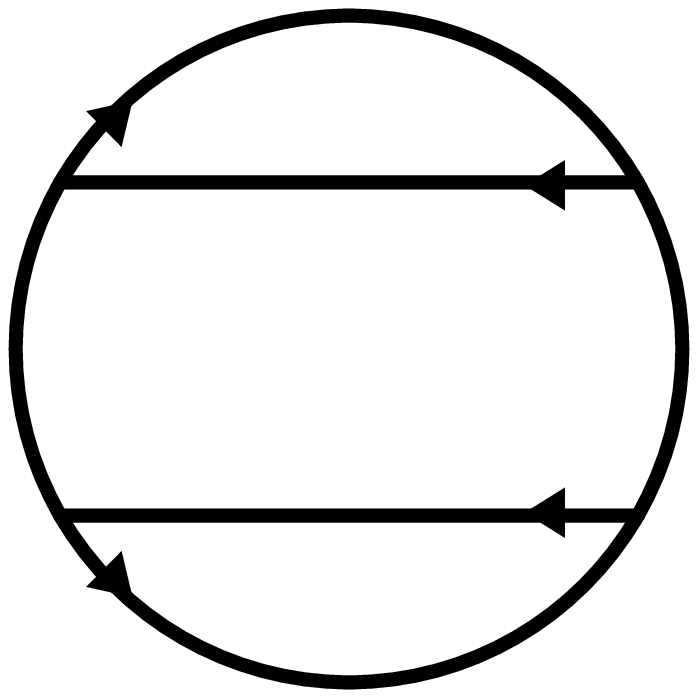}{2.8}\raisebox{0.6cm}
    {\hspace{-0.5cm}$\dagger$\hspace{0.2cm}}
\end{eqnarray*}
\caption{\label{balldagger}Hybride ball graph}
\end{figure}

The fifth graph in the third row of $\Gamma^{(3)}_\text{Ball}$
is a hybrid between a ghost graph and a star graph.
As shown in figure~\ref{balldagger} in the original graph
the lower loop is built up only by two ghost propagators, whereas
the upper loop is built up by two vector propagators with the
star operator acting on it. A graphical
representation and the corresponding bracket notation can be
found in figure~\ref{balldagger}.

\begin{figure}[H]
\begin{eqnarray*}
    \hspace{-0.3cm}\graph{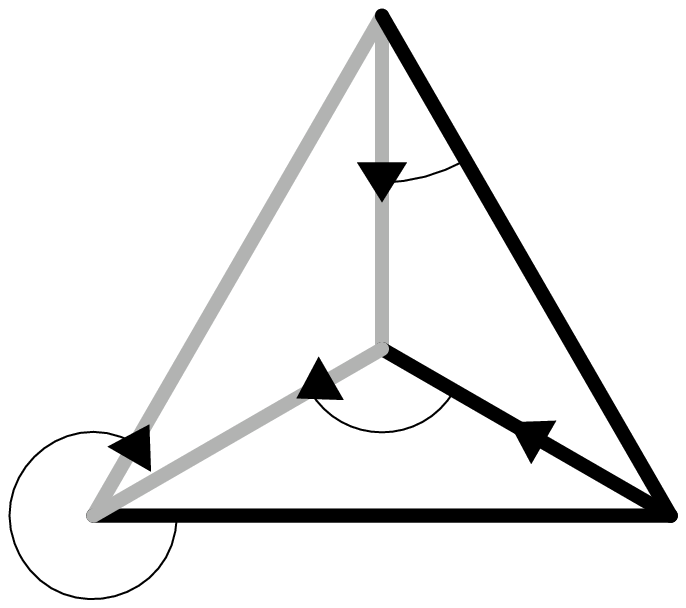}{3.5}\raisebox{0.7cm}
    {\hspace{-1.2cm}$\ast$\hspace{0.5cm}}
&=& \begin{array}{c} 
      \graph{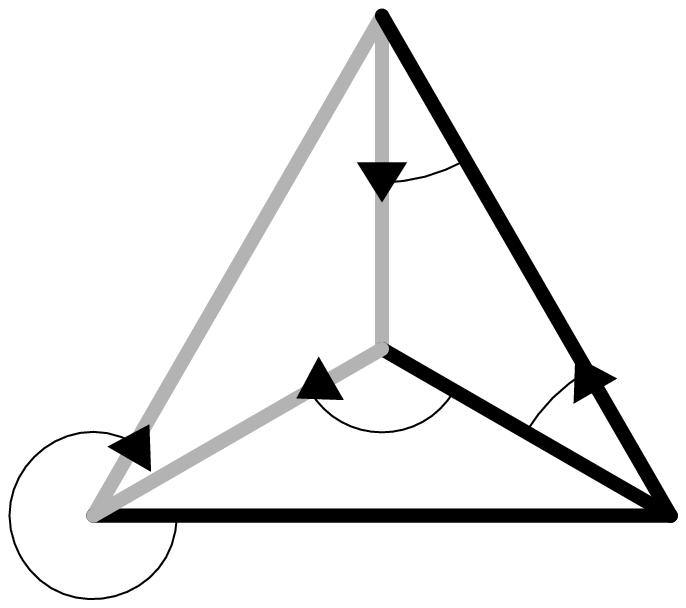}{3.5}\\[-0.9cm]
      \scriptstyle{\bigl(G \overleftarrow{D}_{\mu},D_{\nu} G,
      G_{\nu\ro} \overleftarrow{D}_{\ta}|}\\
      \scriptstyle{\phantom{\bigl(}G_{\si\ta},G_{\ro\mu},
      G \overleftarrow{D}_{\si}\bigr)_2}
    \end{array}  
    -
    \begin{array}{c}  
      \graph{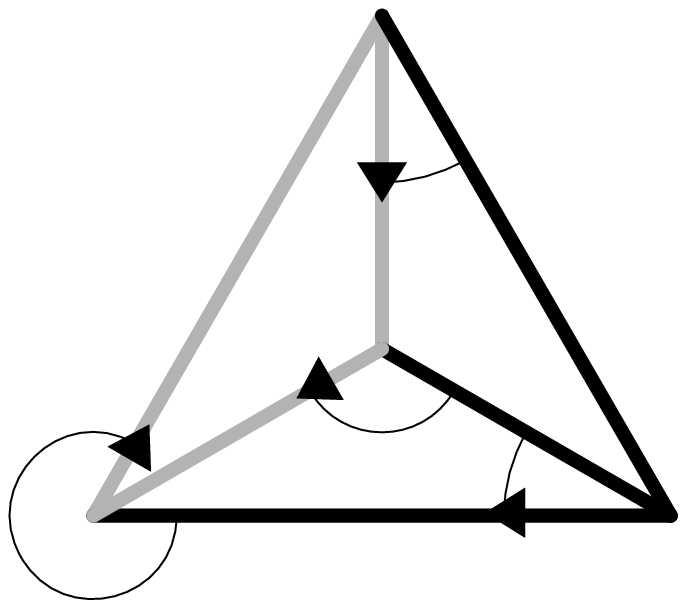}{3.5}\\[-0.9cm]
      \scriptstyle{\bigl(G \overleftarrow{D}_{\mu},D_{\nu} G,
      G_{\nu\ro} \overleftarrow{D}_{\ta}|}\\
      \scriptstyle{\phantom{\bigl(}G_{\si\ta},D_{\ta} G_{\ro\mu},
      G \overleftarrow{D}_{\si}\bigr)_2}  
    \end{array}\\
&=& -
   \begin{array}{c}
     \graph{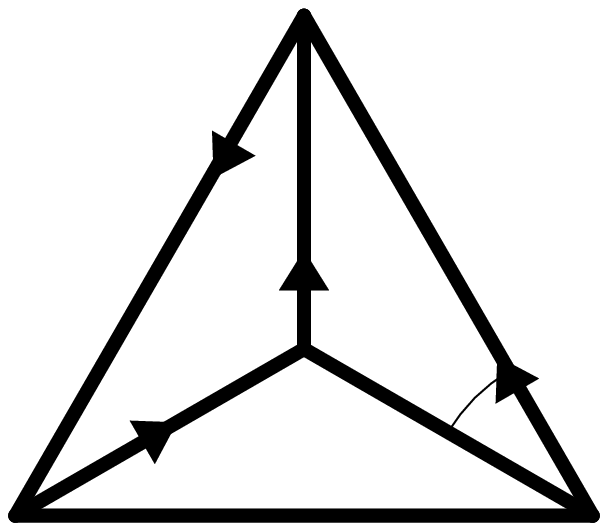}{3.5}\\[-1.0cm]
     \scriptstyle{\bigl(D_{\kappa} G_{\kappa\mu},
     G_{\nu\eta} \overleftarrow{D}_{\eta},
     G_{\nu\ro}\overleftarrow{D}_{\ta}|}\\
     \scriptstyle{\phantom{\bigl(} G_{\si\ta},
     G_{\ro\mu},D_{\la} G_{\la\si}\bigr)_2}
   \end{array}  
   + 
   \begin{array}{c}
     \graph{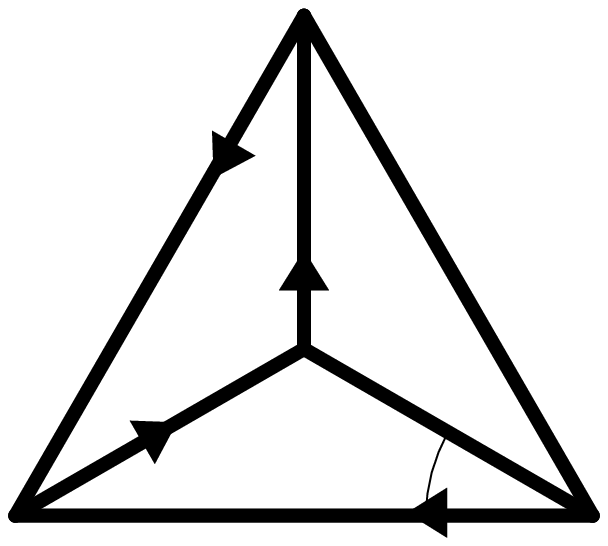}{3.5}\\[-1.0cm]
     \scriptstyle{\bigl(D_{\kappa} G_{\kappa\mu},
     G_{\nu\eta} \overleftarrow{D}_{\eta},G_{\nu\ro}|}\\
     \scriptstyle{\phantom{\bigl(} G_{\si\ta},
     D_{\ta} G_{\ro\mu},D_{\la} G_{\la\si}\bigr)_2}
   \end{array}\\ 
&=&  \graph{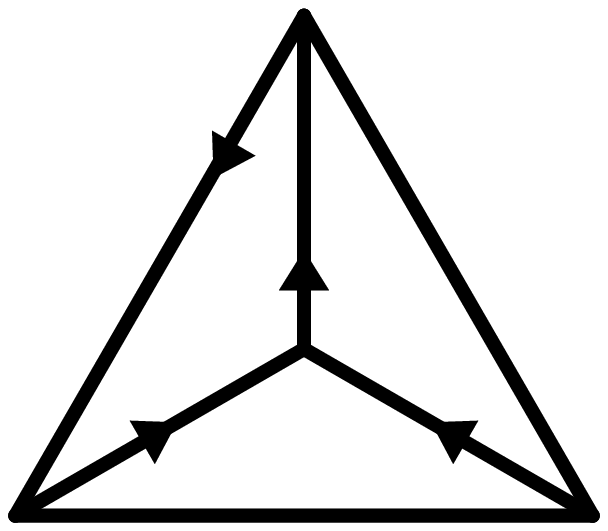}{3.5}\raisebox{0.7cm}
    {\hspace{-1.2cm}$\dagger$\hspace{0.5cm}}
\end{eqnarray*}
\caption{\label{tetdagger}Hybride tetrahedron graph}
\end{figure}

The second graph in the last row of $\Gamma^{(3)}_\text{Tetrahedron}$
in figure~\ref{Tetraeder3loop} needs some explanation also. It is
built up by a ghost loop containing three ghost propagators and
a vertex with a star operator acting on it. A graphical
representation and the corresponding bracket notation can be
looked up in figure~\ref{tetdagger}.

Taking only graphs which contain no $\delta$-function,
we see that the ball graphs can be divided in eleven classes,
whereas the tetrahedron graphs can be divided in six classes.

The graphs can be categorized in accordance with
corresponding primary graphs as shown in the first two
columns of the tables~\ref{ballklassen} and
\ref{tetklassen}. The third column holds the number
of graphs we get when applying the star operator at
each primary graph. By transforming the ghost propagators
of the ghost graphs into a gluon propagator (\ref{PgI})
we obtain primary graphs which after partial integration,
significantly reduce the number of graphs in the associated class.
Regarding classes with no matching ghost graph,
we add a suitable primary graph, perform the partial
integration and eventually substract the same primary graph.
Thus the number of graphs
containing no 4-vertex can by means of this procedure
be reduced to nearly half the number as quoted in the fourth column.
The numbers written in brackets signify the amount of
artificially added primary graphs.

\begin{table}[ H]
\begin{equation*}
\begin{array}{|c|c|c|c||c|c|c|c|}
\hline
  \mbox{class}
& \begin{array}{c} 
    \mbox{primary}\\ 
    \mbox{graph}
  \end{array}
& \begin{array}{c} 
    \mbox{number}\\ 
    \mbox{of graphs}
  \end{array}
& \mbox{reduction}
& \mbox{class}&
  \begin{array}{c} 
    \mbox{primary}\\ 
    \mbox{graph}
  \end{array}
& \begin{array}{c} 
    \mbox{number}\\ 
    \mbox{of graphs}
  \end{array}
& \mbox{reduction}\\
\hline
B.1 & \graph{Ball1}{2} &  7 & 2+(1) &  
B.7 &  \graph{Ball7}{2} &  6 & 3\\
B.2 & \graph{Ball2}{2} & 10 & 4     &  
B.8 &  \graph{Ball8}{2} & 10 & 4\\
B.3 & \graph{Ball3}{2} &  7 & 2     &  
B.9 &  \graph{Ball9}{2} & 10 & 4+(1)\\
B.4 & \graph{Ball4}{2} & 16 & 8+(1) & 
B.10 & \graph{Ball10}{2} &  8 & 4+(1)\\
B.5 & \graph{Ball5}{2} & 16 & 8     & 
B.11 & \graph{Ball11}{2} &  3 & 1+(1)\\
B.6  &  \graph{Ball6}{2} &  6 & 3+(1) &&&&\\
\hline
\multicolumn{4}{c|}{} &&& 99 & 43+(6)\\
\cline{5-8}
\end{array}
\end{equation*}
\caption{Classes of ball graphs}
\label{ballklassen}
\end{table}

Thus, as shown in table~\ref{ballklassen},
just by using the Ward identity respectively adding
some well chosen primary graphs, we reduce the number of
ball graphs (composed of 3-vertices) from originally 99 to 49.
The remaining ball graphs contain
at least one 4-vertex and
therewith at least one $\delta$-function.

\begin{table}[h]
\begin{equation*}
\begin{array}{|c|c|c|c||c|c|c|c|}
\hline
  \mbox{class}
& \begin{array}{c} 
    \mbox{primary}\\ 
    \mbox{graph}
  \end{array}
& \begin{array}{c} 
    \mbox{number}\\ 
    \mbox{of graphs}
  \end{array}
& \mbox{reduction}
& \mbox{class}&\begin{array}{c} 
    \mbox{primary}\\ 
    \mbox{graph}
  \end{array}
& \begin{array}{c} 
    \mbox{number}\\ 
    \mbox{of graphs}
  \end{array}
& \mbox{reduction}\\
\hline
T.1 & \graph{Tetraeder1}{2.2} &  4 & 1+(1) & 
T.4 & \graph{Tetraeder4}{2.2} & 16 & 8+(1)\\
T.2 & \graph{Tetraeder2}{2.2} &  6 & 2     & 
T.5 & \graph{Tetraeder5}{2.2} &  8 & 4+(1)\\
T.3 & \graph{Tetraeder3}{2.2} & 10 & 4+(1) & 
T.6 & \graph{Tetraeder6}{2.2} & 16 & 8\\
\hline
\multicolumn{4}{c|}{} &&& 60 & 27+(4)\\
\cline{5-8}
\end{array}
\end{equation*}
\caption{Classes of tetrahedron graphs}
\label{tetklassen}
\end{table}

As shown in table~\ref{tetklassen} by using the Ward identity
respectively by adding primary graphs, the number of tetrahedron graphs
can be also reduced significantly, namely from 60 to 31.

The remaining ball and tetrahedron graphs contain at least one
$\delta$-function. Therefore, by means of the Jacobi identity
such ball graphs can be transformed in two tetrahedron
graphs as shown graphically in~(\ref{ball2tetgraph}).

\setcoordinatesystem units <1.5 cm , 1.5 cm> point at 0 0
\begin{equation}
\label{ball2tetgraph}
\beginpicture
\setcoordinatesystem units <1.125 cm , 1.125 cm> point at 0 -0.3
\setplotsymbol (.)
\setplotarea x from -1.2 to 1.2, y from -1 to 1
\circulararc 300 degrees from -0.87 -0.5 center at 0 0
\circulararc 5.5 degrees from -0.95 -0.33 center at 0 0
\circulararc 5.5 degrees from -0.99 -0.14 center at 0 0
\circulararc 5.5 degrees from -1     0.05 center at 0 0
\circulararc 5.5 degrees from -0.97  0.24 center at 0 0
\circulararc 5.5 degrees from -0.91  0.42 center at 0 0
\plot -0.87  0.5 0.87  0.5 /
\plot -0.87 -0.5 0.87 -0.5 /
\put {$\scriptscriptstyle{G1}$} at 0 0.85
\put {$\scriptscriptstyle{G2}$} at 0 0.35
\put {$\scriptscriptstyle{G3}$} at 0.75 0
\put {$\scriptscriptstyle{G4}$} at 0 -0.35
\put {$\scriptscriptstyle{G5}$} at 0 -0.85
\put {$\scriptstyle{(\delta,G1,G2|G3,G4,G5)_1}$} at 0 -1.4
\endpicture
=
\beginpicture
\setplotsymbol (.)
\setplotarea x from 2.35 to 4.7, y from -0.7 to 1
\plot 3.5 0 3.5 1 4.36 -0.5 2.64 -0.5 3.5 0 4.36 -0.5 /
\plot 2.64 -0.5 2.69 -0.4 /
\plot 2.75 -0.3 2.81 -0.2 /
\plot 2.86 -0.1 2.92 0 /
\plot 2.98 0.1 3.04 0.2 /
\plot 3.10 0.3 3.15 0.4 / 
\plot 3.21 0.5 3.27 0.6 /
\plot 3.33 0.7 3.38 0.8 /
\plot 3.44 0.9 3.5 1 /
\put {$\scriptscriptstyle{G1}$} at 3.65 0.38
\put {$\scriptscriptstyle{\tilde G4}$} at 4.1 0.4
\put {$\scriptscriptstyle{G3}$} at 3.8 0
\put {$\scriptscriptstyle{G5}$} at 3.5 -0.35
\put {$\scriptscriptstyle{G2}$} at 3.2 0
\put {$\scriptstyle{(\delta,G1,\tilde G4|G3,G5,G2)_2}$} at 3.5 -0.8
\endpicture
+
\beginpicture
\setplotsymbol (.)
\setplotarea x from 5.8 to 8.2, y from -0.7 to 1
\plot 7 0 7 1 7.86 -0.5 6.14 -0.5 7 0 7.86 -0.5 /
\plot 6.14 -0.5 6.19 -0.4 /
\plot 6.25 -0.3 6.31 -0.2 /
\plot 6.36 -0.1 6.42 0 /
\plot 6.48  0.1 6.54 0.2 /
\plot 6.60  0.3 6.65 0.4 / 
\plot 6.71  0.5 6.77 0.6 /
\plot 6.83  0.7 6.88 0.8 /
\plot 6.94  0.9 7    1 /
\put {$\scriptscriptstyle{G1}$} at 7.15 0.38
\put {$\scriptscriptstyle{\tilde G5}$} at 7.6 0.4
\put {$\scriptscriptstyle{G3}$} at 7.3 0
\put {$\scriptscriptstyle{G4}$} at 7 -0.35
\put {$\scriptscriptstyle{G2}$} at 6.7 0
\put {$\scriptstyle{(\delta,G1,\tilde G5|G3,G4,G2 )_2}$} at 7 -0.8
\endpicture\\
\end{equation}

The value of the graphs consisting of structure constants
can easily be determined by use of table~\ref{TabSt}.
In particular we get:
\begin{equation}
\begin{array}{lcr}
(\seteins,\seteins,\seteins|
 \seteins,\seteins,{\bf F}^2)_1&=&
  -\phantom{\frac{1}{2}}\,C_2^3(F_{\mu\nu}^a)^2,\\
(\seteins,\seteins,\seteins|
 {\bf F}^2,\seteins,\seteins)_1&=&
  -\phantom{\frac{1}{2}}\,C_2^3(F_{\mu\nu}^a)^2,\\
(\seteins,\seteins,\seteins|
 \seteins,{\bf F}_{\mu\nu},{\bf F}_{\mu\nu})_1&=&
   \phantom{-}\frac{1}{2}\,C_2^3(F_{\mu\nu}^a)^2,\\
(\seteins,\seteins,\seteins|
 {\bf F}_{\mu\nu},\seteins,{\bf F}_{\mu\nu})_1&=&
  -\frac{1}{2}\,C_2^3(F_{\mu\nu}^a)^2,\\
(\seteins,\seteins,{\bf F}_{\mu\nu}|
 \seteins,\seteins,{\bf F}_{\mu\nu})_1&=&
  -\frac{1}{4}\,C_2^3(F_{\mu\nu}^a)^2,\\
({\bf F}_{\mu\nu},\seteins,\seteins|
 {\bf F}_{\mu\nu},\seteins,\seteins)_1&=&
  -\phantom{\frac{1}{2}}\,C_2^3(F_{\mu\nu}^a)^2;\\
\end{array}
\end{equation}
\begin{equation}
\begin{array}{lcr}
(\seteins,\seteins,\seteins|\seteins,\seteins,{\bf F}^2)_2&=&
\phantom{-}\frac{1}{2}\,C_2^3(F_{\mu\nu}^a)^2,\\
(\seteins,\seteins,\seteins|
 \seteins,{\bf F}_{\mu\nu},{\bf F}_{\mu\nu})_2&=&
-\frac{1}{4}\,C_2^3(F_{\mu\nu}^a)^2,\\
(\seteins,\seteins,{\bf F}_{\mu\nu}|
 \seteins,\seteins,{\bf F}_{\mu\nu})_2&=&0.\\
\end{array}
\end{equation}

Following our prescription of the \mbox{two-loop} calculation we 
expand according to (\ref{Skalarpropagator},\ref{Vektorpropagator})
the propagators in each graph. After discarding all graphs which 
depend on the field strength more than quadratically, we set the 
phasefactors equal to one in those graphs containing already two 
field strength tensors. In the remaining graphs we let the covariant 
derivatives take effect on the phasefactors. Since the 
phasefactors~(\ref{PFSE}) depend on points in coordinate space, 
whereas the propagators depend only on differences of these points 
(translation invariance), we will rewrite the phasefactors in the 
following manner. Take an arbitrary path $x_0,\cdots,x_n$ of a graph 
connecting the point $x_n$ with $x_0$. Thus the phasefactor connecting 
the last two points of this path may be written as
\begin{eqnarray}\label{Phidiff}
\Phi(x_{n-1},x_n)&=&\exp{\frac{1}{2} x^\mu_{n-1} F_{\mu\nu}x^\nu_n}
\nonumber\\
&=& \exp{\frac{1}{2}\Biggl(\sum_{i=0}^{n-1} 
         (x_{n-i}-x_{n-i-1})^\mu F_{\mu\nu} (x_n-x_{n-1})^\nu +}
\nonumber\\ 
& & \phantom{ 
      \exp{\frac{1}{2}\Biggl(\sum_{i=0}^{n-1}}(x_{n-i}-x_{n-i-1})^\mu)
    }
     x_{0}^\mu F_{\mu\nu} (x_n-x_{n-1})^\nu \Biggr)\nonumber\\
&=& 1 + \frac{1}{2}\sum_{i=0}^{n-1} 
         \bigl((x_{n-i}-x_{n-i-1})^\mu F_{\mu\nu} (x_n-x_{n-1})^\nu + 
\nonumber\\
& & \phantom{ 
      \exp{\frac{1}{2}\Biggl(\sum_{i=0}^{n-1}}(x_{n-i}-x_{n-i-1})^\mu)
    }        
     x_{0}^\mu F_{\mu\nu} (x_n-x_{n-1})^\nu \bigr)\nonumber\\
&& \phantom{1} + \frac{1}{8}\sum_{i=0}^{n-1} 
     \bigl((x_{n-i}-x_{n-i-1})^\mu F_{\mu\nu} (x_n-x_{n-1})^\nu +
\nonumber\\
& & \phantom{ 
      \exp{\frac{1}{2}\Biggl(\sum_{i=0}^{n-1}}(x_{n-i}-x_{n-i-1})^\mu)
    }              
      x_{0}^\mu F_{\mu\nu} (x_n-x_{n-1})^\nu \bigr)^2 + \cdots\quad.
\nonumber\\      
\end{eqnarray}
Following the instructions of~\cite{NSVZ84} respectively our 
\mbox{two-loop} calculation, we fix~$x_0$ at the origin of our 
coordinate space by setting its value to zero. As a result the 
phasefactor of equation~(\ref{Phidiff}) depends solely on differences
of points in coordinate space. Using the identities of 
appendix~\ref{IGR} these differences can be absorbed in the Green 
functions of the propagators. Finally, discarding again all graphs not
containing the field strength quadratically, we have decomposed the 
exact graphs into ordinary Feynman graphs each multiplied by a product
of structure constants.

We want to reemphasize, that both, the fixed point and the path, can 
be chosen freely since the divergences of an exact graph do not depend
on the choice of these.

After expanding the exact graphs by means 
of~(\ref{Skalarpropagator},\ref{Vektorpropagator})
we get, after using some obvious identities, about 2000 graphs.
Again, we want to refrain from calculating all these graphs explicitly,
but reduce their number by using similar identities as demonstrated
in the \mbox{two-loop} case.

First we eliminate the $\delta$~tensors by help of the identities
of appendix~\ref{IGR}. By doing so, we have to watch carefully, if the
$\delta$~tensors, building up a dimensional~$d$, belong to a divergent
subloop of the graph.

\begin{figure}[h]
\begin{multline*}
 \graph{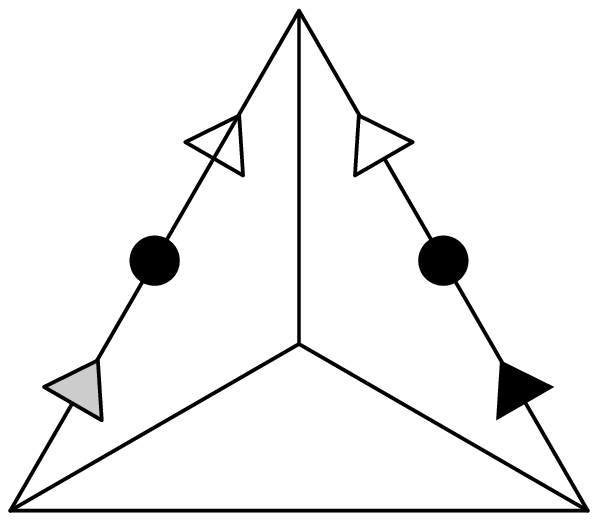}{3.5}
 \raisebox{-0.1cm}{\hspace{-2.9cm}$\scriptstyle{\mu}$}
 \raisebox{0.75cm}{\hspace{0.35cm}$\scriptstyle{\si}$}
 \raisebox{-0.1cm}{\hspace{1.4cm}$\scriptstyle{\nu}$}
 \raisebox{0.75cm}{\hspace{-0.675cm}$\scriptstyle{\ta}$}
  \hspace{+0.725cm}
 (\seteins,\seteins,\seteins|
  \seteins,\seteins,{\bf F}_{\mu\nu}{\bf F}_{\si\ta})_2\\
\quad =\ \phantom{-}\frac{1}{12}
         \left(\hspace{-0.7cm}\graph{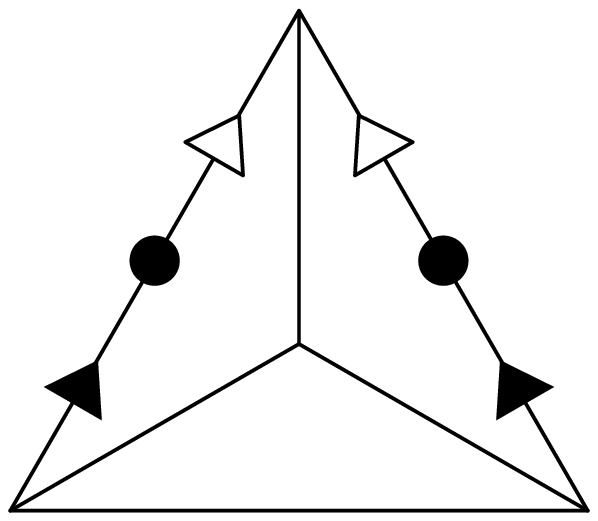}{3.5}\hspace{-0.7cm}
   \ -\ \hspace{-0.7cm}\graph{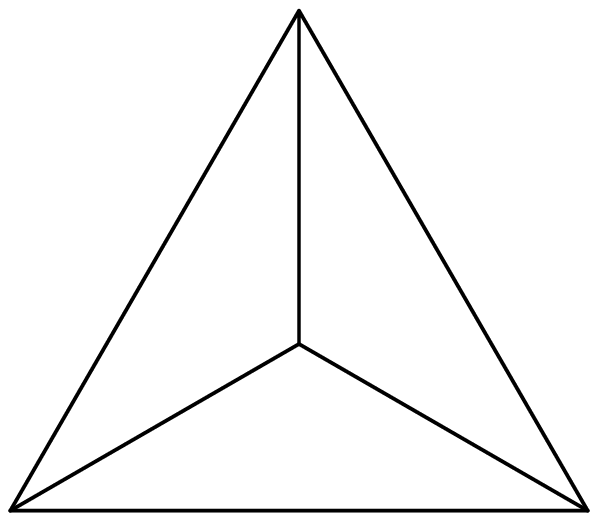}{3.5}\hspace{-0.7cm}\right)
   (\seteins,\seteins,\seteins|\seteins,\seteins,{\bf F}^2)_2 + 
   O(\eps^0)\\
 =\ - \frac{1}{24}\left(\hspace{-0.7cm}\graph{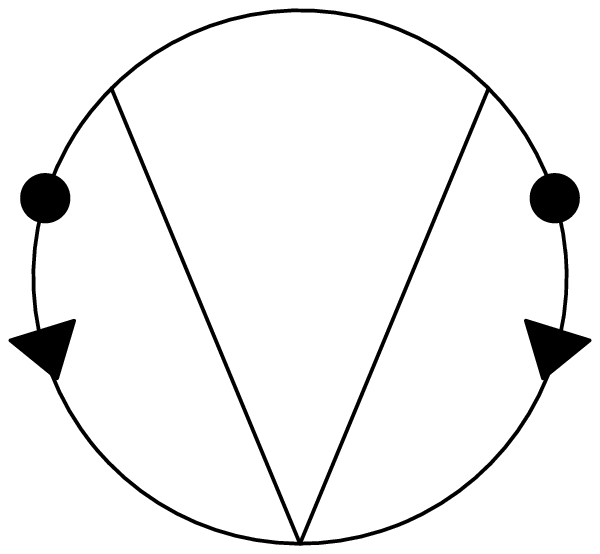}{3.2}
   \hspace{-0.7cm}
   \ +\ \hspace{-0.7cm}\graph{Tetraeder1c}{3.5}\hspace{-0.7cm}\right)
   (\seteins,\seteins,\seteins|\seteins,\seteins,{\bf F}^2)_2 + 
   O(\eps^0)
\end{multline*}
\caption{\label{SimpTet}Calculation of a simple tetrahedron graph}
\end{figure}

Thereafter we gather all tetrahedron graphs containing four derivatives.
Then we build up a basis which contains only graphs without 
subdivergences, like the one given in figure~\ref{SimpTet}. Since 
these graphs have only simple poles, we can use the identity
\begin{equation}\label{FF2F2}
{\mathbf F}_{\mu\nu} {\mathbf F}_{\si\tau} = \frac{1}{d(d-1)}
                        \left(\delta_{\mu\ta}\delta_{\nu\si}
                             -\delta_{\mu\si}\delta_{\nu\ta}
                        \right){\mathbf F}^2
\end{equation}
while neglecting the $\eps$ part of the dimension~$d$, which means
$d$ equals four.

In figure~\ref{SimpTet} we give the most elementary example
of a tetrahedron graph with four derivatives being reduced
to the ordinary tetrahedron graph without derivatives.

Applying a $\delta$ tensor to equation~(\ref{FF2F2}) we get the 
identity 
\begin{equation}\label{F22F2}
{\mathbf F}^2_{\mu\nu} = {\mathbf F}^2 \frac{\delta_{\mu\nu}}{d},
\end{equation}
which allows us to reduce now tetrahedron graphs containing two
derivatives, but no subdivergences to the ordinary tetrahedron graph
without derivatives.

Thus, after some algebra the sum of all tetrahedron graphs should be
simplified in such a way, that only the ordinary tetrahedron graph and
some other non-tetrahedral graphs remain. After doing this calculation
we see, that even the coefficient of the ordinary tetrahedron graph 
vanishes, which explains why the final answer contains 
no~$\zeta(3)$~term.

At present we have to calculate finally 24 Feynman graphs to determine
the pole part of the effective action. Unfortunately, due to the amount
and complexity of the rules needed for doing this task automatically, 
some parts of this reduction were introduced by hand. In the following,
we give our complete answer for the effective action. Each graph is 
illustrated in a pair of brackets. The first row contains the 
graphical notation where the following definitions are used:
\begin{eqnarray*}
\setcoordinatesystem units <0.5 cm , 0.5 cm> point at 0 -0.25
\beginpicture
\setplotarea x from -1 to 0.5, y from -0.5 to 0.5
\plot -1 0 -0.5 0 -0.5 0.2 -0.2 0 -0.5 -0.2 -0.5 0 /
\plot -0.2 0 0.3 0 /
\endpicture
& \Leftrightarrow & \partial_\mu\\
\setcoordinatesystem units <0.5 cm , 0.5 cm> point at 0 -0.25
\beginpicture
\setplotarea x from -1 to 0.5, y from -0.5 to 0.5
\plot -1 0 -0.5 0 -0.5 0.2 -0.2 0 -0.5 -0.2 -0.5 0 0.3 0 /
\endpicture
& \Leftrightarrow & \partial_\nu\\
\bar {\mathbf F}^2_{\mu\nu}
&=& (\seteins,\seteins,\seteins|
     \seteins,\seteins,{\mathbf F}^2_{\mu\nu})_2\\
\bar {\mathbf F}^2_{\phantom{\mu\nu}}
&=& (\seteins,\seteins,\seteins|
     \seteins,\seteins,{\mathbf F}^2)_2 \ .
\end{eqnarray*}
The bracket notation is to be found in the second row.
The divergences of the graph are given in the third row 
where we omitted a factor
$C^3_2/(16\pi^2)^3 S_{cl}$. The definition
of this bracket can be found in appendix~\ref{Nt}.

\begin{multline*}
\Gamma_3^\text{div} = \\
\frac{1}{6}
\begin{pmatrix}
6 \graph{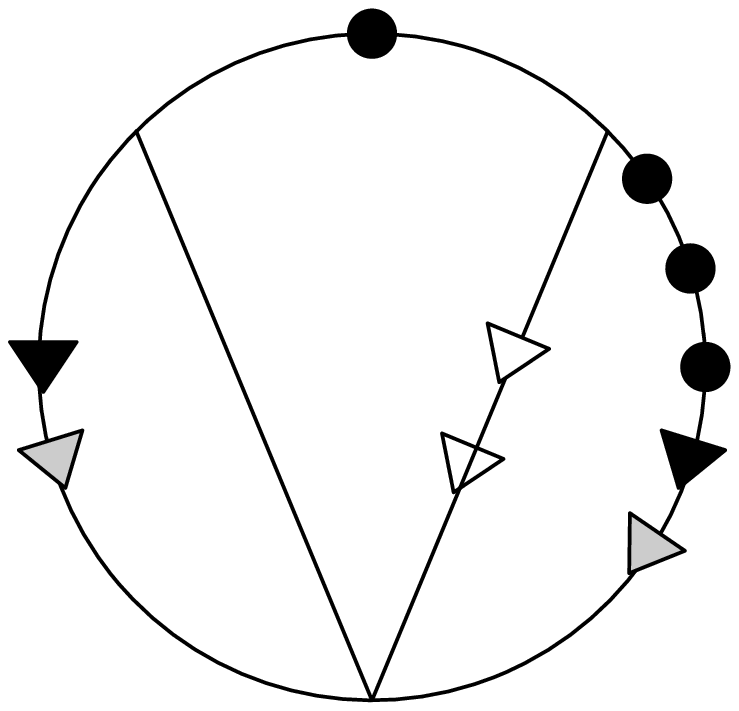}{2.5} \bar {\mathbf F}^2_{\mu\nu}\\
(G_1,G_0,\partial_{\rho\sigma} G_0|\\
\phantom{(}\seteins,\partial_{\mu\nu} G_0,\partial_{\rho\sigma} G_3
{\mathbf F}^2_{\mu\nu})_2\\
\\
\left(0,0,\frac{17}{1152}\right)
\end{pmatrix}
+\frac{7}{3}
\begin{pmatrix}
6 \graph{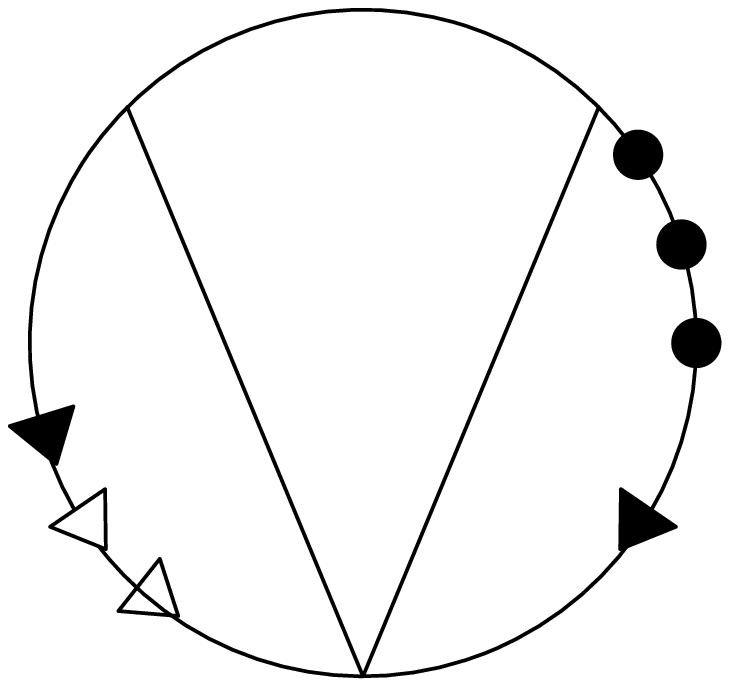}{2.5}\bar {\mathbf F}^2_{\mu\nu}\\
(G_0,G_0,\partial_{\rho} G_0|\\
\phantom{(}\seteins, G_0,\partial_{\rho\mu\nu} G_3
{\mathbf F}^2_{\mu\nu})_2\\
\\
(0,0,\frac{1}{768})
\end{pmatrix}\displaybreak[0]\\
-\frac{3}{2}
\begin{pmatrix}
-6 \graph{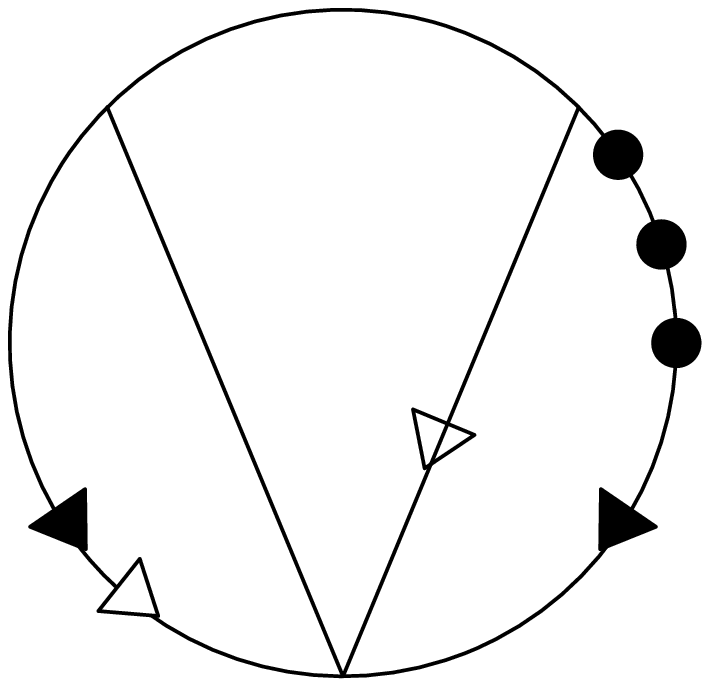}{2.5}\bar {\mathbf F}^2_{\mu\nu}\\
(G_3,G_0,\partial_{\rho\mu} G_0|\\
\phantom{(}\seteins, \partial_\nu G_0,\partial_{\rho} G_0
{\mathbf F}^2_{\mu\nu})_2\\
\\
(-\frac{1}{12},\frac{7}{48},-\frac{67}{576})
\end{pmatrix}
+\left(1-\frac{7}{3}d\right)
\begin{pmatrix}
6 \graph{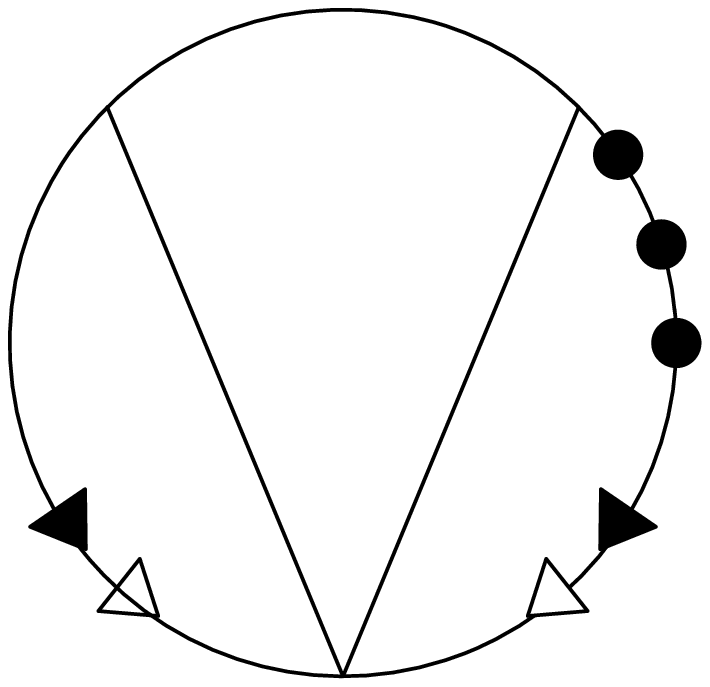}{2.5}\bar {\mathbf F}^2_{\mu\nu}\\
(G_0,G_0,\partial_{\rho\mu} G_0|\\
\phantom{(}\seteins, G_0,\partial_{\rho\nu} G_3
{\mathbf F}^2_{\mu\nu})_2\\
\\
(-\frac{1}{12},\frac{1}{16},-\frac{1}{48})
\end{pmatrix}\displaybreak[0]\\
-\frac{65}{3}
\begin{pmatrix}
6 \graph{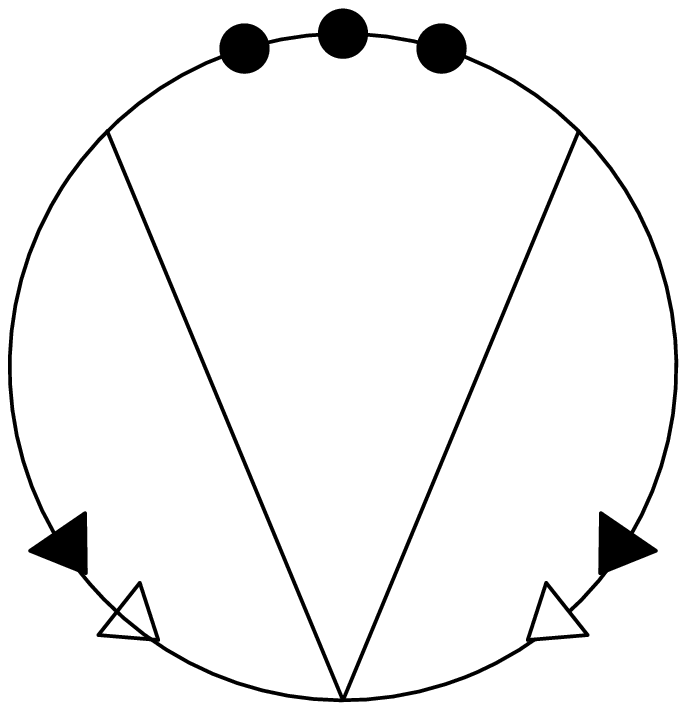}{2.5}\bar {\mathbf F}^2_{\mu\nu}\\
(G_3,G_0,\partial_{\rho\mu} G_0|\\
\phantom{(}\seteins, G_0,\partial_{\rho\nu} G_0
{\mathbf F}^2_{\mu\nu})_2\\
\\
(-\frac{1}{6},\frac{5}{72},\frac{35}{864})
\end{pmatrix}
+\frac{51}{4}
\begin{pmatrix}
2 \graph{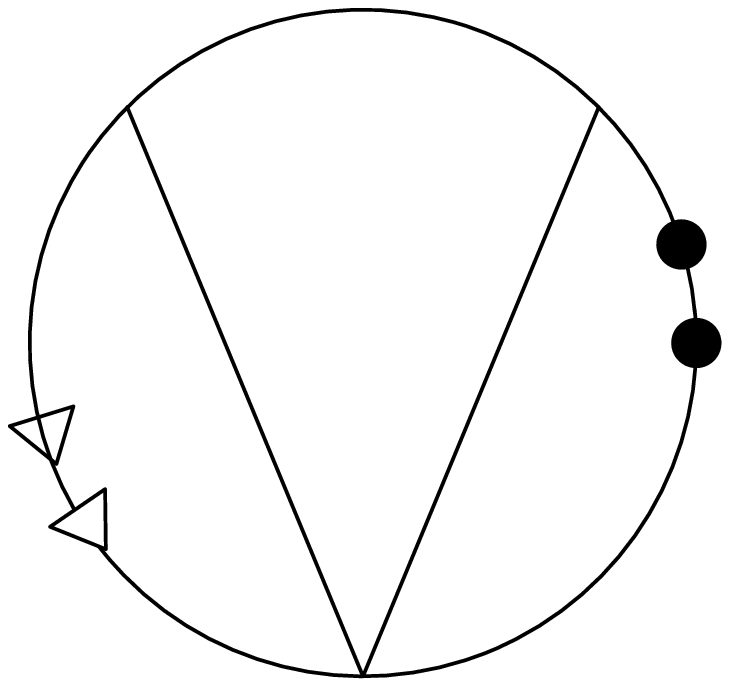}{2.5}\bar {\mathbf F}^2_{\mu\nu}\\
(G_0,G_0,\partial_{\mu\nu} G_0|\\
\phantom{(}\seteins, G_0, G_2
{\mathbf F}^2_{\mu\nu})_2\\
\\
(0,0,\frac{1}{432})
\end{pmatrix}\displaybreak[0]\\
-\frac{961}{12}
\begin{pmatrix}
-2 \graph{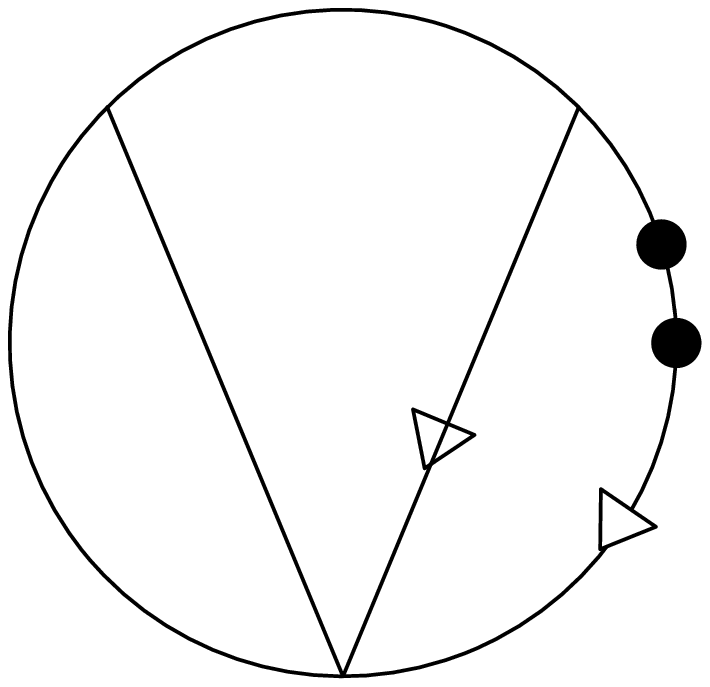}{2.5}\bar {\mathbf F}^2_{\mu\nu}\\
(G_0,G_0,G_0|\\
\phantom{(}\seteins, \partial_{\mu} G_0, \partial_{\nu} G_2
{\mathbf F}^2_{\mu\nu})_2\\
\\
(\frac{1}{6},-\frac{7}{24},\frac{3}{16})
\end{pmatrix}
+\frac{10183}{48}
\begin{pmatrix}
2 \graph{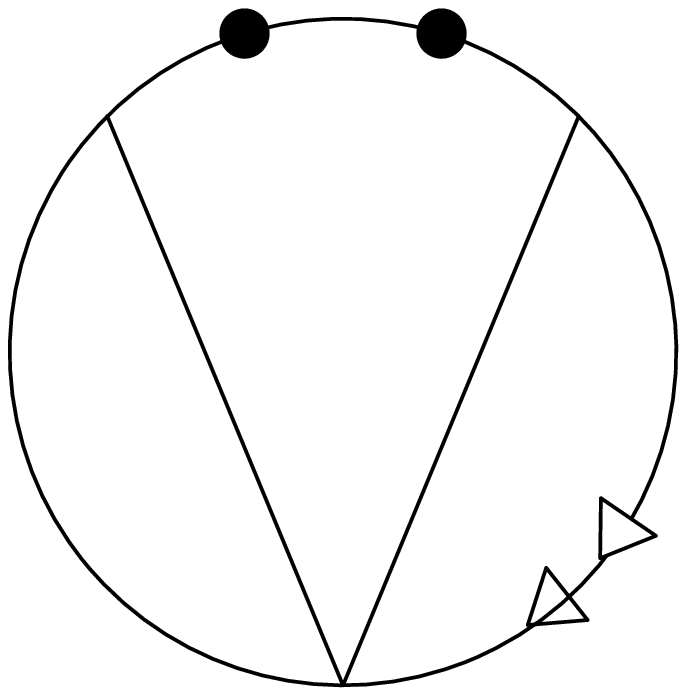}{2.5}\bar {\mathbf F}^2_{\mu\nu}\\
(G_2,G_0, G_0|\\
\phantom{(}\seteins, G_0,\partial_{\mu\nu} G_0
{\mathbf F}^2_{\mu\nu})_2\\
\\
(0,\frac{1}{12},-\frac{1}{48})
\end{pmatrix}\displaybreak[0]\\
+\frac{632491}{4608}
\begin{pmatrix}
 \graph{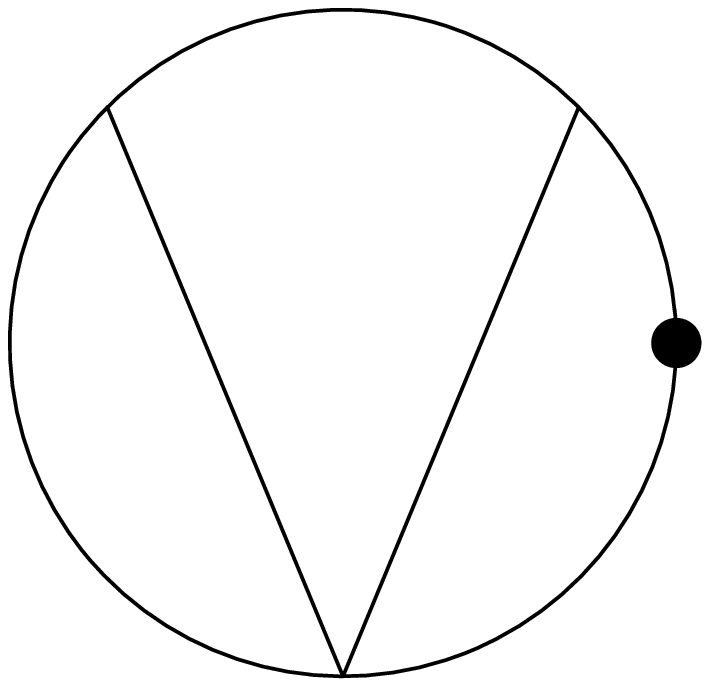}{2.5}\bar {\mathbf F}^2\\
(G_0,G_0,G_0|\\
\phantom{(}\seteins, G_0, G_1
{\mathbf F}^2)_2\\
\\
(-\frac{2}{3},1,-\frac{2}{3})
\end{pmatrix}
-\frac{82699}{2304}
\begin{pmatrix}
 \graph{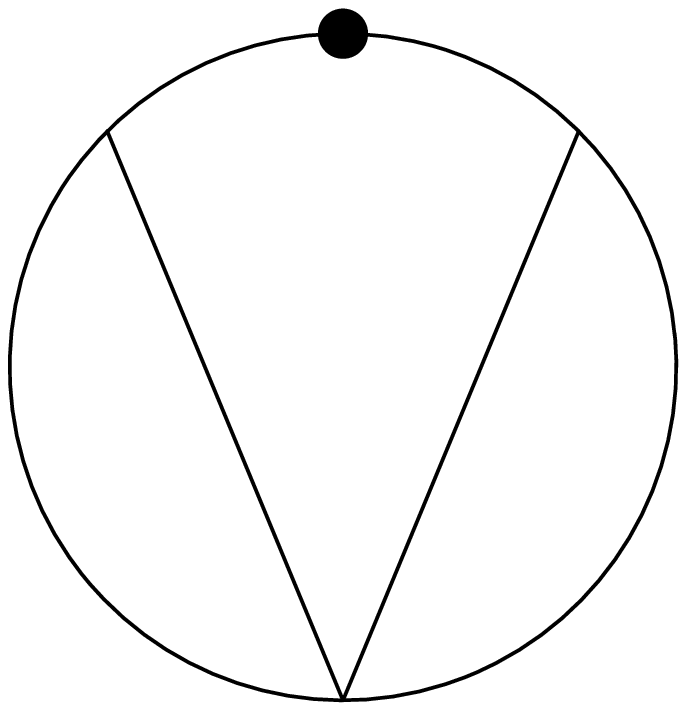}{2.5}\bar {\mathbf F}^2\\
(G_1,G_0, G_0|\\
\phantom{(}\seteins, G_0, G_0{\mathbf F}^2)_2\\
\\
(-\frac{4}{3},\frac{2}{3},\frac{1}{3})
\end{pmatrix}\displaybreak[0]\\
-\frac{6668113}{41472}
\begin{pmatrix}
\graph{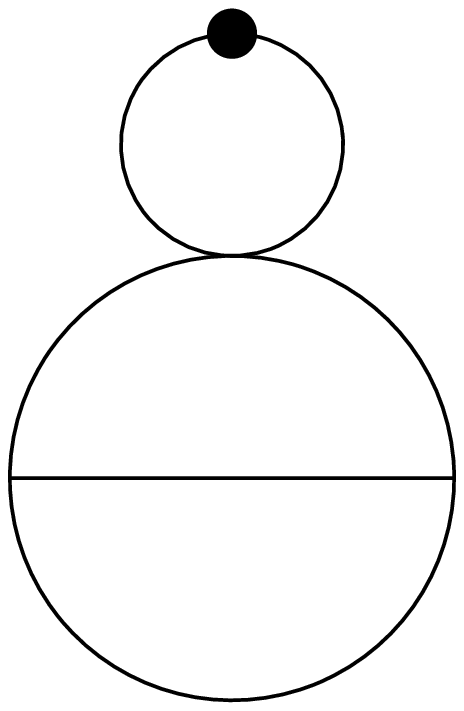}{2.5}\bar {\mathbf F}^2\\
(G_0,G_0, G_1|\\
\phantom{(}\seteins, \seteins, G_1{\mathbf F}^2)_2\\
\\
(-2,1,0)
\end{pmatrix}
+\left(-\frac{16405}{432}-\frac{20651}{5184} d +
        \frac{241}{192} d^2\right)
\begin{pmatrix}
24 \graph{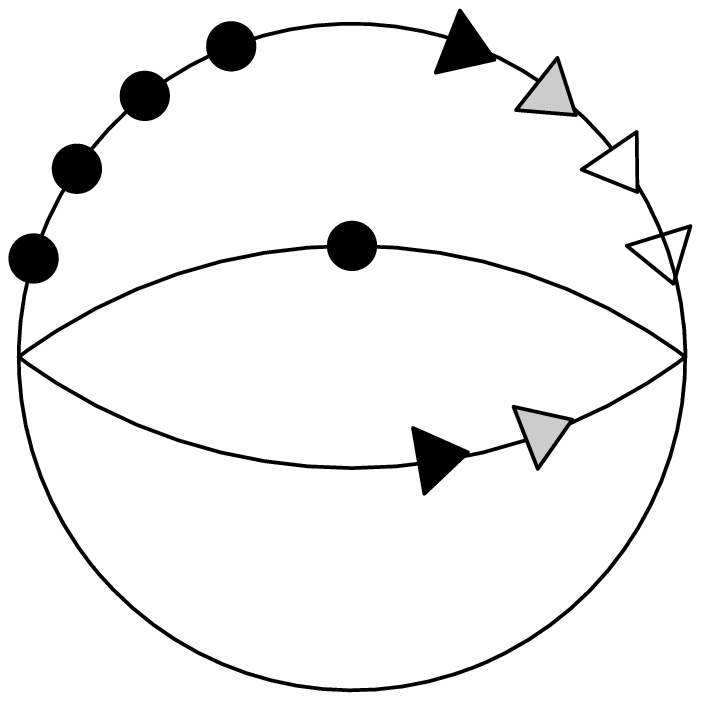}{2.5}\bar {\mathbf F}^2_{\mu\nu}\\
(\seteins,G_0,\partial_{\rho\sigma} G_0|\\
\phantom{(}\seteins, G_1,\partial_{\rho\sigma\mu\nu} G_4
{\mathbf F}^2_{\mu\nu})_2\\
\\
(2,-\frac{2}{3},0)
\end{pmatrix}\displaybreak[0]\\
+\frac{7}{9} d
\begin{pmatrix}
12 \graph{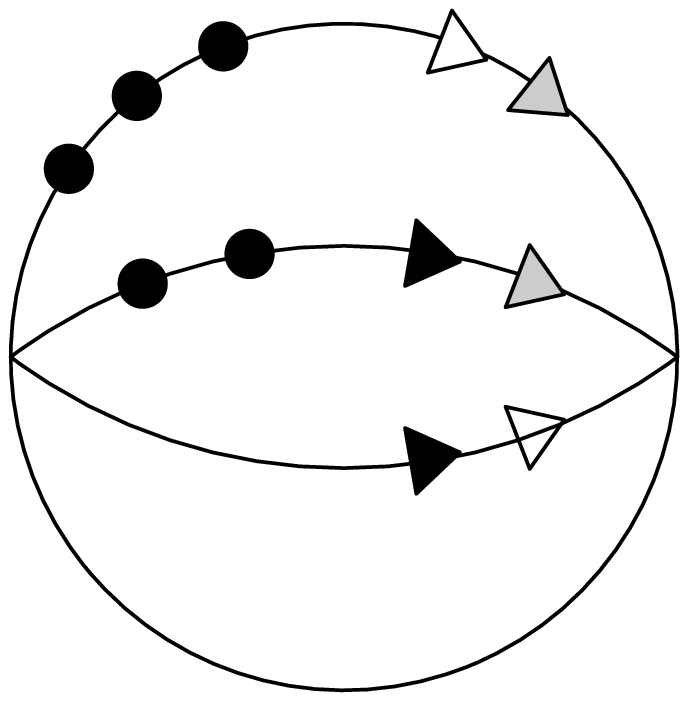}{2.5}\bar {\mathbf F}^2_{\mu\nu}\\
(\seteins,G_0,\partial_{\rho\mu} G_0|\\
\phantom{(}\seteins, \partial_{\rho\sigma} G_2,\partial_{\sigma\nu} G_3
{\mathbf F}^2_{\mu\nu})_2\\
\\
(\frac{1}{4},\frac{7}{24},-\frac{73}{576})
\end{pmatrix}
+\left(-\frac{241}{24}-\frac{241}{96} d\right)
\begin{pmatrix}
12 \graph{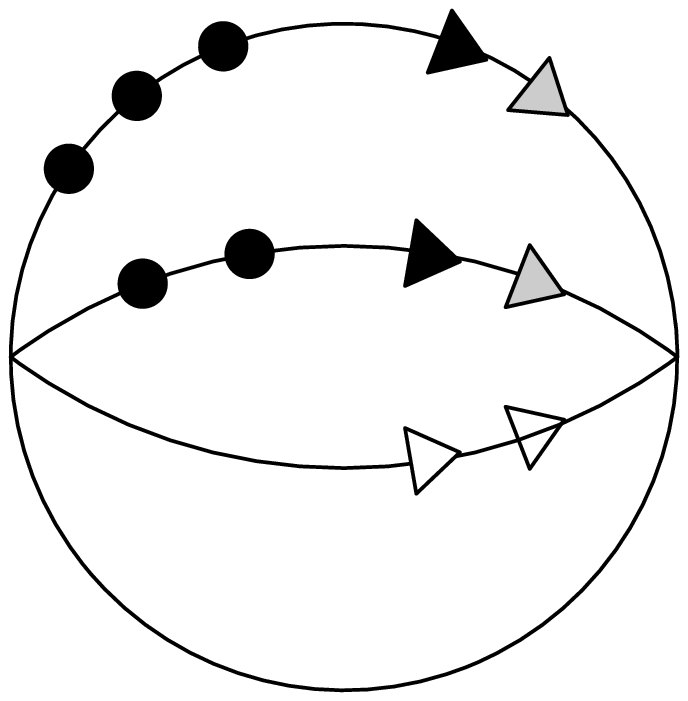}{2.5}\bar {\mathbf F}^2_{\mu\nu}\\
(\seteins,G_0,\partial_{\mu\nu} G_0|\\
\phantom{(}\seteins, \partial_{\rho\sigma} G_2,
\partial_{\rho\sigma} G_3{\mathbf F}^2_{\mu\nu})_2\\
\\
(0,\frac{1}{4},-\frac{5}{288})
\end{pmatrix}\displaybreak[0]\\
+\left(-\frac{6529}{54}-\frac{32081}{5184} d +
        \frac{241}{64} d^2\right)
\begin{pmatrix}
6 \graph{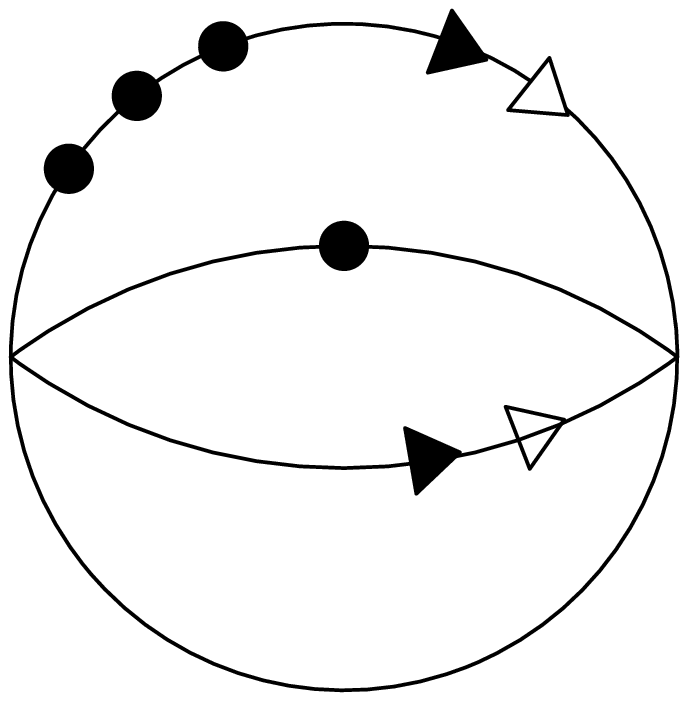}{2.5}\bar {\mathbf F}^2_{\mu\nu}\\
(\seteins,G_0,\partial_{\rho\mu} G_0|\\
\phantom{(}\seteins, G_1,\partial_{\rho\nu} G_3
{\mathbf F}^2_{\mu\nu})_2\\
\\
(-\frac{1}{2},\frac{1}{6},0)
\end{pmatrix}\displaybreak[0]\\
+\left(-\frac{16603}{648}+\frac{7573}{10368} d +
        \frac{499}{1152} d^2\right)
\begin{pmatrix}
6 \graph{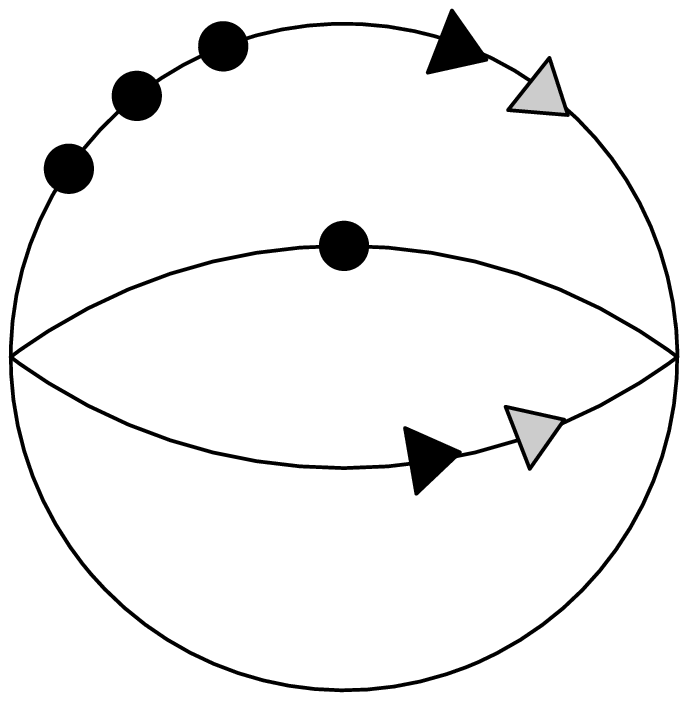}{2.5}\bar {\mathbf F}^2\\
(\seteins,G_0,\partial_{\rho\si} G_0|\\
\phantom{(}\seteins, G_1,\partial_{\rho\si} G_3
{\mathbf F}^2)_2\\
\\
(-2,\frac{7}{6},\frac{1}{48})
\end{pmatrix}\displaybreak[0]\\
-\frac{14}{3}
\begin{pmatrix}
4 \graph{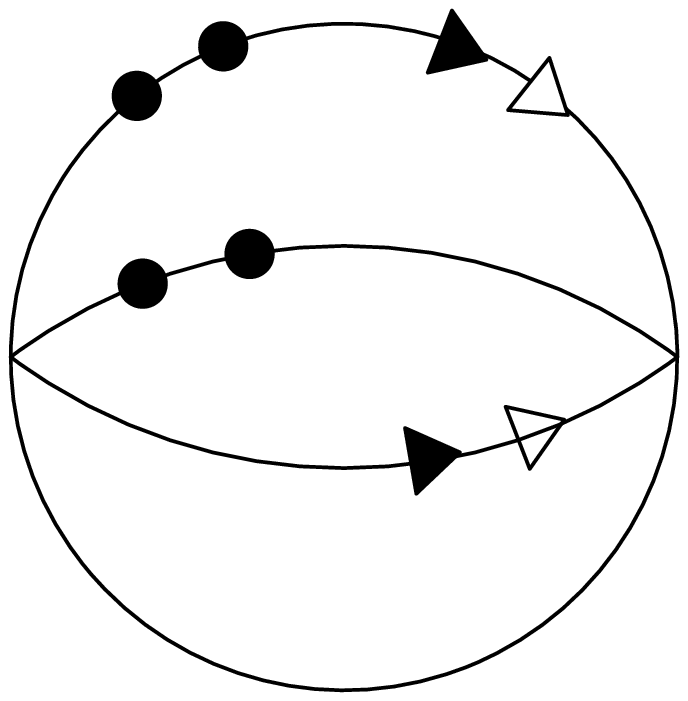}{2.5}\bar {\mathbf F}^2_{\mu\nu}\\
(\seteins,G_0,\partial_{\rho\mu} G_0|\\
\phantom{(}\seteins, G_2,\partial_{\rho\nu} G_2
{\mathbf F}^2_{\mu\nu})_2\\
\\
(0,-\frac{1}{3},\frac{7}{36})
\end{pmatrix}
+\left(\frac{2749}{216}-\frac{7}{36} d\right)
\begin{pmatrix}
4 \graph{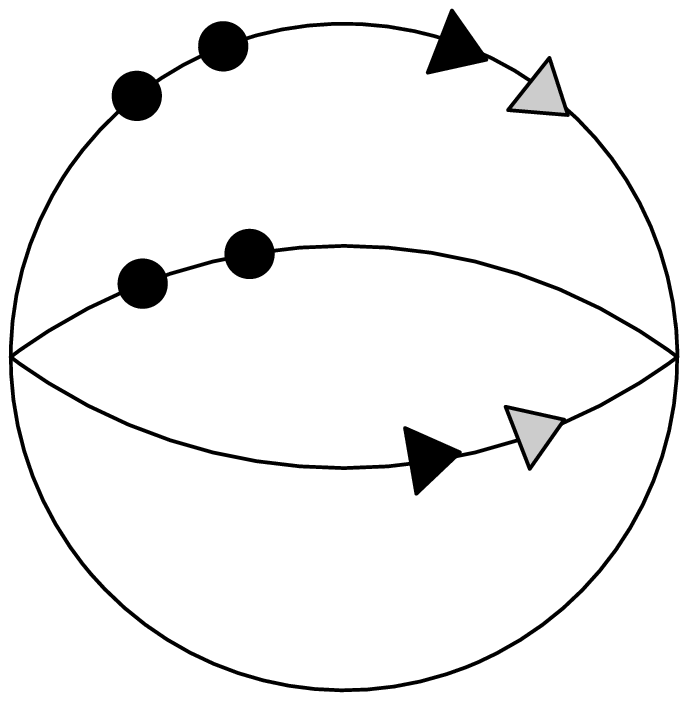}{2.5}\bar {\mathbf F}^2\\
(\seteins,G_0,\partial_{\rho\si} G_0|\\
\phantom{(}\seteins, G_2,\partial_{\rho\si} G_2
{\mathbf F}^2)_2\\
\\
(0,-\frac{4}{3},\frac{19}{24})
\end{pmatrix}\displaybreak[0]\\
+\frac{185}{162}
\begin{pmatrix}
6 \graph{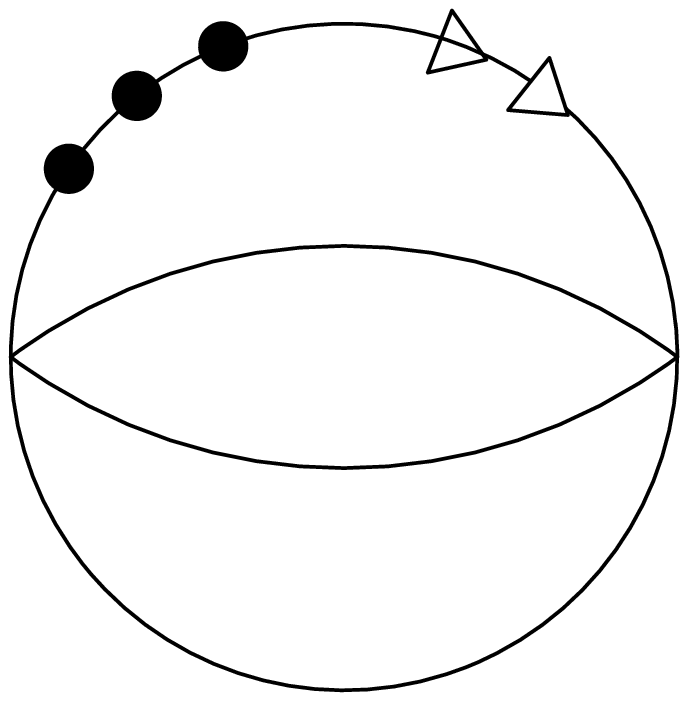}{2.5}\bar {\mathbf F}^2_{\mu\nu}\\
(\seteins,G_0, G_0|\\
\phantom{(}\seteins, G_0,\partial_{\mu\nu} G_3
{\mathbf F}^2_{\mu\nu})_2\\
\\
(0,\frac{1}{2},-\frac{7}{16})
\end{pmatrix}
+\left(\frac{763}{32}+\frac{775}{384} d-\frac{241}{384} d^2\right)
\begin{pmatrix}
6 \graph{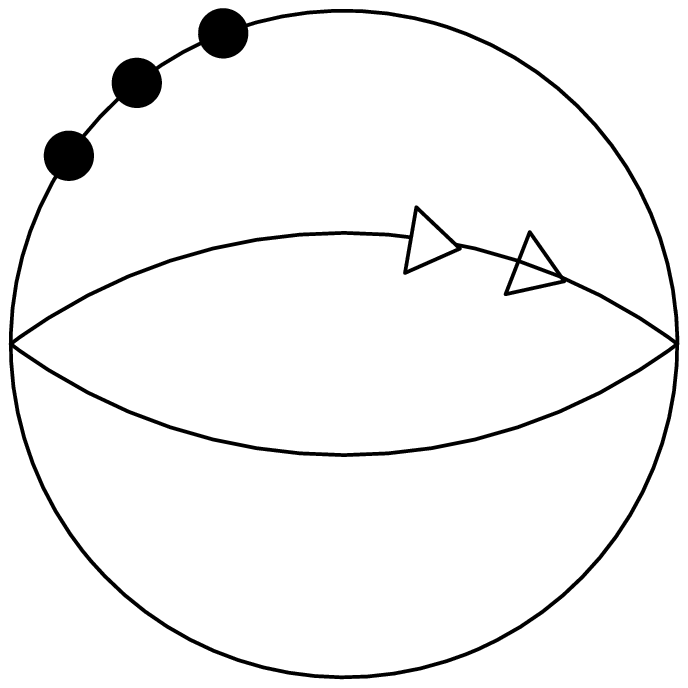}{2.5}\bar {\mathbf F}_{\mu\nu}^2\\
(\seteins,G_0, G_0|\\
\phantom{(}\seteins,\partial_{\mu\nu} G_0, G_3
{\mathbf F}_{\mu\nu}^2)_2\\
\\
(0,0,\frac{1}{32})
\end{pmatrix}\displaybreak[0]\\
+\left(-\frac{3385}{96}-\frac{2017}{1152} d +
        \frac{241}{128} d^2\right)
\begin{pmatrix}
2 \graph{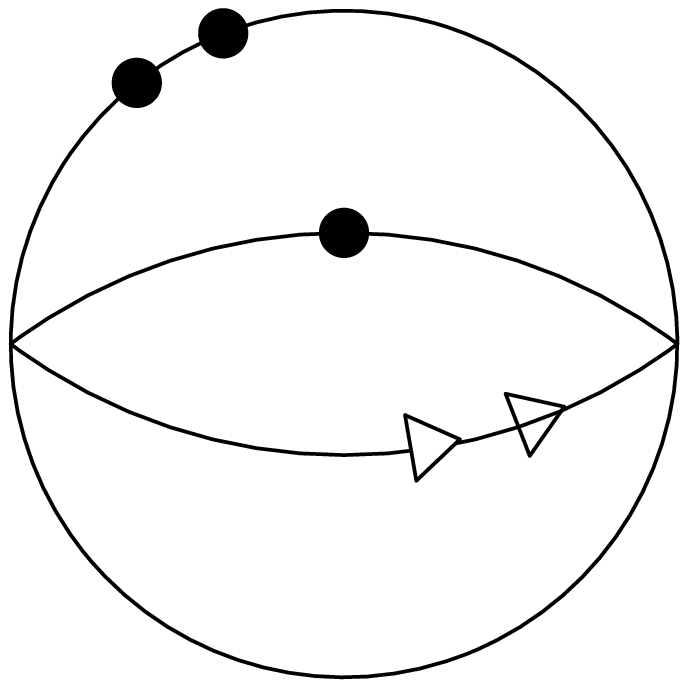}{2.5}\bar {\mathbf F}_{\mu\nu}^2\\
(\seteins,G_0, \partial_{\mu\nu} G_0|\\
\phantom{(}\seteins, G_1, G_2
{\mathbf F}_{\mu\nu}^2)_2\\
\\
(0,\frac{1}{6},-\frac{5}{144})
\end{pmatrix}\displaybreak[0]\\
+\left(\frac{3507715}{5184}-\frac{16281731}{414272} d +
       \frac{18422737}{27648} d^2-\frac{11149}{4608} d^3 -
       \frac{241}{6144} d^4
\right)
\begin{pmatrix}
2 \graph{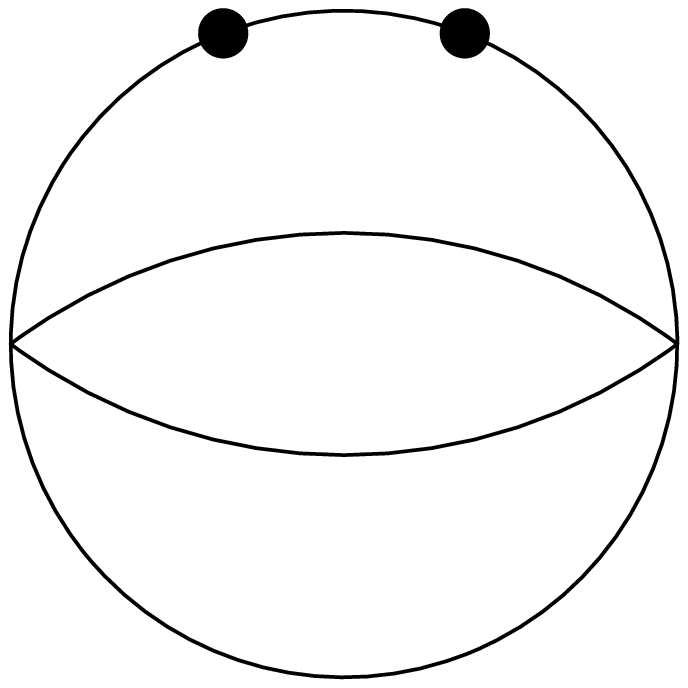}{2.5}\bar {\mathbf F}^2\\
(\seteins,G_0, G_0|\\
\phantom{(}\seteins, G_0, G_2
{\mathbf F}^2)_2\\
\\
(0,-\frac{2}{3},\frac{3}{4})
\end{pmatrix}\displaybreak[0]\\
+\left(\frac{1093061}{4608}-\frac{36075}{1024} d -
       \frac{41537}{9216} d^2 \right)
\begin{pmatrix}
\graph{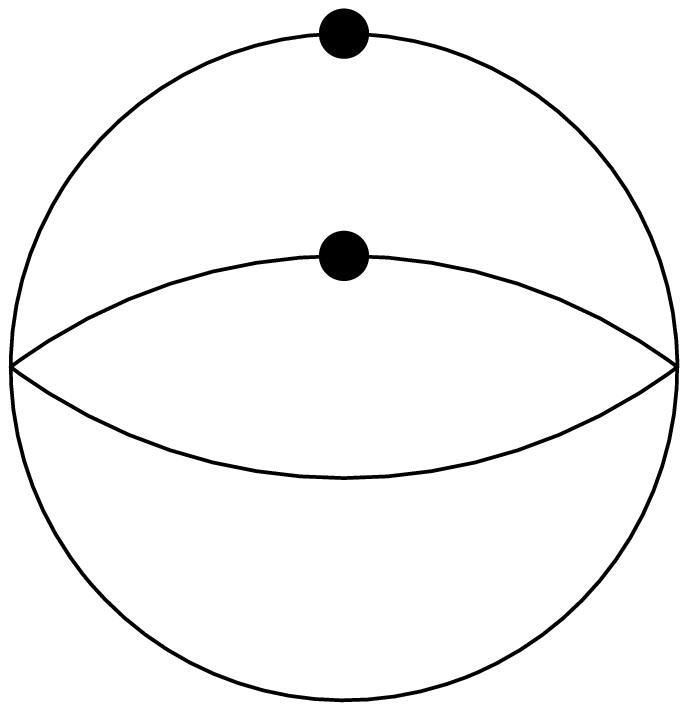}{2.5}\bar {\mathbf F}^2\\
(\seteins,G_0, G_0|\\
\phantom{(}\seteins, G_1, G_1{\mathbf F}^2)_2\\
\\
(-\frac{4}{3},\frac{4}{3},-\frac{1}{3})
\end{pmatrix}\displaybreak[0]\\
+\left(-\frac{1180189}{13824}+\frac{242191}{4608} d -
        \frac{901981}{165888} d^2 +\frac{10009}{27648}d^3\right)
\begin{pmatrix}
\graph{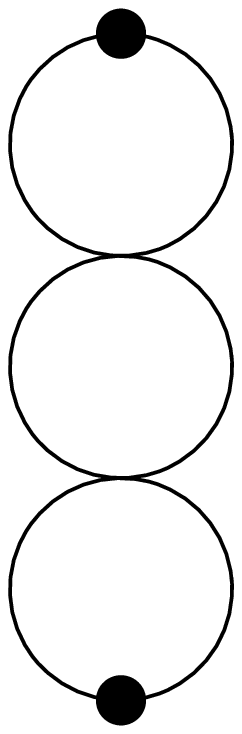}{2.5}\bar {\mathbf F}^2\\
(\seteins,G_1, G_1|\\
\phantom{(}\seteins, \seteins, G_1{\mathbf F}^2)_2\\
\\
(-4,0,0)
\end{pmatrix} + O(\eps^0)\displaybreak[0]\\
\end{multline*}
\begin{equation*}
= \left(-\frac{748}{27\eps^2} + \frac{2857}{81\eps}\right)
  \left(\frac{C_2}{16 \pi^2}\right)^3 S_{cl} + O(\eps^0)
\end{equation*}
Hence the renormalization constant $Z_g$ of section~\ref{BFMinYM}
is given by
\begin{equation}
Z_g^{-1}\ = \ 
    1 + \frac{22}{3\eps} \frac{g^2 C_2}{16 \pi^2}
      + \frac{34}{3\eps} \left(\frac{g^2 C_2}{16 \pi^2}\right)^2
      + \left(\frac{748}{27\eps^2} - \frac{2857}{81\eps}\right)
        \left(\frac{g^2 C_2}{16 \pi^2}\right)^3
\end{equation}
and the corresponding $\beta$-function is of the shape
\begin{eqnarray}\label{betafunc}
\beta(g)&=&-\eps\left(\frac{d}{d g}\ln(Z_g g^2)\right)^{-1}\nonumber\\
&=& -\eps \frac{g}{2} - \frac{11}{3} g \frac{g^2 C_2}{16 \pi^2}
    - \frac{34}{3} g \left(\frac{g^2 C_2}{16 \pi^2}\right)^2
    - \frac{2857}{54} g \left(\frac{g^2 C_2}{16 \pi^2}\right)^3 +
       O(g^9),
\end{eqnarray}
in agreement with former results \cite{TVZ80,LV93,PGJ01}. 

%
%
\section{Conclusion and Discussion}\label{CD}
We have succeeded in computing the \mbox{three-loop} 
\mbox{$\beta$-function} for a nonabelian gauge theory with a simple
gauge group by means of the covariant background field method. Our
results confirm those of~\cite{TVZ80} and~\cite{LV93}, where 
conventional field theory methods were used. We also agree with the 
recent work of~\cite{PGJ01} who employed the noncovariant version of
the background field method and extended the results 
of~\cite{TVZ80,LV93} to the case with Yukawa couplings and fermions.
We are of the opinion that our work demonstrates that it is not 
pedantic or impractical to insist on manifest background field gauge
invariance throughout. On the contrary, it leads to a considerable 
reduction in calculational labor. The extension to fermion and scalar
fields is straightforward (see \eg \cite{NSVZ84} for the exact 
propagator of a quark field) and would make our methods suitable for 
phenomenological applications.

In the covariant background field method there are precisely as many
vertices and propagators as in conventional field theory. We selected
the \mbox{Fock-Schwinger} gauge for the background field because this
field then appears only through its field strength. It follows that 
Feynman graphs are at most overall logarithmically divergent. This
leads to a considerable reduction in labor, especially in higher loop
orders, as compared to the usual way of employing the background field
method where overall quadratically divergent graphs appear. Another
advantage of choosing the \mbox{Fock-Schwinger} gauge is, that in
contradiction to~\cite{JO82,JO82b}, the explicit knowledge of the 
Schwinger phasefactor implies factorization of Lorentz and Lie 
structure of the Feynman graphs. We showed that calculational labor
can be further reduced by first using an identity relating the exact
gluon and ghost propagators. Via such manipulations we could reduce
the full \mbox{two-loop} calculation to a single logarithmically
divergent Feynman graph. At the \mbox{three-loop} level, all
tetrahedral graphs could be removed (see sect.~\ref{YM3Loop}) and
hence simple poles with a residue proportional to $\zeta(3)$ do not
occur in the first place. This improves on earlier observations of
explicit cancellation of such poles~\cite{TVZ80}. In fact, the
efficiency of the covariant background field method allows us to use
a high level language such as {\em Mathematica} to reduce the amount
of Feynman graphs as described in section~\ref{YM3Loop}. 

We now give a brief discussion of the prospects for an eventual
three-loop supergravity calculation along the lines of our work
in Yang-Mills theory. We will use the component field language and
comment afterwards on the use of superfields. It is well-known that
supergravity gives rise to a one- and \mbox{two-loop} finite
S-matrix~\cite{Gr77}. However, at three loops one 
expects~\cite{DKS77} a divergence of the effective action of the 
schematic form
\begin{equation}
\Gamma^{(3)}_\infty = {G^2\over\epsilon} \int C^4  + \ldots  \quad ,
\end{equation}
where~$G$ is Newton's constant, $C^4$ is a scalar consisting of four
Weyl tensors and the dots represent fermionic terms. Only an explicit
calculation can determine the absence or presence of this potential
divergence. Contrary to naive expectations, it was shown recently with
string-theoretical methods~\cite{BDDPR98} that $N=8$ supergravity
probably does not have such a three-loop divergence. In fact, the
first divergence of maximal supergravity is expected to occur at
five-loop order. It would be interesting to confirm this surprising
result with an explicit three-loop calculation and investigate if it
still holds for $N<8$. We shall now sketch how such a calculation would 
proceed by means of both the covariant and noncovariant background
field method. Starting from the classical action for supergravity, one
would select suitable background covariant gauges for the various
quantum fields~\cite{DF76,FNJ80}. For the background field metric one
should work in normal coordinates~\cite{Pa78,BP80,ALN77,MSV99}, this
being the analogue of the \mbox{Fock-Schwinger} gauge for Yang-Mills
fields. Indeed, in this gauge one can express the background metric in
terms of its field strength, \ie the Weyl tensor. In addition we are
free to demand a covariantly constant background Weyl tensor, \ie we
may assume a locally symmetric background space.

\begin{figure}[h]
\setcoordinatesystem units <0.75 cm , 0.75 cm> point at 0 0
\centerline{
\beginpicture
\setplotarea x from -2.5 to 12.5, y from -1.5 to 1
\put {$H$} at -2.15 1.1
\put {$H$} at  0.2 1.1
\put {$H$} at -2.2 -1.1
\put {$H$} at  0.15 -1.1
\setquadratic
\plot -0.646447 0.353553 -0.505025 0.353553 -0.505025 0.494975 
      -0.505025 0.636396 -0.363604 0.636396 -0.222183 0.636396 
      -0.222183 0.777817 -0.222183 0.919239 -0.0807612 0.919239 /
\plot -1.35355 0.353553 -1.49497 0.353553 -1.49497 0.494975 
      -1.49497 0.636396 -1.6364 0.636396 -1.77782 0.636396 
      -1.77782 0.777817 -1.77782 0.919239 -1.91924 0.919239 /
\plot -1.35355 -0.353553 -1.49497 -0.353553 -1.49497 -0.494975 
      -1.49497 -0.636396 -1.6364 -0.636396 -1.77782 -0.636396 
      -1.77782 -0.777817 -1.77782 -0.919239 -1.91924 -0.919239 /
\plot -0.646447 -0.353553 -0.505025 -0.353553 -0.505025 -0.494975 
      -0.505025 -0.636396 -0.363604 -0.636396 -0.222183 -0.636396 
      -0.222183 -0.777817 -0.222183 -0.919239 -0.0807612 -0.919239 /
\setlinear
\circulararc 360 degrees from -0.5 0  center at -1 0
\setquadratic
\setshadegrid span <0.04cm> point at -1 0
\vshade -1.5 0 0 <,z,,> -1 -0.5 0.5 -0.5 0 0 /
\put {Fig. a} at -1.0 -1.8
\put {$C$} at  7.85 1.1
\put {$C$} at  10.2 1.1
\put {$C$} at  7.8 -1.1
\put {$C$} at  10.15 -1.1
\setquadratic
\plot 9.35355 0.353553 9.49497 0.353553 9.49497 0.494975 9.49497 
      0.636396 9.6364 0.636396 9.77782 0.636396 9.77782 0.777817 
      9.77782 0.919239 9.91924 0.919239 /
\plot 8.64645 0.353553 8.50503 0.353553 8.50503 0.494975 8.50503 
      0.636396 8.3636 0.636396 8.22218 0.636396 8.22218 0.777817 
      8.22218 0.919239 8.08076 0.919239 /
\plot 8.64645 -0.353553 8.50503 -0.353553 8.50503 -0.494975 8.50503 
     -0.636396 8.3636 -0.636396 8.22218 -0.636396 8.22218 -0.777817 
      8.22218 -0.919239 8.08076 -0.919239 /
\plot 9.35355 -0.353553 9.49497 -0.353553 9.49497 -0.494975 9.49497 
     -0.636396 9.6364 -0.636396 9.77782 -0.636396 9.77782 -0.777817 
      9.77782 -0.919239 9.91924 -0.919239 /
\setlinear
\circulararc 360 degrees from 9.5 0  center at 9 0
\setquadratic
\setshadegrid span <0.04cm> point at 9 0
\vshade  8.5 0 0 <,z,,> 9 -0.5 0.5 9.5 0 0 /
\put {Fig. b} at 9.0 -1.8
\endpicture
} \caption[Comparison between the ordinary and the covariant 
           background field method in gravity.]
          {Comparison between the ordinary (a) and the covariant 
           (b) background field method in gravity.}
\label{VergleichGrav}
\end{figure}

\begin{figure}[h]
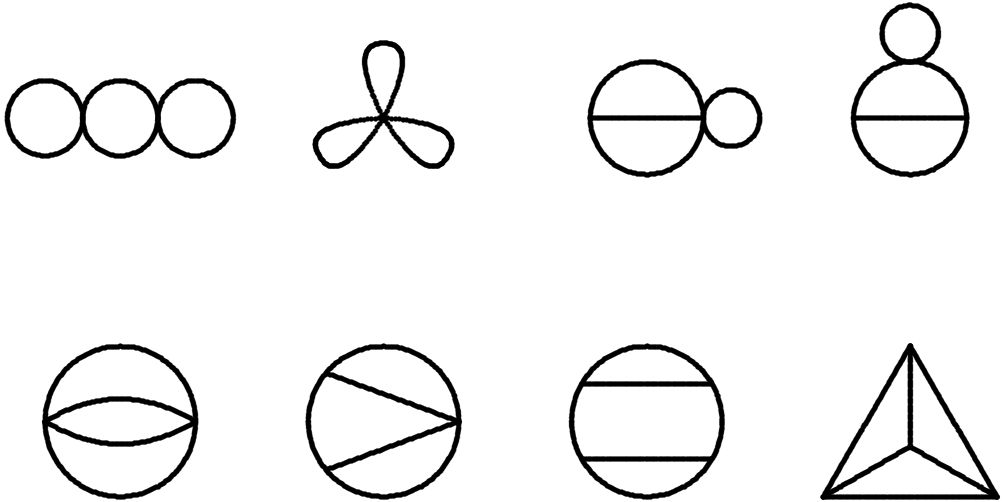

\setcoordinatesystem units <0.5 cm , 0.5 cm> point at 0 0
\centerline{
\beginpicture
\setplotarea x from -3.5 to 3.5, y from -4 to 4
\dickelinie
\circulararc 360 degrees from -1 0  center at -2 0
\circulararc 360 degrees from -1 0 center at 0 0
\circulararc 360 degrees from 1 0 center at 2 0
\endpicture
\beginpicture
\setplotarea x from -3.5 to 3.5, y from -4 to 4
\dickelinie
\setquadratic
\plot 0 0  0.5 1.5 0 2 -0.5 1.5 0 0 /
\plot 0 0 1.55 -0.32 1.71 -1 1.04 -1.19 0 0 /
\plot 0 0 -1.55 -0.32 -1.71 -1 -1.04 -1.19 0 0 /
\setlinear
\endpicture
\beginpicture
\setplotarea x from -3.5 to 3.5, y from -4 to 4
\dickelinie
\circulararc 360 degrees from -1.5 0  center at 0 0
\plot -1.5 0 1.5 0 /
\circulararc 360 degrees from 1.5 0 center at 2.25 0
\endpicture
\beginpicture
\setplotarea x from -3.5 to 3.5, y from -4 to 4
\dickelinie
\circulararc 360 degrees from -1.5 0  center at 0 0
\plot -1.5 0 1.5 0 /
\circulararc 360 degrees from 0 1.5 center at 0 2.25
\endpicture
}
\centerline{
\beginpicture
\setplotarea x from -3.5 to 3.5, y from -4 to 4
\dickelinie
\circulararc 360 degrees from -2 0  center at 0 0
\setquadratic
\plot -2 0 0 0.6 2 0 /
\plot -2 0 0 -0.6 2 0 /
\setlinear
\endpicture
\beginpicture
\setplotarea x from -3.5 to 3.5, y from -4 to 4
\dickelinie
\circulararc 360 degrees from -2 0  center at 0 0
\plot -1.53 1.29 2 0 -1.53 -1.29 /
\endpicture
\beginpicture
\setplotarea x from -3.5 to 3.5, y from -4 to 4
\dickelinie
\circulararc 360 degrees from -2 0  center at 0 0
\plot -1.73 1 1.73 1 /
\plot -1.73 -1 1.73 -1 /
\endpicture
\beginpicture
\setplotarea x from -3.5 to 3.5, y from -4 to 4
\dickelinie
\plot 0 -0.67 0 2 /
\plot 0 -0.67 2.31 -2 /
\plot 0 -0.67 -2.31 -2 /
\plot 0 2 2.31 -2 -2.31 -2 0 2 /
\endpicture
}
\caption[Three-loop graphs in supergravity.]{
Three-loop graphs in supergravity (thick lines represent all quantum
fields to the extent that the corresponding quantum vertices exist).}
\label{SG3}
\end{figure}

In the noncovariant approach~\cite{HV74} one linearizes the background
metric, $g_{\mu\nu}=\eta_{\mu\nu}+H_{\mu\nu}$. This ultimately leads
one to compute three-loop graphs with maximal degree of divergence
equal to eight, see figure~\ref{VergleichGrav}a. On the other hand,
within the covariant approach described above, one must compute only
logarithmically divergent three-loop graphs, see
figure~\ref{VergleichGrav}b. The exact propagators appearing in these
graphs are of the generic form 
\begin{equation}
G(x,x') = \sum_{k=0}^4 C^k G_k(x-x') \quad ,
\end{equation}
where we must include terms of up to fourth order in the Weyl tensor
in order to take account of all divergences. Supergravity has a
non-polynomial action, so that vertices up to sixth order in the
quantum fields must be taken into account, see figure~\ref{SG3}. The
graphs involving fifth and sixth degree vertices are however as easy
to compute as \mbox{two-loop} and \mbox{one-loop} graphs,
respectively. Thus, as in Yang-Mills theory, the most complicated
graphs are tetrahedral. We therefore choose our gauges such that the
number of three-quantum-field vertices is minimized. As was already
shown in~\cite{Ve92}, one can reduce the number of three-graviton
vertices to just two.

Of course, there also exist efficient superspace methods for such
calculations, at least for $N=1$ or 2 supergravity. The background
field method can be extended to superspace~\cite{GGRS83} and used for
calculations at the \mbox{one-loop} level in supergravity~\cite{GrS82}
and \mbox{two-loop} level in super Yang-Mills theories~\cite{GZ85}.
We should mention in particular~\cite{GZ84}, which advocates the use
of gauge-covariant superderivatives instead of ordinary
superderivatives and employs exact super-propagators. This is shown to
lead to improved power counting and hence less work. This is
completely analogous to our work in ordinary space. Just like the
manipulations with gauge-covariant superderivatives directly on
supergraphs, we could manipulate ordinary gauge-covariant derivatives
on ordinary graphs. Selecting suitable normal coordinate gauges in
superspace as in~\cite{OS80} should lead to further improvement since
only the background superfield strength would then appear.

%
%
\appendix
\section{Notation}\label{Nt}
The greek letters $\mu,\nu,\ldots$ are used to indicate lorentzian
indices which can adopt values from $0$ to~$d-1$ with~$d$ being
the dimension of the coordinate space. The colored indices of
SU(N) are indicated by the latin
letters $i,j,\ldots$ in the fundamental and by the latin letters
$a,b,\ldots$ in the adjoint representation.

Following DeWitt the dot product
\begin{equation}
a\cdot b = \int dx\, a^i(x) b_i(x) \label{DeWittProd}
\end{equation}
is defined in such a way, that $i$ represents all indices,
which are in common in $a$ and $b$ and have to be summed up.
Furthermore we write~$dx$ instead of~$d^dx$, as long as the context
allows no confusion.

A bracket underneath a Feynman graph gives the value of
the \mbox{UV-divergences}. They have to be read from the right 
to the left. With every comma the degree of the pole part increases.
The coefficients which are separated by semicolons belong to
the same degree of divergences. They are only specified, if
the divergences contain $\zeta(2 n)$-functions.
The definition is given by
\begin{equation}
(\cdots,a,b,c_1;c_2) \ =\ \cdots + \frac{a}{\eps^3} +
  \frac{b}{\eps^2} + \frac{c_1 + c_2 \zeta(3)}{\eps}\quad .
\end{equation}

\section{Identities of $G_i$ and $R_i$} \label{IGR}
In~$d$ dimensions the functions $G_i$ are defined by
\begin{equation} \label{Gi}
G_{i-1}(x)\ =\ 
  \frac{\Gamma\left(\frac{d}{2} - i\right)}{(4\pi)^\frac{d}{2}}
  \left(\frac{x^2}{4}\right)^{i-\frac{d}{2} }\hspace{2.7cm} 
  \mbox{for } i \geq 1
\end{equation}
Thus, for $d<2$ all functions $G_i$ are free of \mbox{UV-divergences}.
This result can be analytically continued to higher dimensions.
But in contrast for $i > d/2$ the functions $G_i$ show up
IR~divergences, which means that as soon as in $d=n-\eps$
dimensions the inequality $i+1 \ge n/2$ holds the functions
$G_i$ have to be replaced by the following functions
\begin{equation}\label{Ri}
R_{i-1}(x)\ =\ 
  G_{i-1}(x) + 
  \frac{\mu^{-\eps}}{8\pi^2 \eps}\left(-\frac{x^{2}}{4}
  \right)^{i-(d+\eps)/2}
  \qquad \mbox{for } i > d/2 \quad ,
\end{equation}
where the added pole part cancels the IR~divergence of the
function $G_i$.

Concerning the derivatives we get the following identities
\begin{eqnarray*}
\partial_\mu G_{i}(x) &=& -\frac12 x_\mu G_{i-1}(x) \\
\partial_\mu R_{i}(x) &=& -\frac12 x_\mu R_{i-1}(x) \\
\partial^2 G_{i}(x) &=& -i\, G_{i-1}(x)\\
\partial^2 R_{i}(x) &=& -i\, R_{i-1}(x) + \frac{\mu^{-\eps}}{16 \pi^2}
\frac{1}{(i-2)!} \left(\frac{x^2}{4}\right)^{i-2}\\
\partial^2 G_0(x) &=& -\delta^d(x).
\end{eqnarray*}
Using these identities, products of two functions $G_i$  can be
transformed in suited products. Some of the most useful
transformations are listed below
\begin{eqnarray}
G_i\,\partial_\mu G_j
&=& \phantom{+} G_{j-1}\,\partial_\mu G_{i+1}\label{id1}\\
G_i\,\partial_\mu \partial_\nu G_j
&=& \phantom{+} (\partial_\mu G_{i+1})\,\partial_\nu G_{j-1}
   -\frac{1}{2}\,\delta_{\mu\nu}G_{i} G_{j-1} \\
&=& \phantom{+} (\partial_\mu \partial_\nu G_{i+2})\, G_{j-2}
   -\frac{1}{2}\,\delta_{\mu\nu} \left(G_{i} \,  G_{j-1} - 
     G_{i+1}\, G_{j-2}\right),
   \label{id2}
\end{eqnarray}
whereas no index adopts a value less than zero. Several identities
depend on the dimension~$d$. These are for example
\begin{eqnarray}
(d-2-2i)\,G_i(x)\,\partial_\mu G_j(x) 
&=& (d-2-2j)\,G_j(x)\,\partial_\mu G_i(x)\\
\partial_\mu G_0(x)\,\partial^\mu G_{i+1}(x)
&=& \left(\frac{d}{2}-1\right) G_0(x)G_i(x),
\end{eqnarray}
which have to be used carefully, since the value of~$d$ depends on
whether it is part of an UV divergent subloop or not.

The identities presented above deal with products of propagators
depending all on the same endpoints which means that in
coordinate space their values have just to be multiplied.

Regarding propagators concatenated to each other we need
the following property
\begin{eqnarray}\label{Gn2G0}
\int\ dy \frac{1}{(x-y)^{2\alpha}} \frac{1}{(y-z)^{2\beta}}
&=& \pi^{d/2}
\frac{\Gamma(d/2-\alpha)\Gamma(d/2-\beta)\Gamma(\alpha+\beta-d/2)}
     {\Gamma(\alpha)\Gamma(\beta)\Gamma(d-\alpha-\beta)}\times
\nonumber\\     
& &\phantom{\frac{\Gamma(d/2-\alpha)}
     {\Gamma(\alpha)\Gamma(\beta)}}
\frac{1}{(x-z)^{2(\alpha+\beta-d/2)}},
\end{eqnarray}
which leads to
\begin{eqnarray}\label{Gn}
G_i(x_0-x_{i+1}) &=& i! \int dx_1dx_2\cdots dx_{i}\,
G_0(x_0-x_1)\,G_0(x_1-x_2)\cdots G_0(x_i-x_{i+1})\nonumber\\
&=& i!\
\beginpicture
\setcoordinatesystem units <1cm, 1cm>
\setplotarea x from 0 to 4, y from -0.7 to 1
  \linethickness=0.5pt
  \putrule from 0 0 to 4 0
  \put{\mbox{$\scriptstyle{\bullet}$}} at 0   0
  \put{\mbox{$\scriptstyle{\bullet}$}} at 0.5 0
  \put{\mbox{$\scriptstyle{\bullet}$}} at 1   0
  \put{\mbox{$\scriptstyle{\bullet}$}} at 3.5 0
  \put{\mbox{$\scriptstyle{\bullet}$}} at 4.0 0
  \put{\mbox{$\scriptstyle{x_0}$}} [b] <0mm,1.5mm> at 0   0
  \put{\mbox{$\scriptstyle{x_1}$}} [b] <0mm,1.5mm> at 0.5 0
  \put{\mbox{$\scriptstyle{x_2}$}} [b] <0mm,1.5mm> at 1   0
  \put{\mbox{$\scriptstyle{x_i}$}} [b] <0mm,1.5mm> at 3.5 0
  \put{\mbox{$\scriptstyle{x_{i+1}}$}} [b] <0mm,1.5mm> at 4   -0.03
  \put{\mbox{$\cdots$}} <0mm,1.5mm> at 2.2 0
\endpicture
\end{eqnarray}
Thus the convolution of $n+1$ functions $G_0$ multiplied by
a factor $n!$ equals the function $G_i$. This explains why the
graphical representation of the function $G_i$ consists of
a factor $i!$ and a line with $i+1$ dots.

\section{Group Theory}\label{GT}

For a simple compact Lie group $G$ with elements
\begin{equation*}
U(x)\ =\ e^{\Lambda(x)} \quad \mbox{with}\ \Lambda(x)
    \ =\ t_a \Lambda^a(x),
\end{equation*}
where the antihermitian generators $t_a$ with $a = 1,\ldots, \dim(G)$
satisfy the algebra
\begin{equation*}
[t_a,t_b]\ =\ f_{abc}t_c,
\end{equation*}
the Dynkin index and the Casimir operator are defined by
\begin{eqnarray}
\mbox{tr}\left(t_a\,t_b\right) &=& - T(D) \delta_{ab}\\
t_a\,t_a &=& - C_2(D) \seteins \quad .
\end{eqnarray}
This representation is called {\em fundamental} and $T(D) = 1/2$
is a common normalization choice.

Specializing on SU(N) groups the dimension equals
$N^2-1$ and we get
\begin{eqnarray}
t_a\,t_a &=& \frac{1-N^2}{2 N} \,\seteins \\
t_a\,t_b\,t_a &=& \frac{1}{2 N} \,t_b \quad.
\end{eqnarray}
Thus in the fundamental representation the Casimir operator is given
by $(1-N^2)/(2 N)$ multiplied with the unit matrix. Please note,
that in $\delta_{ab}$ the indices run up to $N^2-1$, whereas
in~$\delta_{ij}$ the range of the indices is limited to $N$.

Choosing the adjoint representation
\begin{equation}
(T_a)_{bc} \ =\ -f_{abc}
\end{equation}
we are lead to the following Casimir operator
\begin{equation}
\tr(T_a T_b) \ =\ -N \delta_{ab}
\end{equation}
for the SU(N) groups.

In this paper all terms written in the adjoint
representation are illustrated bold. The Casimir operator remains
undetermined. The adjoint field strength \eg is given by
\begin{eqnarray*}
{\mathbf F}_{\mu\nu} &=& F^a_{\mu\nu} (T^a)^{bc}\ =\ 
   - F^a_{\mu\nu} f^{abc}\\
{\mathbf F}^2_{\mu\nu} &=& {\mathbf F}_{\mu\ro} {\mathbf F}_{\ro\nu}\\
{\mathbf F}^2 &=& \Tr\,{\mathbf F}^2_{\mu\nu}\\
\tr\,{\mathbf F}^2 &=& C_2(F^a_{\mu\nu})^2 ,
\end{eqnarray*}
where $\tr$ respectively $\Tr$ are the shortcuts for the trace over
the group indices respectively Lorentz indices.

To calculate products of structure constants in an effective way
we establish a graphical representation.
Using the definitions
\begin{equation*}
\delta_{ab}\ =\ a\,\rule[1mm]{1.5cm}{0.1mm}\,b;\qquad
  \delta_{aa}\ =\ \mbox{dim}(G)
\end{equation*}
the table~\ref{TabSt} shows some needed identities.

\begin{table}[H]
\[
\begin{array}{|l|l|l|}
\hline
\mbox{structure constants}&\mbox{adjoint representation}&
\mbox{graphical}\\
\hline
\rule[-6mm]{0mm}{1.5cm}
f_{abc}&-(T_a)_{bc}&
  \beginpicture
    \setcoordinatesystem units <0.4 cm , 0.4 cm>
    \plot 0 1 0 0 0.87 -0.5 /
    \plot 0 0 -0.87 -0.5 /
    \put{$\scriptstyle{a}$} at 0 1.4
    \put{$\scriptstyle{b}$} at -1.2 -0.7
    \put{$\scriptstyle{c}$} at 1.2 -0.7
  \endpicture\\
f_{iaj}f_{jbi}\ =\ -C_2\delta_{ab}&\mbox{tr}(T_a T_b)\ 
                =\ -C_2\delta_{ab}&
\beginpicture
    \setcoordinatesystem units <0.4 cm , 0.4 cm>
    \setplotarea x from -1 to 2, y from -1.5 to 1.5
    \circulararc 360 degrees from 0 0.5 center at 0 0
    \plot -0.5 0 -1 0 /
    \plot  0.5 0  1 0 /
    \put{$\scriptstyle{a}$}[b] at -0.9 0.1
    \put{$\scriptstyle{b}$}[b] at  0.9 0.1
    \put{$=\,-C_2 $}[l] at 1.3 0
    \plot 4.5 0 6 0 /
    \put{$\scriptstyle{a}$}[b] at 4.6 0.1
    \put{$\scriptstyle{b}$}[b] at 5.9 0.1
  \endpicture\\
\rule[-6mm]{0mm}{1.5cm}
\begin{array}{rcl}
f_{abi} f_{cdi} &=& f_{bci} f_{dai} -\\
& &f_{bdi} f_{cai}
\end{array}
  &\begin{array}{rcl}f_{abi} (T_i)_{dc}&=&[T_a, T_b]\\
                    &=&(T_a)_{di} (T_b)_{ic} -\\
                    & & (T_b)_{di} (T_a)_{ic} 
   \end{array}&
\beginpicture
    \setcoordinatesystem units <0.4 cm , 0.4 cm>
    \setplotarea x from -1 to 1, y from -1.5 to 1.5
    \plot 0.53 1.28 0 0.75 -0.53 1.28 /
    \plot 0 0.75 0 -0.45 /
    \plot 0.53 -0.98 0 -0.45 -0.53 -0.98 /
    \put{$\scriptstyle{a}$}[t] at 0.83 1.08
    \put{$\scriptstyle{b}$}[t] at -0.83 1.28
    \put{$\scriptstyle{c}$}[b] at -0.83 -0.98
    \put{$\scriptstyle{d}$}[b] at 0.83 -0.98
    \endpicture =
    \beginpicture
    \setcoordinatesystem units <0.4 cm , 0.4 cm>
    \plot 1.28 0.73 0.75 0.2 1.28 -0.33 /
    \plot 0.75 0.2 -0.45 0.2 /
    \plot -0.98 0.73  -0.45 0.2 -0.98 -0.33 /
    \put{$\scriptstyle{a}$}[r] at 1.18 0.83
    \put{$\scriptstyle{d}$}[r] at 1.18 -0.53
    \put{$\scriptstyle{b}$}[l] at -0.88 0.93
    \put{$\scriptstyle{c}$}[l] at -0.88 -0.53
    \endpicture -
    \beginpicture
    \setcoordinatesystem units <0.4 cm , 0.4 cm>
    \plot 1.28 1.03 0.75 0.5 0.35 0.1 /
    \plot -0.05 -0.3 -0.73 -0.88 /
    \plot 0.75 0.5 -0.45 0.5 /
    \plot -0.98 1.03  -0.45 0.5 1.03 -0.88 /
    \put{$\scriptstyle{a}$}[r] at 0.98 0.93
    \put{$\scriptstyle{d}$}[r] at 1.18 -0.43
    \put{$\scriptstyle{b}$}[l] at -0.68 1.03
    \put{$\scriptstyle{c}$}[l] at -0.88 -0.43
    \endpicture\\
\rule[-6mm]{0mm}{1.5cm}
f_{iaj} f_{jbk} f_{kci} \ = \ -\frac12 C_2 f_{abc}
  & \begin{array}{rcl}
           \mbox{tr}(T_a T_b T_c) &=& T_i T_b T_i \\
           &=& -\frac12 C_2(T_b)_{ac}
    \end{array}
  & \beginpicture
    \setcoordinatesystem units <0.4 cm , 0.4 cm>
    \setplotarea x from -1 to 1, y from -1.5 to 1.5
    \plot 0 1 0 0.5 0.43 -0.25 0.87 -0.5 /
    \plot 0 0.5 -0.43 -0.25 -0.87 -0.5 /
    \plot -0.43 -0.25 0.43 -0.25 /
    \put{$\scriptstyle{a}$} at 0 1.4
    \put{$\scriptstyle{b}$} at -1.2 -0.7
    \put{$\scriptstyle{c}$} at 1.2 -0.7
    \endpicture = -\frac12 C_2
    \beginpicture
    \setcoordinatesystem units <0.4 cm , 0.4 cm>
    \plot 0 1 0 0 0.87 -0.5 /
    \plot 0 0 -0.87 -0.5 /
    \put{$\scriptstyle{a}$} at 0 1.4
    \put{$\scriptstyle{b}$} at -1.2 -0.7
    \put{$\scriptstyle{c}$} at 1.2 -0.7
    \endpicture\\
\hline
\end{array}
\]\label{TabSt}
\caption{Identities of the structure constants}
\end{table}

\section{Calculation of two two-loop graphs}\label{2LoopGraph}

In this appendix we will calculate the \mbox{UV-divergences} of the 
last two graphs in the last row of figure~\ref{ur2loopneu} explicitly.
Using
\begin{eqnarray*}
D_\mu \Phi G_0 \overleftarrow D_\nu
&=& - \Phi \partial_\mu \partial_\nu G_0 - \frac{1}{2} \Phi 
       {\mathbf F}_{\mu\nu} G_0
    - \Phi {\mathbf F}_{\mu\la} \partial^\la \partial_\nu G_1
    + \Phi {\mathbf F}_{\nu\la} \partial^\la \partial_\mu G_1 \\
&& \mbox{} + {\mathbf F}_{\mu\la} {\mathbf F}_{\nu\rho} \partial^\la 
       \partial^\rho G_2
    - \frac{1}{2} {\mathbf F}^2_{\mu\nu} G_1 + O({\mathbf F}^3)\\
&& \mbox{and}\\
{\mathbf F}_{\si\tau} D_\mu \Phi G_1 \overleftarrow D_\nu
&=& -\Phi {\mathbf F}_{\si\tau} \partial_\mu \partial_\nu G_1-
    \frac{1}{2} {\mathbf F}_{\si\tau}{\mathbf F}_{\mu\nu}G_1-
    {\mathbf F}_{\si\tau}{\mathbf F}_{\rho\nu} \partial^\rho 
    \partial_\mu G_2 \\
& & \mbox{} + {\mathbf F}_{\si\tau}{\mathbf F}_{\rho\mu} 
    \partial^\rho \partial_\nu G_2 + O({\mathbf F}^3)
\end{eqnarray*}
we get
\begin{eqnarray}
D_\mu G_{\si\tau}\overleftarrow D_\nu
&=& \delta_{\si\tau} D_\mu \Phi G_0 \overleftarrow D_\nu +
    2 {\mathbf F}_{\si\tau} D_\mu \Phi G_1 \overleftarrow D_\nu +
    \frac{1}{4} \delta_{\si\tau} {\mathbf F}^2 
    \partial_\mu \partial_\nu G_2 \nonumber\\
&& \mbox{} +\frac{1}{3} \delta_{\si\tau} {\mathbf F}^2_{\la\rho} 
   \partial^\la \partial^\rho \partial_\mu \partial_\nu G_3 -
    2 {\mathbf F}^2_{\si\tau} \partial_\mu \partial_\nu G_2 
    \nonumber\\
&=& - \Phi \Bigl(\delta_{\si\tau} \partial_\mu \partial_\nu G_0
    + \frac{1}{2}  \delta_{\si\tau} {\mathbf F}_{\mu\nu} G_0
    + 2 {\mathbf F}_{\si\tau} \partial_\mu \partial_\nu G_1 
    \nonumber \\
& & \phantom{-\Phi\ } + 2\delta_{\si\tau} {\mathbf F}_{\mu\la}
    \partial^\la \partial_\nu G_1
    - 2  \delta_{\si\tau} {\mathbf F}_{\nu\la}
    \partial^\la \partial_\mu G_1 \Bigr) \nonumber \\
&&  \mbox{} - {\mathbf F}_{\si\tau}{\mathbf F}_{\mu\nu}G_1
    - {\mathbf F}_{\si\tau}{\mathbf F}_{\rho\nu}
    \partial^\rho \partial_\mu G_2
    + {\mathbf F}_{\si\tau}{\mathbf F}_{\rho\mu}
    \partial^\rho \partial_\nu G_2 \nonumber \\
&&  \mbox{} + \delta_{\si\tau} {
    \mathbf F}_{\mu\la} {\mathbf F}_{\nu\rho}
    \partial^\la \partial^\rho G_2
    - \frac{1}{2}\delta_{\si\tau} {\mathbf F}^2_{\mu\nu} G_1
    + \frac{1}{4} \delta_{\si\tau} {\mathbf F}^2
    \partial_\mu \partial_\nu G_2 \nonumber\\
&&  \mbox{} + \frac{1}{3} \delta_{\si\tau} {\mathbf F}^2_{\la\rho}
    \partial^\la \partial^\rho \partial_\mu \partial_\nu G_3 -
    2 {\mathbf F}^2_{\si\tau} \partial_\mu \partial_\nu G_2 
    \nonumber \\
&& \mbox{} + O({\mathbf F}^3).
\end{eqnarray}
This result combined with the identity $(\seteins,
{\mathbf F}_{\mu\nu},{\mathbf F}_{\rho\si})=
-(\seteins,\seteins,{\mathbf F}_{\mu\nu}{\mathbf F}_{\rho\si})/2$
and the symmetries of the $\Theta$ graphs leads us to
\begin{eqnarray}
\left(G^{\mu\nu}, G^{\si\tau}, 
D_\mu G_{\si\tau}\overleftarrow D_\nu\right)
&=& - \left(\Phi \delta^{\mu\nu} G_0,\Phi \delta^{\si\tau} G_0,
      \Phi \delta_{\si\tau}\partial_\mu\partial_\nu G_0\right)
      \nonumber\\
& & \mbox{} + 6 \left(G_0,G_0,G_1 {\mathbf F}^2 \right)
     -   \left(\delta_{\mu\nu} G_0, G_0, \delta^{\mu\nu} G_1
     {\mathbf F}^2 \right)\nonumber\\
& & \mbox{} -   \frac{1}{2}\left(\delta_{\mu\nu} G_0,\delta^{\mu\nu}
     G_0, G_1 {\mathbf F}^2 \right)\nonumber\\
& & \mbox{} - 2 \left(\delta_{\mu\nu} G_0,\delta^{\mu\nu}G_1,
    \partial^\rho\partial^\si G_1 {\mathbf F}^2_{\rho\si} \right)
    \nonumber\\
& & \mbox{} - 2 \left(\delta_{\mu\nu} G_0,\delta^{\mu\nu}
        \partial^\rho\partial^\si G_0,
         G_2 {\mathbf F}^2_{\rho\si} \right)\nonumber\\
& & \mbox{} - 2 \left(G_0,\delta_{\mu\nu} G_0,
        \delta^{\mu\nu}\partial^\rho\partial^\si G_2
        {\mathbf F}^2_{\rho\si} \right)
     + O({\mathbf F}^3).\label{Grapha2loop}
\end{eqnarray}
As expected, all graphs containing the field strength linearly vanish
due to symmetry reasons. The first term in (\ref{Grapha2loop})
does not contribute either, since contracting the term
$\partial_\mu\partial_\nu G_0$ with $\delta^{\mu\nu}$ of the first
propagator, we get a $\delta$-function, which entails that
initial and end point of the \mbox{$G_0$-function} are identical.
Since $G_0(0)$ equals zero by definition, the first term can be
neglected.

We refrain from contracting those $\delta$-tensors, which lead to an
additional factor of dimension~$d$. Thus, when using the
\mbox{$R^\ast$ method}, we can still decide, if all $\delta$-tensors
building up a factor~$d$ belong to any subgraph and if therefor the
$\eps$ part of the factor~$d$ has to be neglected, while calculating
the divergences of the appropriate subgraph.

In many cases we can eliminate these $\delta$-tensors with the help
of the identity
\begin{equation} \label{xdG}
d\,G_i = \left(2 + 2i - x_\mu \partial^\mu\right)G_i
\end{equation}
unless the $G_i$ function belongs exactly to the same subgraphs as the
$\delta$-tensors do. Applying this identity \eg to the second graph of
(\ref{Grapha2loop}), we get
\begin{eqnarray}
\left(\delta_{\mu\nu} G_0, G_0, \delta^{\mu\nu} G_1 
     {\mathbf F}^2 \right)
&=& 4\left( G_0, G_0, G_1 {\mathbf F}^2 \right)+
    2\left( \partial_\mu G_1, G_0, \partial^\mu G_1 {\mathbf F}^2 
     \right)\nonumber\\
&=& 6\left( G_0, G_0, G_1 {\mathbf F}^2 \right)-
     \left( \seteins, G_1, G_1 {\mathbf F}^2 \right),
\end{eqnarray}
whereas the last equal sign is based on the identity
\begin{equation}\label{d12}
\partial_1\cdot\partial_2\ =\ \frac{1}{2}\partial_3^2-
\frac{1}{2}\partial_1^2-\frac{1}{2}\partial_2^2.
\end{equation}
The numbers illustrate on which propagator of the
3-vertex the derivatives operate. In momentum space
the analogue identity is known as
\begin{equation}
p_1\cdot p_2\ = \ \frac{1}{2}p_3^2-
\frac{1}{2}p_1^2-\frac{1}{2}p_2^2.
\end{equation}

For the second graph of the last row in figure~\ref{ur2loopneu} we
get, by using the identities (\ref{id1}) and (\ref{id2}), the
following result
\begin{eqnarray}\label{2LG1}
\left(G^{\mu\nu}, G^{\si\tau}, D_\mu G_{\si\tau}
   \overleftarrow D_\nu\right)
&=& \phantom{-}\frac{1}{2} \left(\seteins,G_1, G_1 
{\mathbf F}^2 \right)
    + 3  \left(G_0,G_0, G_1 {\mathbf F}^2 \right)\nonumber\\
& & \mbox{} - 4  \left(G_0,G_0, \partial^\mu\partial^\nu G_2 
    {\mathbf F}_{\mu\nu}^2 \right)\nonumber\\
& & \mbox{} - 2  \left(G_0,\partial^\mu\partial^\nu G_0, G_2 
    {\mathbf F}_{\mu\nu}^2 \right)
    + O({\mathbf F}^3),
\end{eqnarray}
whereas the third graph yields
\begin{eqnarray}\label{2LG2}
\left(G^{\mu\nu},G_{\rho\nu}\overleftarrow D^\sigma,D^\rho
G_{\mu\sigma}\right)
&=& - \frac{1}{2}\left(G_0,G_0, G_1 {\mathbf F}^2 \right)
    + \left(G_0,G_0, \partial^\mu\partial^\nu G_2 
    {\mathbf F}_{\mu\nu}^2 \right)
      \nonumber\\
& & \mbox{} + 2\left(G_0,\partial^\mu\partial^\nu G_0, G_2 
    {\mathbf F}_{\mu\nu}^2 \right)
    + O({\mathbf F}^3).
\end{eqnarray}
Next, we want to substitute the last two Feynman graphs
of (\ref{2LG1}) respectively (\ref{2LG2}), both containing
derivatives, with Feynman graphs containing no derivatives.
For this reason we introduce two Feynman graphs without
subdivergences. The first one is given by

\begin{eqnarray}\label{G0G0ddG2s1}
2\beginpicture
    \setcoordinatesystem units <0.4 cm , 0.4 cm>
    \setplotarea x from -1 to 1, y from -1.5 to 1.5
\kreis
\strichLR
\punktO
\arrow <1.75mm> [0.25,0.85] from -1.82 0.8 to -1.62 1.2
\put{$\scriptstyle{\rho}$} at -2.2 1.05
\put{\ma\bullet} at 0 0
\arrow <1.75mm> [0.25,0.85] from -1.32 0 to -0.72 0
\put{$\scriptstyle{\sigma}$} at -1.0 0.5
\put{\ma\bullet} at  1.73 -1
\put{\ma\bullet} at  1.18 -1.62
\put{$\scriptstyle{\rho}$} at -2.3 -1.05
\put{$\scriptstyle{\sigma}$} at -1.9 -1.75
\put{$\scriptstyle{\mu}$} at -1.0 -2.3
\put{$\scriptstyle{\nu}$} at  0.0 -2.4
\arrow <1.75mm> [0.25,0.85] from -1.745 -0.97  to -1.73 -1
\arrow <1.75mm> [0.25,0.85] from -1.195  -1.6   to -1.17 -1.62
\arrow <1.75mm> [0.25,0.85] from -0.45 -1.95  to -0.42 -1.96
\arrow <1.75mm> [0.25,0.85] from  0.38 -1.965  to  0.42 -1.96
\endpicture
\left(\seteins,\seteins,{\mathbf F_{\mu\nu}^2}\right)
&=& \phantom{+} \frac{1}{4}
\beginpicture
    \setcoordinatesystem units <0.4 cm , 0.4 cm>
    \setplotarea x from -1.5 to 1.5, y from -2 to 2
  \circulararc 360 degrees from 0 0 center at 0 -1
  \circulararc 360 degrees from 0 0 center at 0 1
\put{\ma\bullet} at  0 -2
\put{\ma\bullet} at  0  0
\put{\ma\bullet} at  0  2
\endpicture
\left(\seteins,\seteins,{\mathbf F^2}\right)\nonumber\\
& & \mbox{} + 2
\beginpicture
    \setcoordinatesystem units <0.4 cm , 0.4 cm>
\kreis
\strichLR
\put{\ma\bullet} at  1.18 -1.61
\put{\ma\bullet} at  0.435 -1.955
\put{$\scriptstyle{\mu}$} at -1.9 -1.75
\put{$\scriptstyle{\nu}$} at -1.0 -2.3
\arrow <1.75mm> [0.25,0.85] from -1.195  -1.6   to -1.17 -1.62
\arrow <1.75mm> [0.25,0.85] from -0.45 -1.95  to -0.42 -1.96
\endpicture
\left(\seteins,\seteins,{\mathbf F_{\mu\nu}^2}\right)\nonumber\\
\left(\partial_\rho G_1,\partial_\sigma G_1,
\partial^\rho\partial^\sigma \partial^\mu\partial^\nu G_2
 {\mathbf F}_{\mu\nu}^2 \right)
&=&\phantom{+}\frac{1}{4}
   \left(\seteins,G_1, G_1 {\mathbf F}^2 \right)\nonumber\\
& & \mbox{} +
\left(G_0,G_0,\partial^\mu\partial^\nu G_2 {\mathbf F}_{\mu\nu}^2
\right),
\end{eqnarray}
where we added the graphical notation to make things more 
comprehensive. It is easily shown, that this graph, by repetitive use 
of the identity~(\ref{d12}), decomposes in two graphs. Both can be 
found in the equations~(\ref{2LG1}) and~(\ref{2LG2}), whereas the last
one belongs to the group of graphs which we want to eliminate.
Since the introduced graph contains no subdivergences, which means, 
that it has only a simple pole in $\eps$, we are allowed to use the
identity~(\ref{F22F2}) with $d=4$. Therefore, being interested
in the divergent part only, the new graph simplifies to
\begin{eqnarray}\label{G0G0ddG2s2}
2\beginpicture
    \setcoordinatesystem units <0.4 cm , 0.4 cm>
    \setplotarea x from -1 to 1, y from -1.5 to 1.5
\kreis
\strichLR
\punktO
\arrow <1.75mm> [0.25,0.85] from -1.82 0.8 to -1.62 1.2
\put{$\scriptstyle{\rho}$} at -2.2 1.05
\put{\ma\bullet} at 0 0
\arrow <1.75mm> [0.25,0.85] from -1.32 0 to -0.72 0
\put{$\scriptstyle{\sigma}$} at -1.0 0.5
\put{\ma\bullet} at  1.73 -1
\put{\ma\bullet} at  1.18 -1.62
\put{$\scriptstyle{\rho}$} at -2.3 -1.05
\put{$\scriptstyle{\sigma}$} at -1.9 -1.75
\put{$\scriptstyle{\mu}$} at -1.0 -2.3
\put{$\scriptstyle{\nu}$} at  0.0 -2.4
\arrow <1.75mm> [0.25,0.85] from -1.745 -0.97  to -1.73 -1
\arrow <1.75mm> [0.25,0.85] from -1.195  -1.6   to -1.17 -1.62
\arrow <1.75mm> [0.25,0.85] from -0.45 -1.95  to -0.42 -1.96
\arrow <1.75mm> [0.25,0.85] from  0.38 -1.965  to  0.42 -1.96
\endpicture
\left(\seteins,\seteins,{\mathbf F_{\mu\nu}^2}\right)
&=& -\frac{1}{2}
\beginpicture
    \setcoordinatesystem units <0.4 cm , 0.4 cm>
    \setplotarea x from -1 to 1, y from -1.5 to 1.5
\kreis
\strichLR
\punktO
\arrow <1.75mm> [0.25,0.85] from -1.82 0.8 to -1.62 1.2
\put{$\scriptstyle{\rho}$} at -2.2 1.05
\put{\ma\bullet} at 0 0
\arrow <1.75mm> [0.25,0.85] from -1.32 0 to -0.72 0
\put{$\scriptstyle{\sigma}$} at -1.0 0.5
\put{\ma\bullet} at  0 -2
\put{$\scriptstyle{\rho}$} at -2.3 -1.05
\put{$\scriptstyle{\sigma}$} at -1.9 -1.75
\arrow <1.75mm> [0.25,0.85] from -1.745 -0.97  to -1.73 -1
\arrow <1.75mm> [0.25,0.85] from -1.195  -1.6   to -1.17 -1.62
\endpicture
\left(\seteins,\seteins,{\mathbf F^2}\right) + O(\eps^0)\nonumber\\
&=& \phantom{+} \frac{1}{8}
\beginpicture
    \setcoordinatesystem units <0.4 cm , 0.4 cm>
    \setplotarea x from -1.5 to 1.5, y from -2 to 2
  \circulararc 360 degrees from 0 0 center at 0 -1
  \circulararc 360 degrees from 0 0 center at 0 1
\put{\ma\bullet} at  0 -2
\put{\ma\bullet} at  0  0
\put{\ma\bullet} at  0  2
\endpicture
\left(\seteins,\seteins,{\mathbf F^2}\right)\nonumber\\
& &\mbox{} -\frac{1}{4}
\beginpicture
    \setcoordinatesystem units <0.4 cm , 0.4 cm>
\kreis
\strichLR
\put{\ma\bullet} at  0 -2
\endpicture
\left(\seteins,\seteins,{\mathbf F^2}\right)+ O(\eps^0)\nonumber\\
& &\mbox{respectively}\nonumber\\
\left(\partial_\rho G_1,\partial_\sigma G_1,
\partial^\rho\partial^\sigma \partial^\mu\partial^\nu G_2
 {\mathbf F}_{\mu\nu}^2 \right)
&=&-\frac{1}{2}\left(\partial_\rho G_1,\partial_\sigma G_1,
\partial^\rho\partial^\sigma G_1 {\mathbf F}^2 \right)+ O(\eps^0)
\nonumber\\
&=&\phantom{+}\frac{1}{8}
   \left(\seteins,G_1,G_1{\mathbf F^2}\right)\nonumber\\
& &\mbox{} -\frac{1}{4}
   \left(G_0,G_0,G_1 {\mathbf F^2}\right) + O(\eps^0) .
\end{eqnarray}
Finally, by comparing the right sides of the equations
(\ref{G0G0ddG2s1}) and (\ref{G0G0ddG2s2}) we get the identity
\begin{equation}\label{G0G0ddG2}
\left(G_0,G_0,\partial^\mu\partial^\nu G_2 {\mathbf F}_{\mu\nu}^2 
\right)
\ =\ -\frac{1}{8}\left(\seteins,G_1,G_1{\mathbf F^2}\right) -
   \frac{1}{4}\left(G_0,G_0,G_1 {\mathbf F^2}\right) + O(\eps^0),
\end{equation}
which allows us to eliminate one of the said graphs.
In a similar manner we create another graph containing
no subdivergences
\begin{equation}
4\beginpicture
    \setcoordinatesystem units <0.4 cm , 0.4 cm>
    \setplotarea x from -1 to 1, y from -1.5 to 1.5
\kreis
\strichLR
\punktO
\arrow <1.75mm> [0.25,0.85] from -1.82 0.8 to -1.62 1.2
\put{$\scriptstyle{\rho}$} at -2.2 1.05
\put{\ma\bullet} at 0.4 0
\put{\ma\bullet} at 1.2 0
\arrow <1.75mm> [0.25,0.85] from -0.92 0 to -0.12 0
\arrow <1.75mm> [0.25,0.85] from -1.32 0 to -0.62 0
\arrow <1.75mm> [0.25,0.85] from -1.72 0 to -1.12 0
\put{$\scriptstyle{\sigma}$} at -0.9 0.5
\put{$\scriptstyle{\tau}$} at -0.4 0.5
\put{$\scriptstyle{\rho}$} at -1.4 0.5
\put{\ma\bullet} at  1.73 -1
\put{\ma\bullet} at  1.23 -1.62
\put{$\scriptstyle{\sigma}$} at -2.3 -1.05
\put{$\scriptstyle{\tau}$} at -1.9 -1.75
\put{$\scriptstyle{\mu}$} at -1.0 -2.3
\put{$\scriptstyle{\nu}$} at  0.0 -2.4
\arrow <1.75mm> [0.25,0.85] from -1.745 -0.97  to -1.73 -1
\arrow <1.75mm> [0.25,0.85] from -1.195  -1.6   to -1.17 -1.62
\arrow <1.75mm> [0.25,0.85] from -0.45 -1.95  to -0.42 -1.96
\arrow <1.75mm> [0.25,0.85] from  0.38 -1.965  to  0.42 -1.96
\endpicture\ =\
\left(\partial_\rho G_1,\partial^\rho\partial_\sigma\partial_\tau G_2,
\partial^\sigma\partial^\tau \partial^\mu\partial^\nu G_2
 {\mathbf F}_{\mu\nu}^2 \right)
\end{equation}
which leads us to the identity
\begin{equation}\label{G0ddG0G2}
\left(G_0,\partial^\mu\partial^\nu G_0, G_2 {\mathbf F}_{\mu\nu}^2 
\right)
\ =\ -\frac{1}{12}\left(\seteins,G_1,G_1{\mathbf F^2}\right) +
   \frac{1}{6}\left(G_0,G_0,G_1 {\mathbf F^2}\right) + O(\eps^0)\quad .
\end{equation}
Taking in account the equation
\begin{equation}
(G_{\mu\nu},\seteins,G_{\mu\nu})\ =\
2\left(\seteins,G_1,G_1 {\mathbf F^2}\right)+ O(\eps^0)
\end{equation}
eventually only two graphs remain, which contribute to the divergence
for the Yang-Mills-theory up to two loop as presented in
section~\ref{YM2Loop}:
\begin{eqnarray}
\Gamma^\text{div}_2 &=&
(G_{\mu\nu},\seteins,G_{\mu\nu})-
(G_{\mu\nu},G_{\rho\sigma},D_\mu G_{\rho\sigma} \overleftarrow D_\nu)-
2(G_{\mu\nu},G_{\rho\nu} \overleftarrow D_\sigma,D_\rho G_{\mu\sigma})
\nonumber\\
&=&\frac{17}{12}(\seteins,G_1,G_1 {\bf F}^2) - 
   \frac{17}{6} (G_0,G_0,G_1 {\bf F}^2) +
                O(\eps^0).
\end{eqnarray}
The result, which is wellknown from the literature, can be verified
by looking up the divergences of these graphs given in 
appendix~\ref{D2G}.

\section{Divergences of some graphs}\label{D2G}
In this appendix the divergences of some graphs are specified.
Though, for the renormalization up to the order of two-loop, only
the divergences of one two-loop graph namely $\Theta 1.3$ is needed,
the other graphs are required while doing the three-loop
renormalization, since they appear as subgraphs.
\begin{displaymath}
\begin{array}{cccc}
\beginpicture
    \setcoordinatesystem units <0.4 cm , 0.4 cm>
    \setplotarea x from -1 to 1, y from -1.5 to 1.5
\kreis
\strichLR
\endpicture
&
\beginpicture
    \setcoordinatesystem units <0.4 cm , 0.4 cm>
    \setplotarea x from -1 to 1, y from -1.5 to 1.5
\kreis
\strichLR
\punktO
\endpicture\\
\Theta_0&\Theta_1\\
\beginpicture
    \setcoordinatesystem units <0.4 cm , 0.4 cm>
    \setplotarea x from -1 to 1, y from -1.5 to 1.5
\kreis
\strichLR
\punktO
\punktU
\pfeilLUrunter
\dickpfeilRUrunter
\endpicture&
\beginpicture
    \setcoordinatesystem units <0.4 cm , 0.4 cm>
    \setplotarea x from -1 to 1, y from -1.5 to 1.5
\kreis
\strichLR
\punktO
\punktU
\pfeillMrechts
\dickpfeilRUrunter
\endpicture&
\beginpicture
    \setcoordinatesystem units <0.4 cm , 0.4 cm>
    \setplotarea x from -1 to 1, y from -1.5 to 1.5
\kreis
\strichLR
\punktO
\punktU
\pfeilLOhoch
\dickpfeilRUrunter
\endpicture&
\beginpicture
    \setcoordinatesystem units <0.4 cm , 0.4 cm>
    \setplotarea x from -1 to 1, y from -1.5 to 1.5
\kreis
\strichLR
\punktO
\punktU
\pfeillMrechts
\dickpfeilrMlinks
\endpicture\\
\Theta1.1&\Theta1.2&\Theta1.3&\Theta1.4\\
\beginpicture
    \setcoordinatesystem units <0.4 cm , 0.4 cm>
    \setplotarea x from -1 to 1, y from -1.5 to 1.5
\kreis
\strichLR
\punktrM
\punktlM
\pfeilLUrunter
\dickpfeilRUrunter
\endpicture&
\beginpicture
    \setcoordinatesystem units <0.4 cm , 0.4 cm>
    \setplotarea x from -1 to 1, y from -1.5 to 1.5
\kreis
\strichLR
\punktrM
\punktlM
\pfeilLmrechts
\dickpfeilRUrunter
\endpicture&
\beginpicture
    \setcoordinatesystem units <0.4 cm , 0.4 cm>
    \setplotarea x from -1 to 1, y from -1.5 to 1.5
\kreis
\strichLR
\punktrM
\punktlM
\pfeilLOhoch
\dickpfeilRUrunter
\endpicture&
\beginpicture
    \setcoordinatesystem units <0.4 cm , 0.4 cm>
    \setplotarea x from -1 to 1, y from -1.5 to 1.5
\kreis
\strichLR
\punktrM
\punktlM
\pfeilLmrechts
\dickpfeilRmlinks
\endpicture\\
\Theta2.1&\Theta2.2&\Theta2.3&\Theta2.4
\end{array}
\end{displaymath}

Table~\ref{DivTheta} contains two results for
every graph. The first column transcribed
with ${\mathcal K\overline{R}}$ gives the divergences
of the graphs with uncontracted derivatives,
whereas the second column transcribed
with ${\mathcal K\overline{R}}\delta_{\mu\nu}$ gives
the divergences of the graphs with contracted
derivatives. We want to emphasize once more that,
since we work with renormalized graphs, the result
of the second column cannot always be obtained from the
first one by merely multiplying its result with
a factor~$d$. Again we have to watch carefully if
the contracted derivatives belong to a subdivergence.

\begin{table}
\begin{displaymath}
\renewcommand{\arraystretch}{2}
\begin{array}{|l|c|c||l|c|c|}
\hline
 &{{\mathcal K\overline{R}}}
 &{{\mathcal K\overline{R}}\delta_{\mu\nu}}&
 &{{\mathcal K\overline{R}}}
 &{{\mathcal K\overline{R}}\delta_{\mu\nu}}\\
\hline
\Theta_0& \displaystyle  
\left(\phantom{-}0, \phantom{-}\frac{1}{2}\right)\partial^2\, &-&
\Theta_1& \displaystyle  
\left(-2,\phantom{-}1\right)\phantom{\delta_{\mu\nu}} &-\\[1ex]
\hline
\Theta1.1
&\displaystyle 
\left(-\frac{1}{2},\phantom{-}\frac{3}{8}\right)\delta_{\mu\nu}
&\displaystyle \left(-2,\phantom{-}1\right)
&\Theta2.1
&\displaystyle 
\left(\phantom{-}0,-\frac{1}{12}\right)\delta_{\mu\nu}
&\displaystyle \left(\phantom{-}0,\phantom{-}0\right)\\
\Theta1.2
&\displaystyle 
\left(\phantom{-}\frac{1}{2},-\frac{1}{8}\right)\delta_{\mu\nu}
&\displaystyle \left(\phantom{-}2,\phantom{-}0\right)
&\Theta2.2
&\displaystyle 
\left(\phantom{-}\frac{1}{4},-\frac{1}{16}\right)\delta_{\mu\nu}
&\displaystyle \left(\phantom{-}1,-\frac{1}{2}\right)\\
\Theta1.3
&\displaystyle 
\left(\phantom{-}0,-\frac{1}{4}\right)\delta_{\mu\nu}
&\displaystyle \left(\phantom{-}0,-1\right)
&\Theta2.3
&\displaystyle 
\left(-\frac{1}{4},\phantom{-}\frac{7}{48}\right)\delta_{\mu\nu}
&\displaystyle \left(-1,\phantom{-}\frac{1}{2}\right)\\
\Theta1.4
&\displaystyle \left(-1,\phantom{-}\frac{1}{4}\right)\delta_{\mu\nu}
&\displaystyle \left(-4,\phantom{-}0\right)
&\Theta2.4
&\displaystyle 
\left(-\frac{1}{2},\phantom{-}\frac{1}{8}\right)\delta_{\mu\nu}
&\displaystyle \left(-2,\phantom{-}1\right)\\[0.75ex]
\hline
\end{array}
\end{displaymath}
\label{DivTheta}
\caption{Divergences of some two-loop graphs}
\end{table}

Concerning the divergences of the graphs, table~\ref{DivTheta} shows,
that the partial integration holds, since for $i=1,2$ the equations
\begin{eqnarray*}
(\Theta i.1) + (\Theta i.2) + (\Theta i.3) &=& 0\\
2\,(\Theta i.2) + (\Theta i.4 ) &=& 0
\end{eqnarray*}
are satisfied for both columns.

The following identities do not apply to graphs with contracted
derivatives. They are established with the help of some identities
of appendix~\ref{IGR} and are given by
\begin{eqnarray*}
(\Theta1.2) - 2\,(\Theta2.2) &=& 0\\
(4-d) (\Theta1.2) - (2-d) (\Theta1.3) &=& 0,
\end{eqnarray*}
whereas \eg the equation
\begin{equation*}
(6-d) (\Theta2.2) - (2-d) (\Theta2.3) \ =\ 0\\
\end{equation*}
does not hold, due to the reasons mentioned above.

\bibliographystyle{h-physrev}
\bibliography{ref}

\end{document}